\documentclass[mnsc,nonblindrev]{informs3}

\OneAndAHalfSpacedXI

\let\theoremstyle\relax
\usepackage[utf8]{inputenc}

\usepackage{amssymb,amsmath,amsthm,amsbsy}

\makeatletter
\gdef\th@TH{%
  \def\theorem@headerfont{\normalfont\TheoremHeaderFont}%
  \thm@preskip=6pt plus 3pt minus 2pt\relax
  \thm@postskip=6pt plus 3pt minus 2pt\relax
  \if@MOOR\labelsep1em\else\labelsep0.5em\fi
  \def\@begintheorem##1##2[##3]{\Hy@raisedlink{\hyper@anchorstart{\@currentHref}\hyper@anchorend}\normalfont\TheoremTextFont
    \def\@tempa{##3}%
    \item[\if@OPRE\else\hspace*{1em}\fi\hskip\labelsep
      \theorem@headerfont ##1\ ##2%
      \ifx\@tempa\@empty .\else\ {\bfseries(##3)}.\fi]}%
}
\gdef\th@THkey{%
  \def\theorem@headerfont{\normalfont\TheoremHeaderFont}%
  \thm@preskip=6pt plus 3pt minus 2pt\relax
  \thm@postskip=6pt plus 3pt minus 2pt\relax
  \if@MOOR\labelsep1em\else\labelsep0.5em\fi
  \def\@begintheorem##1##2[##3]{\Hy@raisedlink{\hyper@anchorstart{\@currentHref}\hyper@anchorend}\normalfont\TheoremTextFont
    \def\@tempa{##3}%
    \item[\if@OPRE\else\hspace*{1em}\fi\hskip\labelsep
      \theorem@headerfont \ifx\@tempa\@empty ##1\ ##2.\else {##3}\fi]}%
}
\gdef\th@EX{%
  \def\theorem@headerfont{\normalfont\ExampleHeaderFont}%
  \if@OPRE
    \thm@preskip=6pt plus 3pt minus 2pt\relax
    \thm@postskip=6pt plus 3pt minus 2pt\relax
  \else
    \thm@preskip=0pt\relax
    \thm@postskip=0pt\relax
  \fi
  \if@MOOR\labelsep1em\else\labelsep0.5em\fi
  \def\@begintheorem##1##2[##3]{\Hy@raisedlink{\hyper@anchorstart{\@currentHref}\hyper@anchorend}\normalfont\ExampleTextFont
    \def\@tempa{##3}%
    \item[\if@OPRE\else\hspace*{1em}\fi\hskip\labelsep
      \theorem@headerfont ##1\ ##2%
      \ifx\@tempa\@empty .\else\ {(##3)}.\fi]}%
}
\gdef\th@EXkey{%
  \def\theorem@headerfont{\normalfont\ExampleHeaderFont}%
  \if@OPRE
    \thm@preskip=6pt plus 3pt minus 2pt\relax
    \thm@postskip=6pt plus 3pt minus 2pt\relax
  \else
    \thm@preskip=0pt\relax
    \thm@postskip=0pt\relax
  \fi
  \if@MOOR\labelsep1em\else\labelsep0.5em\fi
  \def\@begintheorem##1##2[##3]{\Hy@raisedlink{\hyper@anchorstart{\@currentHref}\hyper@anchorend}\normalfont\ExampleTextFont
    \def\@tempa{##3}%
    \item[\if@OPRE\else\hspace*{1em}\fi\hskip\labelsep
      \theorem@headerfont \ifx\@tempa\@empty ##1\ ##2.\else {##3}\fi]}%
}
\makeatother

\usepackage{mathtools}
\usepackage{natbib}
\bibpunct[, ]{(}{)}{,}{a}{}{,}%
\def\bibfont{\small}%

\TheoremsNumberedThrough
\ECRepeatTheorems

\EquationsNumberedThrough

\MANUSCRIPTNO{}

\usepackage{graphicx,paralist}
\usepackage{thmtools}
\usepackage{tikz}
\usepackage{pgfplots}
\pgfplotsset{compat=1.10}
\usepgfplotslibrary{fillbetween}
\usetikzlibrary{backgrounds}
\usetikzlibrary{patterns}
\usetikzlibrary{shapes,decorations,arrows,calc,arrows.meta,fit,positioning,automata}
\tikzset{
    -Latex,auto,node distance =1 cm and 1 cm,semithick,
    state/.style ={ellipse, draw, minimum width = 0.7 cm,rounded corners},
    point/.style = {circle, draw, inner sep=0.04cm,fill,node contents={}},
    bidirected/.style={Latex-Latex,dashed},
    el/.style = {inner sep=2pt, align=left, sloped}
}
\usepackage{algorithm}
\usepackage[algo2e]{algorithm2e}
\usepackage[noend]{algpseudocode}
\usepackage{lscape}
\usepackage{float}
\usepackage{varwidth}
\usepackage{lipsum}
\usepackage{multirow}
\usepackage{xcolor}
\usepackage{makecell}
\usepackage{subcaption}
\usepackage{booktabs}
\usepackage{textcomp}
\usepackage{breqn}
\usepackage{bbm}
\usepackage[shortlabels]{enumitem}
\usepackage[normalem]{ulem}
\usepackage{hyperref}
\hypersetup{
    colorlinks,
    linkcolor={red!50!black},
    citecolor={blue!50!black},
    urlcolor={blue!80!black}
}
\graphicspath{{./}{pics/}}

\makeatletter
\newcommand{\AppendixNumberWithin}[1]{%
  \numberwithin{#1}{section}%
  \@ifundefined{theH#1}{}{%
    \expandafter\renewcommand\csname theH#1\endcsname{\theHsection.\arabic{#1}}%
  }%
}
\newcommand{\AppendixTheoremNumbering}{%
  \renewcommand{\theHsection}{appendix.\thesection}%
  \renewcommand{\theHsubsection}{\theHsection.\arabic{subsection}}%
  \renewcommand{\theHsubsubsection}{\theHsubsection.\arabic{subsubsection}}%
  \AppendixNumberWithin{theorem}%
  \AppendixNumberWithin{lemma}%
  \AppendixNumberWithin{proposition}%
  \AppendixNumberWithin{corollary}%
  \AppendixNumberWithin{claim}%
  \AppendixNumberWithin{conjecture}%
  \AppendixNumberWithin{hypothesis}%
  \AppendixNumberWithin{assumption}%
  \AppendixNumberWithin{remark}%
  \AppendixNumberWithin{example}%
  \AppendixNumberWithin{problem}%
  \AppendixNumberWithin{definition}%
  \AppendixNumberWithin{question}%
  \AppendixNumberWithin{answer}%
  \AppendixNumberWithin{exercise}%
  \AppendixNumberWithin{assump}%
  \AppendixNumberWithin{nrm}%
  \AppendixNumberWithin{np}%
  \AppendixNumberWithin{cb}%
  \AppendixNumberWithin{om}%
  \AppendixNumberWithin{of}%
  \AppendixNumberWithin{sdp}%
  \AppendixNumberWithin{ra}%
  \AppendixNumberWithin{fa}%
  \AppendixNumberWithin{ca}%
  \AppendixNumberWithin{sufcond}%
  \AppendixNumberWithin{obs}%
  \AppendixNumberWithin{exmp}%
}
\newcommand{\AppendixTableOfContents}{%
  \section*{Appendix Table of Contents}%
  \@starttoc{aptoc}%
}
\newcommand{\AppendixContentsOn}{%
  \let\AppendixOriginalAddcontentsline\addcontentsline
  \renewcommand{\addcontentsline}[3]{%
    \AppendixOriginalAddcontentsline{##1}{##2}{##3}%
    \def\AppendixContentsFile{##1}%
    \def\AppendixTocFile{toc}%
    \ifx\AppendixContentsFile\AppendixTocFile
      \AppendixOriginalAddcontentsline{aptoc}{##2}{##3}%
    \fi
  }%
}
\makeatother

\newcommand\norm[1]{\left\lVert#1\right\rVert}

\newlength{\algofontsize}
\newcommand{\one}{\ensuremath{\mathbf{1}}}

\newcommand{\E}{\mathbb{E}}
\newcommand{\bE}{\mathbb{E}}

\newcommand{\R}{\mathbb{R}}

\newcommand{\Var}{\mathrm{Var}}

\renewcommand{\mathbf}{\boldsymbol}
\providecommand{\argmax}{\mathop{\mathrm{arg\,max}}}

\newcommand{\cost}{\textnormal{Cost}}
\newcommand{\costcn}{\textnormal{Cost(CN)}}
\newcommand{\costdcn}{\textnormal{Cost(DCN)}}
\newcommand{\costnb}{\textnormal{Cost(NB)}}
\newcommand{\costsg}{\textnormal{Cost(SG)}}
\newcommand{\poa}{\textnormal{PoA}}

\newcommand{\ip}[2]{\left\langle #1,#2\right\rangle}

\begin{document}

\RUNAUTHOR{Anunrojwong et al.}
\RUNTITLE{Battery Operations in Electricity Markets}

\TITLE{Battery Operations in Electricity Markets: Strategic Behavior and Distortions}

\ARTICLEAUTHORS{%
\AUTHOR{Jerry Anunrojwong}
\AFF{Yale School of Management, \EMAIL{jerry.anunrojwong@yale.edu}}
\AUTHOR{Santiago R. Balseiro}
\AFF{Columbia University, Graduate School of Business, \EMAIL{srb2155@columbia.edu}}
\AUTHOR{Omar Besbes}
\AFF{Columbia University, Graduate School of Business, \EMAIL{ob2105@columbia.edu}}
\AUTHOR{Bolun Xu}
\AFF{Columbia University, Earth and Environmental Engineering, \EMAIL{bx2177@columbia.edu}}
}

\ABSTRACT{Battery storage can reduce electricity generation costs by shifting energy across time, but as privately owned batteries become large, they may also be able to exert market power. We study how this market power distorts storage decisions in a two-settlement electricity market with stochastic demand and heterogeneous generator flexibility. We compare centralized battery operations, which minimize generation cost, with decentralized battery operations, in which each battery maximizes its own profit. For a baseline model with linear inverse supply curves, we characterize equilibrium battery policies and generation costs in closed form.

Relative to centralized operations, a strategic battery distorts storage decisions in three ways: it withholds discharge, shifts participation from the day-ahead market to the real-time market, and responds too weakly to real-time demand fluctuations. These distortions raise generation cost, but the resulting efficiency loss admits tight, distribution-free bounds. We measure the resulting efficiency loss through the Price of Anarchy metric, which compares the cost reduction achieved by centralized batteries to that achieved by strategic batteries. For a single battery (without competition), the Price of Anarchy lies between $9/8$ and $4/3$; with $n$ competing batteries, the Price of Anarchy is bounded above by $1+1/(n(n+2))$. Similar bounds continue to hold in richer settings with capacity constraints, battery inefficiency, and virtual bidding. We also show why market power mitigation is subtle: interventions that target one distortion can backfire by redirecting behavior toward another and increasing system cost. Numerical experiments calibrated to California and Texas markets show that losses from a single strategic battery are meaningful but moderate, and that even limited battery competition brings the Price of Anarchy close to one across the specifications we study.
}


\maketitle

\section{Introduction}

Battery storage is becoming an increasingly important part of electricity markets. In renewable-heavy systems, batteries can shift energy across time, charging when renewable output is abundant and discharging when demand is high. In doing so, they can reduce reliance on expensive fast-ramping generators, smooth net demand, and help integrate renewable energy into the grid. As battery capacity grows, however, batteries may no longer behave as negligible price takers. In markets such as California and Texas, grid-scale batteries are often privately owned, and large batteries may acquire market power. This raises a natural question:
\begin{quote}
    \textit{How do batteries operate in electricity markets, and how does the strategic behavior of decentralized batteries distort decisions relative to centralized batteries?}
\end{quote}

The question is especially important because renewable generation creates a growing mismatch between the timing of electricity supply and demand. Solar output is highest around midday, whereas demand often peaks in the evening, when people return home and solar generation declines. Figure~\ref{fig:duck-curve} illustrates this pattern using California data. It plots average hourly \textit{net demand}---defined as electricity demand minus renewable production---for 2019--2023. Net demand is lowest around noon and peaks around 7--8PM, and the gap between the trough and the evening peak has widened over time as solar capacity has expanded. This increasingly steep evening ramp is the well-known ``duck curve.'' Meeting that ramp requires flexible resources that can respond quickly, and batteries are a natural candidate.

\begin{figure}[h!]
    \centering

    \pgfplotstableread[col sep=space]{
    hour y2019 y2020 y2021 y2022 y2023
    0 20.334 20.594 20.317 20.909 20.613
    1 19.356 19.601 19.394 19.962 19.706
    2 18.741 18.955 18.797 19.293 19.069
    3 18.499 18.690 18.551 18.992 18.773
    4 18.807 18.923 18.801 19.214 19.000
    5 19.893 19.810 19.728 20.099 19.880
    6 21.195 20.666 20.673 20.938 20.647
    7 20.068 19.038 18.876 18.590 18.035
    8 16.969 15.795 15.051 14.079 13.150
    9 15.130 13.958 12.767 11.576 10.348
    10 14.387 13.233 11.738 10.527 8.890
    11 14.168 13.137 11.408 10.147 8.194
    12 14.317 13.471 11.531 10.275 8.152
    13 14.908 14.194 12.073 10.932 8.670
    14 15.766 15.269 13.092 12.073 9.846
    15 17.373 17.244 15.010 14.189 12.109
    16 19.833 20.178 18.135 17.533 15.603
    17 22.808 23.362 21.721 21.338 19.615
    18 25.526 25.970 24.913 24.821 23.431
    19 27.009 27.235 26.505 26.704 25.689
    20 26.823 26.828 26.220 26.489 25.663
    21 25.650 25.615 25.147 25.534 24.865
    22 23.735 23.809 23.430 23.921 23.347
    23 21.756 21.951 21.623 22.169 21.698
    }\duckcurvedata

    \begin{tikzpicture}[every path/.style={-}]
    \begin{axis}[
        width=0.78\linewidth,
        height=0.50\linewidth,
        title={California's Hourly Net Demand By Year},
        title style={font=\large},
        xlabel={Hour of the Day},
        ylabel={Mean Net Demand (GW)},
        xlabel style={font=\large},
        ylabel style={font=\large},
        tick label style={font=\large},
        xmin=-0.2, xmax=23.2,
        ymin=7, ymax=28,
        xtick={0,4,8,12,16,20},
        xticklabels={00:00,04:00,08:00,12:00,16:00,20:00},
        xticklabel style={rotate=45, anchor=east},
        axis background/.style={fill=white},
        axis line style={black, -},
        tick style={black, -, line width=0.6pt},
        major tick style={black, -, line width=0.6pt},
        minor tick style={black, -, line width=0.6pt},
        grid=major,
        major grid style={dashed, gray!55, -},
        tick pos=left,
        legend style={
            at={(0.97,0.03)},
            anchor=south east,
            fill=white,
            draw=gray!60,
            font=\large,
            cells={align=left}
        },
        every axis plot/.append style={
            line width=1.1pt,
            mark=none,
            -
        }
    ]

    \addplot[blue, dashed] table[x=hour, y=y2019] {\duckcurvedata};
    \addlegendentry{2019}

    \addplot[green!60!black, dotted] table[x=hour, y=y2020] {\duckcurvedata};
    \addlegendentry{2020}

    \addplot[cyan!70!black, dashdotted] table[x=hour, y=y2021] {\duckcurvedata};
    \addlegendentry{2021}

    \addplot[magenta, dashdotted] table[x=hour, y=y2022] {\duckcurvedata};
    \addlegendentry{2022}

    \addplot[red, solid] table[x=hour, y=y2023] {\duckcurvedata};
    \addlegendentry{2023}

    \end{axis}
    \end{tikzpicture}

    \caption{California's ``duck curve,'' hourly mean net demand by year, 2019--2023. The net demand is the energy demand minus renewable production. (Source: CAISO)}
    \label{fig:duck-curve}
\end{figure}

Battery deployment is now large enough that strategic behavior is no longer merely hypothetical.
California and Texas had 15.07 GW and 15.75 GW of battery capacity, respectively, as of March 2026 \citep{battery-capacity-eia-860m}.\footnote{For context, California and Texas electricity demand on a typical day is roughly 20--40 GW and 40--70 GW, respectively, so battery capacity is already a substantial fraction of load in some hours.} System operators have also begun to observe strategic battery behavior in practice. The Australian Energy Regulator documented strategic rebidding by a 100MW/150MWh battery during tight market conditions on March 16--17, 2023 \citep{australia-strategic-battery-report,parkinson-big-strategic-battery}. After a generator outage and a change in forecast price, the battery rebid from the price floor up to \$10{,}000/MWh and \$15{,}000/MWh, respectively, and set the market price. The regulator concluded that the episode highlighted the market power batteries may be able to exercise at certain times. A related day-ahead-to-real-time pattern also appears in routine market data. California's special report on battery storage documents that average battery discharge bids lie far above prevailing prices in the day-ahead market but much closer to prices in real time \citep{caiso-battery-report}. This pattern is consistent with batteries shifting participation from day-ahead to real time; our model shows how such a shift can raise generation cost.

Electricity markets are especially susceptible to market power because supply and demand must balance in real time at each location. Transmission constraints fragment the grid into local markets, so even in a system with many batteries, an individual battery can wield significant market power in its region. California's grid operator, CAISO, approved a \$7.3 billion transmission plan to integrate new renewable generation while maintaining reliability \citep{caiso-transmission-plan-2023}. The economic importance of these constraints is visible in the large price differences observed across locations: on May 27, 2024 at noon, the California real-time ``base'' price was about \$4/MWh, while congestion prices in some regions reached \$120/MWh.

Batteries also differ from conventional generators in a way that complicates market-power monitoring. A generator's bid is largely disciplined by observable physical and operational constraints, such as fuel costs, heat rates, and startup costs. A battery's bid, by contrast, is shaped not only by physical constraints but also by intertemporal opportunity cost: charging or discharging now changes the value of future actions. As a result, it is less obvious what constitutes a ``reasonable'' battery bid, what form strategic behavior will take, and how such behavior affects system performance.

We study these questions in a tractable model of a two-settlement electricity market with stochastic demand and heterogeneous generator ramp speeds. We compare three regimes: no battery, a centralized battery operated to minimize total generation cost, and a decentralized battery operated to maximize profit. This comparison reveals how strategic battery behavior distorts both day-ahead planning and real-time balancing relative to the system-optimal benchmark. We then quantify the resulting inefficiency, study how competition among batteries mitigates it, analyze when market power mitigation can backfire, and calibrate the model using market data from California and Texas.

\subsection{Summary of Main Contributions}
\label{subsec:summary-main-contributions}

Our contributions are fourfold.

\paragraph{Model and Mechanism.}
We develop a stochastic $T$-period model of battery market power in a two-settlement electricity market. Hourly net demand may follow an arbitrary joint distribution, and real-time battery decisions are nonanticipative: they can depend on the demand history observed up to that point, but not on future realizations. Market-clearing prices are determined endogenously from the supply of slow generators, which commit in the day-ahead market, and fast generators, which can also adjust in real time. Under linear inverse supply, we derive the centralized and decentralized battery policies in closed form (Theorems~\ref{thm:no-battery}--\ref{thm:sb-battery}).

Comparing the two operating regimes identifies three distinct effects of battery market power. Relative to centralized operation, a strategic battery withholds total discharge, shifts participation from the day-ahead market toward the real-time market, and responds too weakly to realized demand shocks. The composition of these distortions depends on generator flexibility. When fewer generators can adjust in real time, quantity withholding is relatively more important; when more generation is flexible, the distortion shifts toward delaying participation from day-ahead to real time. Thus, beyond showing that battery market power raises generation cost, the model identifies how it distorts physical battery operation and participation across settlements.

\paragraph{Tight Welfare Bounds and Competition.}
We quantify the resulting efficiency loss using the Price of Anarchy (PoA), defined as the ratio of the generation-cost reduction achieved by centralized batteries to that achieved by strategic batteries, relative to the no-battery benchmark. Decentralized battery operation weakly lowers generation cost relative to the no-battery benchmark but cannot achieve a greater cost reduction than centralized operation. With one strategic battery, we establish in Theorem~\ref{thm:cost-comparison} the tight, distribution-free bounds
\[
    \frac{9}{8}\leq \poa\leq\frac{4}{3}.
\]
These bounds accommodate arbitrary correlation in net demand and history-dependent real-time policies. The proof decomposes the value created by storage into predictable intertemporal variation and real-time variation that can be smoothed subject to the information structure. An orthogonality property of the centralized real-time policy then reduces the welfare comparison to two nonnegative components, and the PoA becomes a convex combination of their corresponding inefficiency ratios. This reduction yields the tight bounds without imposing a particular distribution of demand.

Equivalently, a strategic battery achieves between $75\%$ and $88.9\%$ of the generation-cost savings achieved under centralized operation. The loss created by battery market power is therefore meaningful but bounded, even in the most concentrated case.

We then introduce competition among $n$ batteries and derive the unique equilibrium in closed form (Theorem~\ref{thm:competition}). The corresponding PoA satisfies
\[
1+\frac{1}{n(n+1)(n^2+n+2)}
\leq \poa \leq
1+\frac{1}{n(n+2)}.
\]
The worst-case excess PoA therefore declines at rate $1/n^2$. Even limited competition sharply reduces the efficiency loss: competition disciplines quantity withholding, the shift from day-ahead to real time, and insufficient real-time responsiveness simultaneously.

\paragraph{Market Power Mitigation and Robustness.}
We study two natural market power mitigation policies, each targeting a different strategic distortion. The first requires each battery's expected real-time discharge to be zero, directly eliminating the observable shift from day-ahead to real time. This intervention backfires: batteries substitute toward greater quantity withholding, battery profits fall, and generation cost rises (Theorem~\ref{thm:da-rt-backfire}). The second policy subsidizes expected discharge in an effort to reduce the incentive for quantity withholding. At every equilibrium, however, any reduction in physical generation cost is no greater than the subsidy expenditure (Theorem~\ref{thm:subsidy-discharge}). These results show why mitigating battery market power is difficult: a policy that targets one strategic distortion need not reduce the others. By contrast, competition reduces all three distortions, making it the more robust mitigation mechanism in our model.

We also establish that the bounded-inefficiency conclusion survives in several richer operational and strategic settings. With heterogeneous battery capacities, the PoA upper bound depends on the Herfindahl index of battery-capacity shares and reduces to the baseline competition bound when capacities are equal (Theorem~\ref{thm:hetero-capacity-bound}). The baseline PoA bounds continue to hold under imperfect round-trip efficiency, subject to a common partition of charging and discharging periods (Theorem~\ref{thm:inefficiency}). With virtual bidders, the baseline upper bound continues to hold, although additional virtual bidding can redirect batteries away from physical intertemporal arbitrage and increase PoA (Theorem~\ref{thm:sb-battery-competition}). Appendix~\ref{app:sec:more-extensions} further shows that joint day-ahead--real-time balance leaves the equilibrium and PoA bounds unchanged, ramping costs preserve the competition result and PoA bounds, and strategic generator bidding raises true generation cost without changing the battery's operating schedule (Theorems~\ref{thm:joint-balance}, \ref{thm:ramping-costs}, and~\ref{thm:strategic-generators}, respectively). Together, these results demonstrate that the main conclusions do not depend on the most stylized features of the baseline model.

\paragraph{Data-Driven Numerical Experiments.}
Finally, we conduct numerical experiments using 2024 hourly price and net demand data from California and Texas. We estimate each market's day-ahead inverse supply curve and incremental real-time price response and fit a joint distribution to complete 24-hour net demand profiles. In the baseline linear-supply setting, PoA with one strategic battery is $1.177$ in California and $1.267$ in Texas. With five competing batteries, these values fall to $1.008$ and $1.019$, respectively. In these calibrated settings, single-battery losses are therefore meaningful but moderate, and even limited competition removes most of the inefficiency.

We then impose battery power and state-of-charge constraints and replace linear inverse supply with monotone convex cubic splines. Competition remains highly effective in both experiments. For the parameter values and demand distributions induced by the California and Texas calibrations, allowing the fitted inverse supply curves to be nonlinear changes PoA only modestly. Across these richer specifications, the main quantitative conclusion is unchanged in both calibrated markets: the efficiency loss from one strategic battery is moderate, and even limited competition brings PoA close to one.

\subsection{Related Work}\label{subsec:related-work}

Our paper connects to several strands of literature at the intersection of electricity markets, storage, and market power.

\paragraph{Sequential Markets and Market Power.} Our paper is closest to the literature on market power in sequential markets, beginning with the seminal work of \cite{allaz-vila}. In that setting, producers use forward commitments strategically because the forward market changes marginal revenue in the spot market; under market power, forward trading can therefore improve both private profit and social welfare by reducing withholding. \cite{ito-reguant-aer-sequential} extend the static Cournot framework of \cite{bushnell-aer-2008} to sequential electricity markets and quantify market power with limits to arbitrage in the Iberian market. \cite{borenstein-et-al-2008-electricity} and \cite{saravia2003} study related market-power issues; \cite{strategic-load-da-rt} study strategic load allocation across settlements, while \cite{dvorkin-regression-equilibrium} studies renewable producers' equilibrium choice of forecasting models in two-stage markets. Our setting adds intertemporal storage: a battery arbitrages across both settlements and periods under stochastic demand, so strategic behavior distorts day-ahead scheduling and real-time balancing.

\paragraph{Storage, Renewables, and Ownership Structure.} A large operations literature studies renewable energy and storage; see, for example, the surveys by \cite{agrawal-yucel-renewable-energy-sourcing,parker-electric-power-industry-review,sunar-swaminathan-pom-survey}. Within that literature, \cite{sioshansi-2010-welfare-storage-ownership} shows that storage-induced price smoothing can create welfare gains even when private incentives are not aligned with system-cost minimization, and \cite{sioshansi-2014-storage-reduces-welfare} shows that storage can reduce welfare when conventional generators are strategic. \cite{peura-bunn-renewable-forward-markets} studies intermittent renewable production in the presence of a forward market, highlighting how forward commitments shape prices and welfare even without storage. \cite{acemoglu2017competition}, \cite{genc-reynolds-own-renewable}, and \cite{bahn-market-power-renewables-ownership} study how ownership structure affects competition and market power in renewable generation. \cite{zhou-negative-price-storage} studies storage and energy disposal under negative prices, and \cite{cruise-control-storage-market-impact} studies storage control with market impact. Our emphasis is different: rather than studying storage value or ownership in a broader aggregate environment, we isolate how a strategically operated battery in a two-settlement market departs from the system optimum and how large that efficiency loss can be.

\paragraph{Battery Bidding and Market Power in Power Systems.} Several papers optimize battery bidding across sequential or multiple electricity markets, including multimarket coordination \citep{lohndorf2023value}, hour-ahead bidding under uncertainty via approximate dynamic programming \citep{jiang2015optimal}, joint bidding and operations across multiple temporal energy markets \citep{akhavan2015joint}, and adaptive trading in continuous intraday markets \citep{bertrand2019adaptive}. Another line studies price-maker batteries in energy and ancillary-service markets using bilevel optimization, often with degradation and reserve or regulation products \citep{khalilisenobari2022optimal,garcia2025optimal,mohsenian-rad-storage-price-maker,bjorndal-storage-market-power,hartwig-kockar-strategic-storage,huang-market-mechanisms-storage,schill-kemfert-strategic-storage-hydro}. These papers are close in motivation and institutional setting, but they are mostly computational and focus on detailed bidding algorithms for a storage owner. By contrast, we use a tractable equilibrium model to isolate the mechanism: decentralized battery operations distort quantity, market timing, and real-time responsiveness, and those distortions admit exact welfare comparisons and PoA bounds.

\paragraph{Investment, Coordination, and Other Flexibility Resources.}
There is also a related operations management literature on storage investment, flexibility resources, and electricity-system coordination. \cite{wu-distributed-storage} studies centralized versus distributed storage siting, sizing, and operations in a distribution system, highlighting the distinction between operational pooling benefits and investment incentives. \cite{kaps-off-grid-sun-renewable-storage-investment}, \cite{peng-renewable-flexible-storage-friends-foes}, and \cite{kaps-marinesi-dual-storage-load-shifting} study broader joint investment-and-operations problems involving renewables, flexible conventional generation, and multiple storage technologies. These papers emphasize how different flexibility resources interact operationally and economically, including substitution and complementarity in capacity choices, coordination across technologies, and the role of operating rules in shaping investment incentives. There is also related work on flexibility beyond grid-scale batteries: \cite{agrawal-yucel-demand-response} studies demand-response program design; \cite{fattahi-peak-load,fattahi_flatten_energy_curve} study direct-load-control contracts for peak shaving and load smoothing; and \cite{ev-charging-msom,leann-perakis-gm} study how electric vehicles can provide flexibility to the grid through smart charging and charging/discharging decisions. Among these OM papers, \cite{gao-aggregating-der} is closest in spirit to ours on the market-design side: it studies how a distributed energy resource aggregator can participate in wholesale electricity markets through efficient aggregation mechanisms, and shows that such aggregation can preserve full market efficiency while reducing the market power of conventional generators. Our paper is complementary to this literature. It focuses on grid-scale battery storage, one of the main flexibility technologies currently being deployed at scale, in a two-settlement wholesale electricity market. Even when we study investment and operations, we do so through the lens of strategic wholesale-market behavior---battery arbitrage, intertemporal opportunity costs, quantity withholding, competition, mitigation, and welfare. That focus lets us characterize the equilibrium distortions created by strategic battery operations and derive sharp efficiency bounds.

\paragraph{Empirical Electricity Market Power.} There is also a large empirical literature in economics measuring market power in electricity markets; see \cite[Section 4.2]{kellogg-reguant-energy-io-survey} for a survey, and \cite{market-power-mitigation-graf} for a review of market power mitigation mechanisms. This literature mostly focuses on conventional generators. The closest battery papers are \cite{karaduman-storage} and \cite{butters-soaking-sun}. \cite{karaduman-storage} studies a single strategic storage owner calibrated to the South Australian market and, like us, documents a discrepancy between private and social incentives, but does not model the same day-ahead/real-time sequential-clearing structure. \cite{butters-soaking-sun} studies equilibrium battery investment and alternative incentive policies, using a competitive-storage baseline and also examining battery market power under monopoly and duopoly. We complement this empirical literature by providing a benchmark theory of battery market power: a tractable model that can be calibrated and that yields closed-form distortions, exact PoA bounds, and centralized-versus-decentralized welfare comparisons.

\section{Model}

This section introduces the market environment, the battery's decision problem, and the price formation process. The model is designed to be rich enough to capture the main operational features behind battery market power in two-settlement electricity markets while remaining analytically tractable.

\paragraph{Demand Process.} There are $T$ time periods in a day, indexed by $t \in \{1,2,\dots,T\}\equiv[T]$. For each period $t$, let $D_t$ denote \textit{net demand}: total electricity demand minus renewable production, before any battery charge or discharge. This is the quantity that must be met by conventional generation and battery operations. The daily net demand vector $D=(D_1,\dots,D_T)$ is drawn from a known joint distribution $\pi$, which allows both intraday uncertainty and cross-period correlation. For each $t$, define the demand history $D_{1:t}\equiv(D_1,\dots,D_t)$, the unconditional mean net demand $\mu_t\equiv\bE[D_t]$, the unconditional variance $\sigma_t^2\equiv\Var(D_t)$, and the daily average mean net demand $\bar\mu\equiv(\mu_1+\cdots+\mu_T)/T$. We assume $\mu_t>0$ for every $t$. For $t'<t$ and realized history $d_{1:t'}$, let $\mu_{t\mid d_{1:t'}}\equiv\bE[D_t\mid D_{1:t'}=d_{1:t'}]$ denote the conditional mean of period-$t$ net demand given the information available through period $t'$.

\paragraph{Two-Settlement Market.} We study the standard two-settlement market used in U.S. wholesale electricity markets. The system operator clears forecast net demand $\mu_t\equiv\bE[D_t]$ in the \textit{day-ahead} (DA) market and subsequently clears the realized deviation $D_t-\mu_t$ in the \textit{real-time} (RT) market. RT demand can be positive or negative. If it is positive, fast generators are called on to increase production; if it is negative, they reduce production. The battery also adjusts in real time after period-$t$ demand is realized.

\paragraph{Generators.} We model two types of conventional generators. \textit{Slow} generators can participate only in DA, whereas \textit{fast} generators can participate in both DA and RT. We assume a continuum of infinitesimal generators that bid their true marginal costs and follow the system operator's dispatch instructions. This modeling choice is intended to capture the shape of the inverse supply curve and its implications for clearing prices and generation cost, while abstracting from non-convexities such as start-up and no-load costs. Those omitted features are represented in reduced form through the distinction between slow and fast generators. Our generator model is adapted from \cite{strategic-load-da-rt}. Let $G_s(\lambda)$, respectively $G_f(\lambda)$, denote the mass of slow, respectively fast, generators with cost at most $\lambda$. The supply functions $G_s(\cdot)$ and $G_f(\cdot)$ are primitives of the model and are assumed to be continuous and strictly increasing.

\paragraph{Battery Participation.} We begin with a single battery, the case in which market power is strongest. Throughout, a battery denotes a single strategic decision-making unit and may represent one physical battery, a battery farm, or a portfolio of storage resources operated under common control within the same market. This benchmark is useful precisely because it is the extreme case: if inefficiency is bounded here, it will be smaller once we allow competition between batteries, which we study in Section~\ref{subsec:competition}.

Before the operating day begins, the battery chooses a DA discharge quantity $z_t^{DA}$ for each period $t \in [T]$. These are scalar decision variables. After demand is realized in period $t$, the battery chooses an \textit{incremental} RT discharge quantity $z_t^{RT}(D_{1:t})$, so that total discharge in period $t$ is $z_t^{DA} + z_t^{RT}(D_{1:t})$. Positive $z$ denotes discharge and negative $z$ denotes charge. The DA decisions are quantities, while the RT decisions are policies indexed by realized demand histories.

In the baseline model, the battery has no state-of-charge limit and no efficiency loss. We relax these assumptions in Sections~\ref{subsec:battery-inefficiency} and \ref{subsec:battery-capacity}. We start with this base here both for tractability and because it makes the battery as powerful as possible, which is the appropriate benchmark for studying market power. Although we describe the model as a single day with multiple time periods, it is best interpreted as a representative day in steady state. Given this interpretation, we will assume that
\begin{align}\label{eqn:da-rt-balance}
    \sum_{t'=1}^{T} z_{t'}^{DA} = 0, \qquad
    \sum_{t'=1}^{T} z_{t'}^{RT}(D_{1:t'}) = 0 \quad \text{for every demand realization in the support.}
\end{align}
These constraints require that the aggregate charge and aggregate discharge of the battery are equal each day, in DA and RT, which allow us to decouple the analysis of battery operations across representative days. These assumptions are also justified by the observed operations of batteries; see Section~\ref{subsec:discussion-assumptions}. We also study the joint DA+RT balance requirement and find that at optimum, both DA and RT are balanced separately, as in \eqref{eqn:da-rt-balance}; see Appendix~\ref{subsec:alternative-balance}.

\paragraph{Price Formation Process.} In period $t$, the DA demand cleared by the system operator is the forecasted mean $\mu_t$. After battery participation, the day-ahead quantity served by conventional generators and the incremental real-time adjustment they must supply are
\begin{align*}
    d_t^{DA} &= \mu_t - z_t^{DA}, \\
    d_t^{RT}(D_{1:t}) &= D_t - \mu_t - z_t^{RT}(D_{1:t}).
\end{align*}
Here $d_t^{DA}$ is the quantity of conventional generation cleared in the day-ahead market after accounting for the battery's day-ahead action, and $d_t^{RT}$ is the incremental adjustment made by fast generators in real time after accounting for the battery's real-time action.

The DA price $\lambda_t^{DA}$ is the market-clearing price at which generators with cost below the price exactly meet DA net demand:
\begin{align}\label{eqn:da-fn-price-eq-demand}
    G_s(\lambda_t^{DA}) + G_f(\lambda_t^{DA}) = d_t^{DA} .
\end{align}
In RT, slow generators cannot adjust output, so only fast generators move. The RT price $\lambda_t^{RT}$ therefore satisfies
\begin{align}\label{eqn:rt-fn-price-eq-demand}
    G_s(\lambda_t^{DA}) + G_f(\lambda_t^{RT}) = d_t^{DA} + d_t^{RT} .
\end{align}
Equations \eqref{eqn:da-fn-price-eq-demand} and \eqref{eqn:rt-fn-price-eq-demand} map these quantities into DA and RT prices. Generators are paid the market-clearing price for the quantities cleared in each settlement. We do not write these payments explicitly because they cancel as transfers in aggregate welfare; see the Generation Cost discussion below and Appendix~\ref{app:sec:welfare-accounting}. The RT price $\lambda_t^{RT}(D_{1:t})$ is a random variable that depends on realized demand and battery decisions. We omit the dependence on $D_{1:t}$ to simplify the notation. This pricing rule is not ad hoc; it is a stylized reduced-form version of standard U.S. locational marginal pricing. In our single-node setting without network congestion or losses, the DA and RT prices coincide with the marginal costs of serving day-ahead demand and real-time imbalances, respectively. When $d_t^{RT}=0$, the DA and RT prices coincide. Positive RT demand raises the RT price above the DA price, whereas negative RT demand lowers it.

\paragraph{Generation Cost.} Slow generators are dispatched in DA and fast generators are adjusted in RT. Total generation cost is therefore obtained by integrating marginal costs up to the relevant clearing prices:
\begin{align}\label{eqn:gen-cost}
    \sum_{t=1}^{T} \left( \int_{\lambda \leq \lambda_t^{DA}} \lambda \, dG_s(\lambda) + \bE_D \left[ \int_{\lambda \leq \lambda_t^{RT}(D_{1:t})} \lambda \, dG_f(\lambda) \right] \right),
\end{align}
where the expectation is taken with respect to the random demand. Because only fast generators can respond in RT, real-time prices are more sensitive to demand shocks than day-ahead prices. This is why volatility matters for costs in our model. The baseline expression is separable across periods, but Section~\ref{subsec:ramping-costs} extends the model to include explicit ramping costs, which break this separability.

Throughout, our objective is the physical production cost incurred by conventional generators, not market payments. Under perfectly inelastic demand, minimizing generation cost is equivalent to maximizing social welfare because the gross benefit from serving load is fixed. Market-clearing prices therefore affect transfers across participants but not total welfare in our setting. This objective is also consistent with standard day-ahead unit commitment and real-time economic dispatch formulations, which minimize system operating cost subject to physical constraints \citep{kirschen2018fundamentals, creti-economics-electricity}.\footnote{With price-elastic demand, welfare would equal consumers' gross utility minus generation cost. With perfectly inelastic demand and fixed served load, the utility term is constant and can be omitted.} Appendix~\ref{app:sec:welfare-accounting} derives the generation-cost expression from the supply functions and shows explicitly why market payments cancel from aggregate welfare.

\paragraph{Battery Operations Regimes.} We compare three regimes.
\begin{itemize}
    \item \textbf{No battery:} the system operates without storage.
    \item \textbf{Centralized battery:} the battery is controlled by the system operator to minimize generation cost.
    \item \textbf{Decentralized battery:} the battery is independently operated to maximize its own profit.
\end{itemize}
In the latter two regimes, DA and RT prices are endogenous because battery charge and discharge decisions affect market clearing.

\paragraph{Day-Ahead and Real-Time Inverse Supply Curves.} Equations \eqref{eqn:da-fn-price-eq-demand} and \eqref{eqn:rt-fn-price-eq-demand} allow flexible price-demand relationships through the primitive supply functions $G_s$ and $G_f$. To obtain tractable closed-form results, we impose additional structure. We assume that at every price $\lambda$, a fraction $k_f$ of generators are fast and a fraction $k_s = 1-k_f$ are slow. This can be interpreted both as a reduced-form description of generator technology and as a stylized representation of reserve procurement: some fast generators are kept available day-ahead so that they can move in RT if needed.

Let $G(\lambda) = G_s(\lambda) + G_f(\lambda)$ denote total supply, so that $G_s(\lambda) = k_s G(\lambda)$ and $G_f(\lambda) = k_f G(\lambda)$. Then \eqref{eqn:da-fn-price-eq-demand} and \eqref{eqn:rt-fn-price-eq-demand} imply
\begin{align}
    \lambda_t^{DA} &= G^{-1}(d_t^{DA}), \label{eqn:lmbd-DA} \\
    \lambda_t^{RT} &= G^{-1}\!\left(d_t^{DA} + \frac{1}{k_f} d_t^{RT}\right). \label{eqn:lmbd-RT}
\end{align}
We assume the inverse supply curve is linear:
\begin{align}\label{eqn:supply-curve-linear}
    G^{-1}(q) = \alpha + \beta q,
\end{align}
where $\alpha,\beta \ge 0$ are constants. Here $\alpha$ is the intercept and $\beta$ is the slope. The linear inverse-supply specification is common in the literature and is the key step that yields closed-form equilibrium expressions in our setting \citep{sioshansi-2010-welfare-storage-ownership,sioshansi-2014-storage-reduces-welfare,ito-reguant-aer-sequential}. It implies the linear pricing system
\begin{equation}\label{eqn:linear-pricing-system}
\begin{aligned}
    \lambda_t^{DA} &= \alpha + \beta d_t^{DA}, \\
    \lambda_t^{RT} &= \lambda_t^{DA} + \frac{\beta}{k_f} d_t^{RT}.
\end{aligned}
\end{equation}
This implies that RT prices are more sensitive to incremental demand shocks than DA prices because only the fast-generator share $k_f$ can adjust in real time. This is the same two-stage pricing form derived by \citet[Equations (5) and (8)]{strategic-load-da-rt}: the DA price is linear in the DA-cleared quantity, and the RT price equals the DA price plus a linear adjustment for the RT imbalance.

Under the linear inverse supply curve assumption \eqref{eqn:supply-curve-linear}, the generation cost
\eqref{eqn:gen-cost} can be written as
\begin{align}
    \sum_{t=1}^{T} k_s \left( \alpha d_t^{DA} + \frac{\beta}{2} (d_t^{DA})^2 \right) + k_f \bE_D\left[ \alpha \left( d_t^{DA} + \frac{d_{t}^{RT}}{k_f} \right) + \frac{\beta}{2} \left( d_t^{DA} + \frac{d_{t}^{RT}}{k_f} \right)^2 \right] . \label{eqn:gen-cost-linear}
\end{align}

\subsection{Discussion of Model Assumptions}\label{subsec:discussion-assumptions}

\paragraph{Net Demand Distribution.} $(D_1,\dots,D_T) \sim \pi$ is a model of \textit{net demand}, defined as total demand minus renewable generation, but does not include battery charge/discharge. Therefore, the uncertainty described in the (potentially correlated) joint distribution $\pi$ reflects both load variation and renewable production. We treat both as \textit{exogenous} and assume demand is perfectly \textit{inelastic}. This is a standard benchmark in wholesale electricity-market models because most end-use load still faces retail tariffs that do not track real-time wholesale prices, and observed short-run electricity demand is typically very inelastic \citep{joskow2006competitive,borenstein-holland-2005,csereklyei2020}. The formulation is flexible. For example, we can take $T=24$ so each period is one hour, matching day-ahead market timing; finer RT intervals could also be incorporated, but we use hourly periods for simplicity. In our model, we assume that everyone has a distributional forecast of the net demand, and this forecast is correct.

\paragraph{Non-Strategic Generators.} The assumption that conventional generators are non-strategic is empirically plausible. System operators and regulators observe engineering characteristics and fuel costs and can therefore estimate conventional generators' marginal costs reasonably well; they can also cap and penalize excessive markups. For example, California's Department of Market Monitoring reports average price-cost markups of 3.6\%, 3.1\%, and 2.5\% in 2023, 2022, and 2021, respectively \citep{caiso-annual-market-perf-2023,caiso-annual-market-perf-2022,caiso-annual-market-perf-2021}. Batteries differ in an important way: the ``correct'' charge and discharge bids depend not only on operating costs, but also on intertemporal opportunity costs and price forecasts. This also makes battery-side market-power mitigation fundamentally harder to design than generator-side mitigation: a direct prohibition on ``withholding'' would require the regulator to know the relevant intertemporal opportunity cost or reference discharge path, not just current marginal cost and available capacity. We therefore model generators as non-strategic and the battery as strategic in order to isolate the effects of battery market power. Nevertheless, we relax this assumption in Section~\ref{subsec:strategic-generators}, where we allow conventional generators to bid strategically and characterize how generator market power interacts with battery market power.

\paragraph{Battery Participation.} A within-day energy balancing constraint for the battery is a natural baseline because grid-scale batteries overwhelmingly arbitrage across hours of the same day rather than across many days. For each year, we calculate each day's average net battery output from CAISO's five-minute systemwide Batteries series, take its absolute value, average across days, and normalize by CAISO's reported aggregate participating battery power capacity. The resulting shares were only 2.9\%, 0.9\%, and 1.3\% in 2022, 2023, and 2024, respectively \citep{caiso-todays-outlook-supply,caiso-battery-capacity-2023,caiso-battery-special-report-2025}.\footnote{CAISO notes that before February 1, 2023, the Batteries trend included battery storage and all hybrids, including renewable components; afterward most hybrid resources were moved to separate Hybrid charts. The 2022 statistic is therefore not strictly comparable with the 2023 and 2024 statistics.} This focus on within-day operation is also consistent with the duration of recent U.S. utility-scale storage deployments: among electrical energy-storage capacity of at least 1 MW completed in the United States during 2010--2022, less than 7\% had a duration exceeding 4 hours \citep{nrel-4-hour-battery}.

Furthermore, although we model the battery as directly choosing quantities, this formulation also covers common market implementations. Under self-scheduling, the battery explicitly chooses quantities in each period. Under economic bidding, the battery submits price-quantity schedules, and with the battery as the only strategic player these bids can be chosen to implement any desired quantity outcome.

\begin{figure}[htbp]
    \centering
    \includegraphics[width=0.74\textwidth]{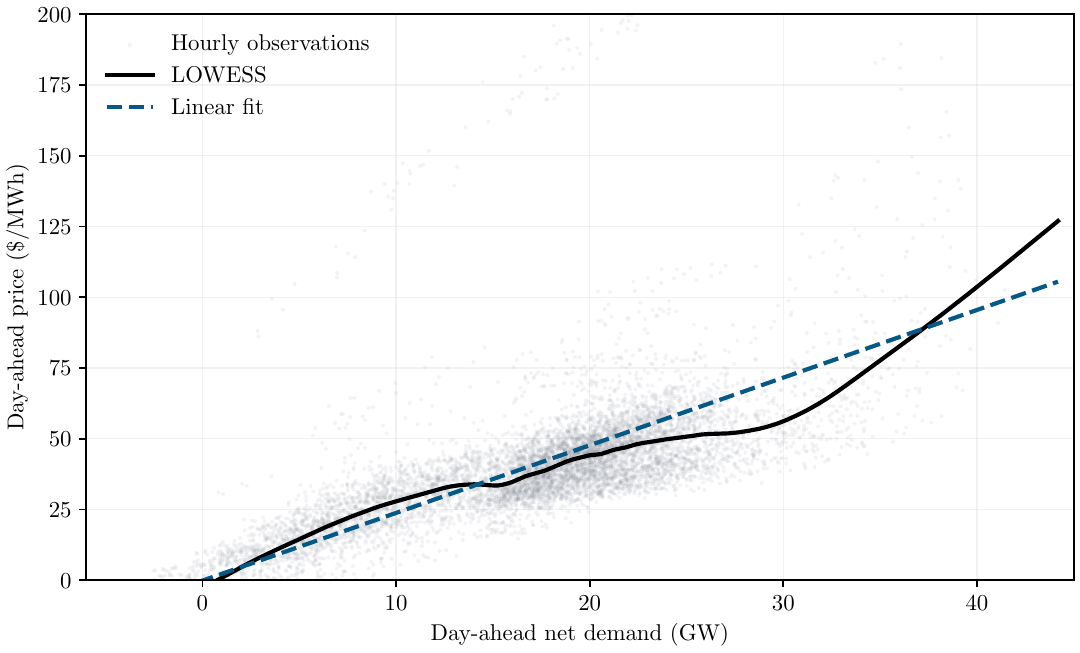}
    \caption{California inverse supply curve using 8,781 matched hourly CAISO observations from 2024. The vertical axis is truncated at \$200/MWh for readability; the LOWESS and ordinary least-squares fits use the full sample. The relationship is approximately linear over most demand levels observed in the data, but the nonparametric fit steepens in periods with unusually high demand.}
    \label{fig:demand_curve_caiso}
\end{figure}

\paragraph{Linearity of the Inverse Supply Curve.}
Figure~\ref{fig:demand_curve_caiso} shows that the California inverse supply curve is close to linear over most of the observed net demand range. The fitted curve steepens in the upper tail, so we do not treat linearity as an exact empirical description. Instead, the linear model is a parsimonious benchmark whose efficiency implications can be checked against richer supply curves. Section~\ref{sec:calibration} performs this check using convex cubic splines fitted to California and Texas data. For the parameter values and demand distributions induced by these calibrations, allowing the fitted inverse supply curve to be nonlinear changes PoA only modestly. Thus, the linear model provides a useful quantitative benchmark for the calibrated settings we study.

\section{Three Modes of Battery Operations}\label{sec:three-modes-battery}

In this section, we characterize battery operations and the resulting generation costs under three regimes: no battery (\S\ref{subsec:main-no-battery}), a centralized cost-minimizing battery (\S\ref{subsec:main-fb-battery}), and a decentralized profit-maximizing battery (\S\ref{subsec:main-sb-battery}). The results hold for any fixed distribution $\pi$ over $(D_1,\dots,D_T)$.

\subsection{No Battery Baseline}\label{subsec:main-no-battery}

In the no-battery regime, we have:

\begin{theorem}[No Battery]\label{thm:no-battery}
The generation cost under no battery $\costnb$ is given by
\begin{align*}
    \costnb =  \sum_{t=1}^{T} \left( \alpha \mu_t + \frac{\beta}{2} \mu_t^2 + \frac{\beta}{2k_f} \Var(D_t) \right) .
\end{align*}
\end{theorem}
The proof is presented in Appendix~\ref{app:subsec:main-no-battery}. This theorem provides the baseline against which we compare the centralized and decentralized battery regimes. The no-battery generation cost depends only on the marginal means $\mu_t \equiv \bE[D_t]$ and variances $\Var(D_t)$, not on the full joint distribution or intertemporal correlation of demand. Correlation is irrelevant because, without a battery, each period clears independently. Demand variability raises cost because generation cost is quadratic in demand. The parameter $k_f$ appears only through the variance term: a larger share of fast generators allows more real-time variability to be absorbed by low-cost flexible generation. Accordingly, when demand is deterministic ($\Var(D_t)=0$ for all $t$), the generation cost no longer depends on $k_f$.

\subsection{Centralized Battery Operations}\label{subsec:main-fb-battery}
We now consider centralized battery operations, in which the system operator directly dispatches the battery to minimize total generation cost. The decision variables are the day-ahead and real-time discharges $z_t^{DA}$ and $z_t^{RT}(D_{1:t})$ for each realization of the period-$t$ demand history $D_{1:t} \equiv (D_1,\dots,D_t)$, and the system operator minimizes the generation cost \eqref{eqn:gen-cost-linear} subject to  constraints $\sum_{t=1}^{T} z_t^{DA} = \sum_{t=1}^{T} z_t^{RT}(D_{1:t}) = 0$. For each $t' < t$, define $\mu_{t|d_{1:t'}} = \bE[D_t|D_{1:t'} = d_{1:t'}]$ and recall that $\mu_{t} \equiv \bE[D_t]$, $\bar{\mu} \equiv \frac{1}{T}\sum_{t=1}^{T} \mu_t$.

\begin{theorem}[Centralized Battery]\label{thm:fb-battery}
The centralized day-ahead battery discharge decisions are given by $z_t^{DA,CN} = \mu_{t} - \bar{\mu}$ for $1 \leq t \leq T$. The centralized real-time battery discharge policies are given recursively by, for each period $1 \leq t \leq T-1$,
\begin{align}\label{eqn:z-RT-CN}
    z_t^{RT,CN}(D_{1:t})
=
\frac{T-t}{T-t+1}(D_t-\mu_t)
- \frac{1}{T-t+1}\sum_{i=t+1}^{T} (\mu_{i\mid D_{1:t}}-\mu_i)
+ \frac{1}{T-t+1} B_t(D_{1:t-1}), 
\end{align}
and $z_T^{RT,CN}(D_{1:T}) = -\sum_{s=1}^{T-1} z_s^{RT,CN}(D_{1:s})$, where $B_t(D_{1:t-1}) \equiv -\sum_{s=1}^{t-1} z_s^{RT,CN}(D_{1:s})$ is the remaining real-time balance requirement entering period $t$.
\end{theorem}

The proof of Theorem~\ref{thm:fb-battery} is given in Appendix~\ref{app:sec:fb-battery} and leverages that the centralized problem is a convex quadratic optimization problem, so the optimizer is unique and can be characterized from the first-order conditions. 

Theorem~\ref{thm:fb-battery} formalizes the intuition that a centrally controlled battery smooths demand as much as the information structure allows. In day-ahead, the battery chooses $z_t^{DA,CN} = \mu_t - \bar{\mu}$, so net day-ahead demand in every period becomes $\mu_t - z_t^{DA,CN} = \bar{\mu}$. The battery therefore fully smooths predictable differences across periods by shifting energy from expected peak periods to expected off-peak periods. Real-time smoothing is more subtle because the operator chooses $z_t^{RT}(D_{1:t})$ after observing current and past demand, but before seeing future realizations. At time $t$, conditional on the observed demand $D_{1:t}$, the remaining predictable imbalance across periods $t,\dots,T$ consists of the current demand surprise $D_t-\mu_t$, the forecast updates for future periods $\mu_{i\mid D_{1:t}}-\mu_i$, and the remaining battery imbalance $B_t(D_{1:t-1})$. The centralized battery chooses $z_t^{RT,CN}(D_{1:t})$ so that the current period's post-battery residual demand equals the conditional average remaining imbalance:
\begin{align*}
    (D_t-\mu_t)-z_t^{RT,CN}(D_{1:t})
=
\frac{(D_t-\mu_t)+\sum_{i=t+1}^{T} (\mu_{i\mid D_{1:t}}-\mu_i)-B_t(D_{1:t-1})}{T-t+1}.
\end{align*}
A recursion for $z_t^{RT,CN}$ is given in \eqref{eqn:z-RT-CN}. It has the following operational interpretation. The first term absorbs most, but not all, of the current demand surprise $D_t-\mu_t$: the factor $\frac{T-t}{T-t+1}$ reflects that this surprise can still be smoothed across the current period and the $T-t$ future periods. The second term is a forecast-correlation correction. If the observed history predicts higher future demand, so that $\sum_{i=t+1}^{T}(\mu_{i\mid D_{1:t}}-\mu_i)$ is positive, the battery discharges less today and preserves energy for later; if the history instead predicts lower future demand, the battery discharges more today. The third term is an inventory or balance correction. If the battery has already discharged heavily in earlier real-time periods, then $B_t(D_{1:t-1})$ is negative and the battery discharges less today; if it has charged earlier, then $B_t(D_{1:t-1})$ is positive and the battery can discharge more today.

While we give a recursion for $z_t^{RT,CN}$ in the theorem statement because it is more interpretable, we can solve the recursion explicitly and give $z_t^{RT,CN}$ in closed form: for $1 \leq t \leq T-1$,
\begin{align}\label{eqn:z-RT-CN-closed-form}
z_t^{RT,CN}(D_{1:t}) &= \frac{(T-t)}{(T-t+1)}(D_t-\mu_t) - \sum_{t'=1}^{t-1} \frac{1}{(T-t'+1)} (D_{t'}-\mu_{t'}) \nonumber \\ &- \frac{1}{(T-t+1)} \sum_{i=t+1}^{T} (\mu_{i|D_{1:t}}-\mu_{i}) 
    + \sum_{t'=1}^{t-1} \sum_{i=t'+1}^{T} \frac{1}{(T-t')(T-t'+1)} (\mu_{i|D_{1:t'}}-\mu_{i}) .    
\end{align}

Under centralized control, predictable demand is fully smoothed in the day-ahead market. In real time, future demand remains uncertain, so realized prices need not be equal across periods. The centralized policy instead smooths each observed demand shock as much as the available information and the remaining daily balance constraint allow. This policy yields zero expected real-time arbitrage profit; together with day-ahead price equalization, it implies that the battery earns zero expected profit overall. The centralized operating policy is socially optimal, but it is not aligned with the objective of a decentralized profit-maximizing battery.

\subsection{Decentralized Battery Operations}\label{subsec:main-sb-battery} %

We now turn to a decentralized battery operated by an independent owner who chooses its day-ahead and real-time charge/discharge decisions to maximize profit rather than minimize total generation cost. As in the centralized benchmark, the battery chooses a day-ahead schedule $z_t^{DA}$ and a real-time adjustment policy $z_t^{RT}(D_{1:t})$ for each period $t$. The battery pays when it charges and is paid when it discharges, both at the market-clearing prices. Its problem is therefore
\begin{align}\label{def:zt-DCN}
    \max_{ (z_t^{DA}, z_{t}^{RT}(\cdot) )_{t=1}^{T} } \sum_{t=1}^{T} \lambda_t^{DA} z_t^{DA} + \bE\left[ \sum_{t=1}^{T} \lambda_t^{RT} z_t^{RT}  \right] ,
\end{align}
where the day-ahead and real-time prices are given by \eqref{eqn:lmbd-DA} and \eqref{eqn:lmbd-RT}, and the inverse supply curve is given by \eqref{eqn:supply-curve-linear}. The following theorem characterizes the decentralized battery policy.

\begin{theorem}[Decentralized Battery]\label{thm:sb-battery}
The decentralized battery discharge decisions are given by, for each period $t$,
\begin{align*}
    z_t^{DA,DCN} &= \frac{(2-k_f)}{(4-k_f)} (\mu_t - \bar{\mu}) \\
    z_t^{RT,DCN}(D_{1:t}) &= \frac{k_f}{(4-k_f)} (\mu_t - \bar{\mu})  + \frac{1}{2} z_t^{RT,CN}(D_{1:t})
\end{align*}
\end{theorem}

The decentralized problem is also a convex quadratic optimization problem, so the optimizer is unique and can be characterized from the first-order conditions. The proof is given in Appendix~\ref{app:sec:sb-battery}. Relative to the centralized benchmark, the theorem shows that decentralized battery operations are distorted in three ways: quantity withholding, a shift from day-ahead to real-time, and reduced real-time responsiveness. To interpret these distortions, it is useful to decompose the decentralized policy into a predictable component and a shock-responsive component. Quantity withholding and the shift from day-ahead to real-time arise in the predictable component of battery operations, whereas reduced real-time responsiveness arises in the shock-responsive component.

\paragraph{Predictable Component of Discharge.}
The first distortion is \textit{quantity withholding}. In period $t$, the decentralized battery's total expected discharge is
\[
    z_t^{DA,DCN} + \bE\left[z_t^{RT,DCN}\right] = \frac{2}{4-k_f}(\mu_t-\bar\mu),
\]
which is strictly smaller than the centrally optimal discharge $\mu_t-\bar\mu$. Intuitively, the centralized battery ``discharges too much'' from the perspective of a profit-maximizing owner: it fully smooths predictable demand differences across periods, equalizes prices, and therefore earns zero profit. The decentralized battery withholds quantity in order to soften its price impact. This withholding raises generation cost because it leaves peak demand too high and off-peak demand too low; with quadratic generation costs, that makes the predictable demand profile unnecessarily expensive.

We measure the extent of quantity withholding by the percentage shortfall of decentralized expected discharge relative to the centralized benchmark. By construction, this measure would be  zero if the decentralized battery discharges as much as the centralized battery, and one if it does not discharge at all. Based on Theorem \ref{thm:sb-battery}, it is given by
\begin{align}\label{eqn:withhold-quantity}
    \text{quantity withholding} \equiv 1 - \frac{z_t^{DA,DCN} + \bE\left[z_t^{RT,DCN}\right]}{z_t^{DA,CN} + \bE\left[z_t^{RT,CN}\right]} = \frac{2-k_f}{4-k_f} .
\end{align}
 Quantity withholding is decreasing in $k_f$. When more generators are fast, the battery can discharge more with less real-time price impact, so it needs to withhold less in order to maximize profit. The withholding percentage is $1/2=50\%$ when generators are mostly slow ($k_f\approx 0$) and $1/3\approx 33.3\%$ when generators are mostly fast ($k_f\approx 1$).

The second distortion is the \textit{shift from day-ahead to real-time}. The decentralized battery carries a positive amount of its expected discharge into real time,
\[
    \bE\left[z_t^{RT,DCN}\right] = \frac{k_f}{4-k_f}(\mu_t-\bar\mu),
\]
whereas the centralized battery has zero expected real-time discharge. Once the battery has already committed some quantity in the day-ahead market, its residual market power in real time changes, so splitting output across the two settlements can increase profit. Put differently, because the two markets clear separately, the battery benefits from spreading its quantity across them and thereby reducing the adverse price impact in each one. This is analogous to the role of forward trading in \cite{allaz-vila}, but here the mechanism operates through battery participation across the day-ahead and real-time electricity markets.

This shift to real time is a structural consequence of sequential market clearing itself: it arises even without demand randomness and even without assuming different demand elasticities across the two markets. What $k_f$ changes is the \emph{extent} of the shift. We measure it by the share of expected discharge that occurs in real time:
\begin{align}\label{eqn:withhold-time}
    \textnormal{shift from day-ahead to real-time} \equiv \frac{\bE\left[z_t^{RT,DCN}\right]}{ z_t^{DA,DCN} + \bE\left[z_t^{RT,DCN}\right]}  = \frac{k_f}{2} .
\end{align}
The shift from day-ahead to real-time is therefore increasing in $k_f$. The shift percentage is $0\%$ when generators are mostly slow ($k_f\approx 0$) and $50\%$ when generators are mostly fast ($k_f\approx 1$). Intuitively, when more generators are fast, the real-time price impact is smaller, so real-time participation becomes more attractive. If almost all generators are slow, real-time price impact is so large that the battery prefers to exercise market power mainly through quantity withholding instead.

Taken together, these two predictable distortions already appear even without demand uncertainty. A decentralized battery both withholds total quantity and strategically delays part of its discharge to real time. More fast generators tilt this trade-off toward shifting across markets, whereas more slow generators tilt it toward withholding total quantity. Table~\ref{table:non-random-sb-battery} summarizes these distortions, together with the real-time responsiveness distortion discussed next.

\begin{table}[ht]
\centering
\begin{tabular}{ c || c | c | c | c   }
 \textbf{regime} & \begin{tabular}{@{}c@{}}generator\\ composition\end{tabular}  & \begin{tabular}{@{}c@{}}quantity\\withholding\end{tabular} & \begin{tabular}{@{}c@{}}shift from\\DA to RT\end{tabular} & \begin{tabular}{@{}c@{}}reduction in\\RT responsiveness\end{tabular} \\ \hline 
 \multirow{4}{*}{\textbf{decentralized}} 
 & \begin{tabular}{@{}c@{}} slow gen. \\ dominate \\ ($k_f \approx 0$)\end{tabular}  & $50\%$ & $0\%$ & $50\%$  \\  \cline{2-5}
 & \begin{tabular}{@{}c@{}} fast gen. \\ dominate \\ ($k_f \approx 1$)\end{tabular}  & $33.3\%$ & $50\%$ & $50\%$ \\ \hline
 \textbf{centralized} & \begin{tabular}{@{}c@{}} centrally \\ optimal \end{tabular} & $0\%$ & $0\%$ & $0\%$   \\
\end{tabular}
\caption{Strategic distortions of the battery as a function of generation composition.}
\label{table:non-random-sb-battery}
\end{table}

\paragraph{Shock-Responsive Component of Discharge.}
Theorem~\ref{thm:sb-battery} also makes the real-time responsiveness distortion transparent. The shock-responsive component of the decentralized real-time policy is
\[
    z_t^{RT,DCN}(D_{1:t}) - \bE\left[z_t^{RT,DCN}(D_{1:t})\right]
    = \frac{1}{2}\left(z_t^{RT,CN}(D_{1:t}) - \bE\left[z_t^{RT,CN}(D_{1:t})\right]\right).
\]
The decentralized battery therefore responds to realized demand fluctuations with exactly one half of the centralized responsiveness. Equivalently, every realized-demand term and every correlation-correction term in the centralized real-time policy appears with coefficient $1/2$ in the decentralized policy. We therefore define \textit{reduction in real-time responsiveness} as
\begin{align}\label{eqn:withhold-responsiveness}
    \textnormal{reduction in real-time responsiveness} \equiv 1 - \frac{z_t^{RT,DCN} - \bE[z_t^{RT,DCN}] }{z_t^{RT,CN} - \bE[z_t^{RT,CN}]} = \frac{1}{2} .
\end{align}
So decentralized operations always reduce real-time responsiveness by exactly $50\%$, regardless of the generator composition $k_f$. Intuitively, this is the real-time analogue of quantity withholding: the battery provides too little balancing service in response to realized shocks. In particular, it responds only half as strongly both to the current demand surprise and to the forecast-update terms $\mu_{i\mid D_{1:t}}-\mu_i$, which capture what current demand reveals about future demand. As in the centralized case, these forecast-update terms vanish when future demand is conditionally independent of current demand.

\section{Comparing Generation Costs Across Different Regimes}\label{sec:compare-gencost}

We now compare generation costs across the three regimes studied so far: no battery, a centralized battery, and a decentralized battery. Let $\costnb$, $\costcn$, and $\costdcn$ denote the corresponding generation costs. These comparisons quantify both the value created by battery storage relative to the no-battery benchmark and the efficiency loss created by strategic battery operations relative to centralized control.

We measure this efficiency loss using a Price of Anarchy metric defined by the relative cost reduction from the no-battery benchmark under centralized versus decentralized battery operations:
\begin{align}\label{eqn:poa}
    \poa \equiv \frac{\costnb - \costcn}{\costnb - \costdcn} . \tag{PoA}
\end{align}
We use this metric only when centralized battery operation creates strictly positive value relative to the no-battery benchmark, $\costnb>\costcn$; otherwise, there is no storage-created value to compare.
This normalization is useful because some generation cost remains unavoidable even under centralized control: the battery can reduce system cost, but it cannot eliminate it. The numerator and denominator in \eqref{eqn:poa} therefore isolate the cost reductions generated by battery operations relative to the status quo without storage. A priori, it is not obvious that the denominator is always positive, since a strategic battery might in principle raise cost relative to the no-battery benchmark. The next theorem shows that this does not happen: decentralized battery operations still lower generation cost relative to no battery, although they do so less than centralized operations. Consequently, whenever $\costnb>\costcn$, the denominator in \eqref{eqn:poa} is positive and $\poa\ge1$.

\begin{theorem}[Cost Comparisons]\label{thm:cost-comparison}
We have $\costnb \geq \costdcn \geq \costcn$. If $\costnb>\costcn$, then $\costnb>\costdcn$ and
\begin{align*}
    \frac{9}{8} \leq \poa \leq \frac{4}{3} ,
\end{align*}
and $\poa$ is decreasing in $k_f$. Both bounds are tight.
\end{theorem}

\begin{proof}[Proof of Theorem~\ref{thm:cost-comparison}]
The proof separates the value of the battery into two sources: predictable variation in average demand across periods and real-time variation that the centralized battery can smooth. Define
\[
    A:=\sum_{t=1}^{T}(\mu_t-\bar\mu)^2,
    \qquad
    S:=\sum_{t=1}^{T}\E\!\left[(D_t-\mu_t)z_t^{RT,CN}(D_{1:t})\right].
\]
We first prove that $S\ge0$. After substituting $z_t^{DA,CN} = \mu_t - \bar{\mu}$, the centralized RT policy minimizes
\[
    \sum_{t=1}^{T}\E\!\left[\left(D_t-\mu_t-z_t^{RT}(D_{1:t})\right)^2\right]
\]
subject to the pathwise balance constraint $\sum_{t=1}^{T}z_t^{RT}(D_{1:t})=0$. Scaling the optimal policy by any scalar $\theta$ preserves feasibility, so the first-order condition along this direction gives
\begin{align}\label{eqn:centralized-rt-orthogonality}
    S=\sum_{t=1}^{T}\E\!\left[(D_t-\mu_t)z_t^{RT,CN}(D_{1:t})\right]
    =
    \sum_{t=1}^{T}\E\!\left[(z_t^{RT,CN}(D_{1:t}))^2\right]\ge 0.
\end{align}

Now substitute the policies from Theorems~\ref{thm:no-battery}--\ref{thm:sb-battery} into the generation cost \eqref{eqn:gen-cost-linear}. For the decentralized regime,
\[
    z_t^{DA,DCN}=\frac{2-k_f}{4-k_f}(\mu_t-\bar\mu),
    \qquad
    z_t^{RT,DCN}=\frac{k_f}{4-k_f}(\mu_t-\bar\mu)+\frac{1}{2}z_t^{RT,CN}(D_{1:t}).
\]
Using $\sum_t(\mu_t-\bar\mu)=0$, $\E[D_t-\mu_t]=0$, $\E[z_t^{RT,CN}(D_{1:t})]=0$, and \eqref{eqn:centralized-rt-orthogonality}, the cross terms vanish and the two cost reductions simplify to
\begin{align}
    \costnb-\costcn
    &=
    \beta\left[
        \frac{1}{2}A+\frac{1}{2k_f}S
    \right],
    \label{eqn:cost-gap-cn}\\
    \costnb-\costdcn
    &=
    \beta\left[
        \frac{12-5k_f+k_f^2}{2(4-k_f)^2}A+\frac{3}{8k_f}S
    \right].
    \label{eqn:cost-gap-dcn}
\end{align}
These two identities are the core of the comparison. Since $A\ge0$ and $S\ge0$, \eqref{eqn:cost-gap-dcn} implies $\costnb\ge\costdcn$. Also, centralized control minimizes generation cost over the same feasible battery policies, so $\costdcn\ge\costcn$.

If $A=S=0$, the battery creates no cost-reduction value. Otherwise, let
\[
c_A:=\frac{12-5k_f+k_f^2}{2(4-k_f)^2},\quad
c_S:=\frac{3}{8k_f},\quad
w_A:=\frac{c_A A}{c_A A+c_S S},\quad
w_S:=\frac{c_S S}{c_A A+c_S S}.
\]
Then \eqref{eqn:cost-gap-cn}--\eqref{eqn:cost-gap-dcn} imply
\[
\poa
=
w_A\frac{1/2}{c_A}
+
w_S\frac{1/(2k_f)}{c_S}.
\]
Since $A,S\ge0$, the weights $w_A,w_S$ are nonnegative and sum to one. Thus $\poa$ is a convex combination of $(4-k_f)^2/(12-5k_f+k_f^2)$ and $4/3$.
The first ratio lies between $9/8$ and $4/3$ for $k_f\in(0,1)$, so $9/8\le\poa\le4/3$. The same expression also shows that $\poa$ is weakly decreasing in $k_f$.

Both bounds are best possible. For the upper bound, take equal means, so $A=0$, and let there be nonzero real-time variation across periods, so $S>0$; then $\poa=4/3$. For the lower bound, take deterministic but unequal means, so $A>0$ and $S=0$; then $\poa=(4-k_f)^2/(12-5k_f+k_f^2)$, which approaches $9/8$ as $k_f\uparrow1$.
\end{proof}

Theorem~\ref{thm:cost-comparison} gives the welfare ranking and shows that the efficiency loss from battery market power is bounded in a distribution-free way. The Price of Anarchy always lies between $9/8$ and $4/3$, so strategic battery operations always create inefficiency, but the worst-case loss remains limited. Equivalently, relative to the no-battery benchmark, decentralized battery operations capture between $75\%$ and $88.9\%$ of the cost savings delivered by centralized operations. $\poa$ is decreasing in $k_f$: when more generators are fast, real-time battery withholding has less price impact, so the misalignment between private incentives and system costs becomes smaller. Sections~\ref{subsec:competition}~and~\ref{sec:extensions} show that bounded inefficiency continues to hold in a range of richer settings.

\subsection{Extension to the Case of Competing Batteries}\label{subsec:competition}

So far, we have focused on a single battery in order to isolate the basic mechanism of battery market power. We now allow $n$ batteries to compete in the same market. Let $\mathcal Z$ denote the set of admissible battery policies satisfying the balance constraints in \eqref{eqn:da-rt-balance}: a policy $z_b\in\mathcal Z$ consists of DA quantities $z_{b,t}^{DA}$ and nonanticipative RT maps $z_{b,t}^{RT}(D_{1:t})$. Formally, for fixed $n$, each battery $b\in\{1,\ldots,n\}$ is a player with strategy set $\mathcal Z$; this is a Cournot game in policies because each strategy specifies day-ahead and real-time discharge quantities. For a policy profile $z=(z_1,\ldots,z_n)$, aggregate battery discharge enters the residual DA and RT demands,
\[
d_t^{DA}=\mu_t-\sum_{b=1}^n z_{b,t}^{DA},
\qquad
d_t^{RT}=D_t-\mu_t-\sum_{b=1}^n z_{b,t}^{RT}(D_{1:t}),
\]
and prices $\lambda_t^{DA}$ and $\lambda_t^{RT}(D_{1:t})$ are determined by \eqref{eqn:lmbd-DA}--\eqref{eqn:lmbd-RT}. Battery $b$'s payoff is
\[
\Pi_b(z_b,z_{-b})
=
\sum_{t=1}^{T}\lambda_t^{DA} z_{b,t}^{DA}
+
\E\!\left[\sum_{t=1}^{T}\lambda_t^{RT}(D_{1:t}) z_{b,t}^{RT}(D_{1:t})\right],
\]
where prices are evaluated at the full profile $(z_b,z_{-b})$. A Nash equilibrium is a profile $z^*$ such that, for every battery $b$ and every alternative policy $z_b\in\mathcal Z_b$,
\[
\Pi_b(z_b^*,z_{-b}^*)\ge \Pi_b(z_b,z_{-b}^*).
\]

\begin{theorem}\label{thm:competition}
Consider $n$ competing batteries. There is a unique equilibrium given by, for each battery $1 \leq b \leq n$, and for $1 \leq t \leq T-1$,
\begin{align*}
z_{b,t}^{DA,DCN} &= \frac{(n+1-k_f)}{((n+1)^2- n k_f)}(\mu_{t} - \bar{\mu})  \\
z_{b,t}^{RT,DCN}(D_{1:t}) &= \frac{k_f}{((n+1)^2-n k_f)} (\mu_{t}-\bar{\mu})   + \frac{1}{(n+1)} z_t^{RT,CN}(D_{1:t})
\end{align*}
For every market, we have the PoA bounds
\begin{align*}
1 + \frac{1}{n(n+1)(n^2+n+2)} \leq \textnormal{PoA} \leq 1 + \frac{1}{n(n+2)}  .
\end{align*}
\end{theorem}

Theorem~\ref{thm:competition} extends the single-battery benchmark: setting $n=1$ recovers the decentralized single-battery formulas in Theorems~\ref{thm:sb-battery} and~\ref{thm:cost-comparison}. The equilibrium exists, is unique, and is symmetric, so each battery chooses the same day-ahead and real-time quantities. Because the model has no battery capacity constraints, the centralized benchmark is unchanged as $n$ varies; competition affects the decentralized equilibrium incentives and how aggregate discharge is divided across batteries.

By symmetry, aggregate decentralized discharge equals $n\left(z_{1,t}^{DA,DCN} + \bE\left[z_{1,t}^{RT,DCN}\right]\right)$. Comparing this with the centralized benchmark $z_t^{DA,CN} + \bE\left[z_t^{RT,CN}\right]$ yields the corresponding distortions:
\begin{align*}
\textnormal{quantity withholding} &= \frac{n+1-nk_f}{(n+1)^2-nk_f}, \\
\textnormal{shift from day-ahead to real-time} &= \frac{k_f}{n+1}, \\
\textnormal{reduction in real-time responsiveness} &= \frac{1}{n+1}.
\end{align*}
All three distortions therefore decay at rate $1/n$ as the number of competing batteries increases. The PoA bounds sharpen this point further: the worst-case efficiency loss decays at rate $1/n^2$. Even a moderate amount of competition therefore goes a long way toward eliminating the efficiency loss created by battery market power. The obvious caveat is that competition also reduces battery profit, which may weaken entry incentives.

\section{Market Power Mitigation Mechanisms}\label{sec:market-power-mitigation}

\subsection{Regulating Day-Ahead versus Real-Time Discrepancy}\label{subsec:da-vs-rt}

\begin{figure}[htbp]
    \centering
    \begin{subfigure}{0.48\textwidth}
        \centering
        \includegraphics[width=\linewidth]{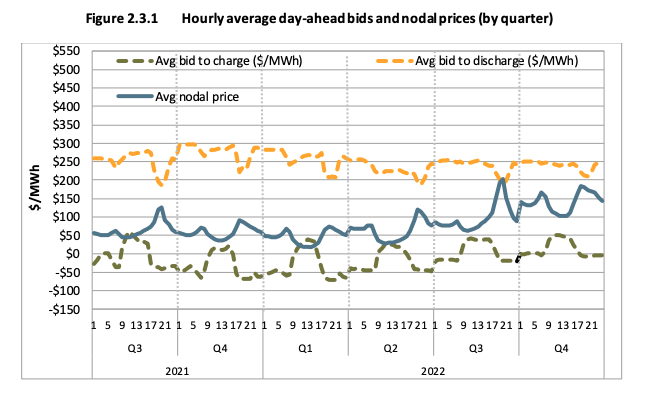}
        \caption{Hourly average DA battery bids and prices.}
        \label{fig:battery-bids-da}
    \end{subfigure}\hfill
    \begin{subfigure}{0.48\textwidth}
        \centering
        \includegraphics[width=\linewidth]{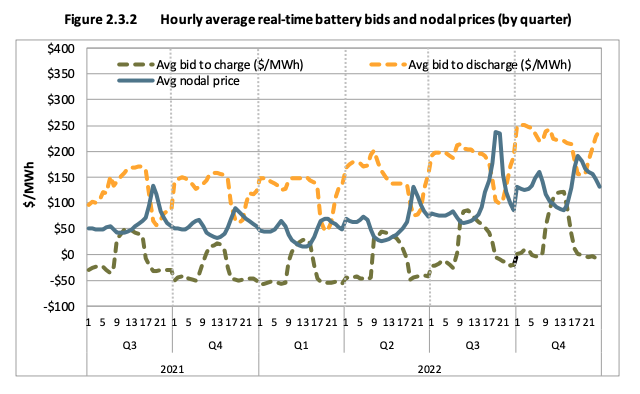}
        \caption{Hourly average RT battery bids and prices.}
        \label{fig:battery-bids-rt}
    \end{subfigure}
    \caption{California battery bids and nodal prices by market. In day-ahead, discharge bids are far above prices; in real-time, bids track prices much more closely. Source: \citet{caiso-battery-report}.}
    \label{fig:battery-bids}
\end{figure}

Of the three distortions identified above, the shift from day-ahead to real-time is perhaps the easiest to see in market data. California's special report on battery storage \citep{caiso-battery-report} documents hourly average battery bids and nodal prices in both the day-ahead and real-time markets; Figure~\ref{fig:battery-bids} reproduces the corresponding patterns. In day-ahead, average discharge bids lie far above prevailing nodal prices, helping batteries avoid being scheduled. In real-time, those bids move much closer to real-time prices, so the same capacity becomes available later. This pattern is consistent with the day-ahead-to-real-time shift highlighted by our model.

A natural question is whether a regulator could directly prohibit battery ``withholding,'' much as organized markets mitigate conventional-generator market power. CAISO already applies local market power mitigation to batteries, but its Department of Market Monitoring reports that these procedures currently have minimal impact on battery dispatch and recommends that default energy bids vary by hour to reflect discharge opportunity costs \citep{caiso-dmm-update-2025}. This experience illustrates why battery mitigation is difficult. As CAISO's 2024 Special Report explains, ``Battery resources do not submit energy price bids solely based on the actual costs of providing energy'' \citep{caiso-battery-special-report-2025}; their bids must also reflect the intertemporal opportunity cost of using limited stored energy now rather than preserving it for a higher-valued hour. A direct no-withholding rule would therefore require the regulator to know the relevant reference discharge path, which is exactly the difficult object in this environment. We therefore study a stylized restriction that directly eliminates predictable shifting from day-ahead to real-time rather than imposing a hard no-withholding constraint itself. The rule requires expected real-time discharge to equal zero but leaves batteries free to respond in real time to unexpected demand fluctuations. Theorem~\ref{thm:da-rt-backfire} shows that this restriction is counterproductive.

\begin{theorem}\label{thm:da-rt-backfire}
Consider $n$ competing batteries and the regulatory intervention that requires each battery's expected real-time discharge to be zero in every period, namely, $\bE[z_{b,t}^{RT}] = 0$ for every battery $b$ and every period $t$. This intervention increases quantity withholding, lowers each battery's profit, and raises system cost.
\end{theorem}

The reason is that strategic batteries can exercise market power in multiple ways. If the regulator compresses the day-ahead-to-real-time distortion, batteries substitute toward the more harmful distortion of pure quantity withholding. Reducing one visible distortion therefore worsens another, and the net effect is lower welfare and lower battery profit.

\subsection{Battery Discharge Subsidy}\label{subsec:discharge-subsidy}

A second policy takes a qualitatively different approach. Rather than restricting market timing, it changes batteries' financial incentive to discharge through an ex ante expected-discharge subsidy. Before demand is realized, the regulator commits to time-varying subsidy rates $s_t\ge 0$ and pays battery $b$
\[
s_t[p_{b,t}]_+,
\qquad
p_{b,t}:=z_{b,t}^{DA}+\bE[z_{b,t}^{RT}],
\]
where $[x]_+:=\max\{x,0\}$.

\begin{theorem}\label{thm:subsidy-discharge}
Consider $n$ competing batteries and any nonnegative subsidy vector. At every equilibrium of the subsidized game, total financial cost---defined as system generation cost plus subsidy payments---is no lower than under no subsidy.
\end{theorem}

Here total financial cost is a fiscal measure: it adds subsidy expenditure to physical generation cost. Subsidy payments remain transfers in aggregate welfare, but they must be financed by the regulator or load. The theorem therefore asks whether the reduction in generation cost can exceed the fiscal cost of the subsidy. At any equilibrium, it cannot. This does not mean such a subsidy can never be desirable: it may still be justified by objectives outside our baseline criterion, such as emissions reductions.

Together, these results show that battery mitigation must account for the multiple ways in which batteries can exercise market power. Restricting predictable real-time discharge can redirect behavior toward greater quantity withholding, while a discharge subsidy cannot lower generation cost by more than its fiscal outlay. By contrast, competition reduces all three strategic distortions simultaneously, making it the more robust mitigation mechanism in our model.

\section{Extensions}\label{sec:extensions}

Our baseline model isolates the core economics of battery market power. We now show that its main insights survive in a range of richer environments. We highlight three extensions in the main text: battery capacity, battery inefficiency, and virtual bidders. They are considered one at a time: each subsection starts from the baseline model with $n$ competing batteries and introduces a single additional feature. Battery capacity and battery inefficiency are robustness checks on the storage technology itself. The virtual bidder case instead adds a market design feature: financial participants who arbitrage predictable DA--RT price gaps.

The common message is that the inefficiency from strategic battery operations remains sharply bounded and falls rapidly with competition. In the inefficiency and virtual bidder cases, the same baseline upper bound,
$\poa \leq 1+1/n(n+2)$, continues to hold. In the capacity case, the bound depends on how capacity is distributed across batteries and reduces to the same expression when batteries have identical capacities. For each extension, the main text explains why the extension matters, what modeling feature is added, and the resulting PoA implication; Appendix~\ref{app:sec:theorem-extensions} gives the full theorem statements and proofs. We defer other extensions to Appendix~\ref{app:sec:more-extensions}.

\subsection{Battery Capacity}\label{subsec:battery-capacity}

We first impose explicit state-of-charge capacity constraints in the baseline model with $n$ competing batteries. This is the most direct physical constraint omitted from the baseline: real batteries cannot shift arbitrary amounts of energy within the day. The extension therefore asks whether the PoA bound relies on treating storage as an unconstrained intertemporal technology. For each battery $b$ with capacity $C_b$, a feasible policy consists of DA/RT dispatch variables $z_b=(z_{b,t}^{DA},z_{b,t}^{RT}(D_{1:t}))_{t=1}^T$, an initial state $s_{b,0}\in[0,C_b]$, and an adapted state-of-charge process $s_b=(s_{b,t}(D_{1:t}))_{t=1}^T$. As in the baseline model, the dispatch variables satisfy the separate day-ahead and real-time energy-balance constraints,
\[
\sum_{t=1}^T z_{b,t}^{DA}=0,
\qquad
\sum_{t=1}^T z_{b,t}^{RT}(D_{1:t})=0
\quad \forall D .
\]
The state variables satisfy the state-of-charge dynamics
\[
s_{b,t}(D_{1:t})
=
s_{b,t-1}(D_{1:t-1})
-z_{b,t}^{DA}
-z_{b,t}^{RT}(D_{1:t}),
\qquad t=1,\ldots,T,
\]
and the capacity constraint
\[
0\le s_{b,t}(D_{1:t})\le C_b
\qquad \forall t,\ D .
\]
Positive values of $z_{b,t}^{DA}+z_{b,t}^{RT}(D_{1:t})$ denote physical discharge, and negative values denote charge. Capacity therefore restricts the realized physical state of charge, not the two settlement positions separately. Apart from this restriction, the two-settlement DA/RT structure and the nonanticipativity of real-time policies are unchanged.

Appendix~\ref{app:subsec:battery-capacity} proves that the capacity-constrained game has a unique equilibrium dispatch and establishes the corresponding capacity-dependent PoA upper bound. With heterogeneous capacities $C_1,\dots,C_n$, the PoA satisfies
\[
\poa \le \frac{2n+3}{2n+4} + \frac{1}{2}\sum_{b=1}^n \left(\frac{C_b}{\sum_{j=1}^n C_j}\right)^2.
\]
The sum in the second term is the Herfindahl index, i.e. the sum of squared capacity shares, which is a standard measure of concentration in industrial organization \citep{herfindahl-1950-concentration}. It is minimized at $1/n$ when capacities are equal and increases toward $1$ as capacity becomes concentrated in a single battery. The PoA upper bound is lower when capacity is more evenly distributed across batteries, reflecting stronger effective competition among batteries. In the case with identical battery capacities, $C_1=\cdots=C_n$, this bound reduces to the baseline competition bound from Theorem~\ref{thm:competition}:
\[
\poa \le 1 + \frac{1}{n(n+2)}.
\]

\subsection{Battery Inefficiency}
\label{subsec:battery-inefficiency}

We next allow imperfect round-trip efficiency. With efficiency $\eta\in(0,1]$, one unit of charge can support only $\eta$ units of later discharge. This changes the energy-balance constraint because charging and discharging are no longer treated symmetrically. To keep the extension analytically tractable, we specify the sets of charging and discharging periods exogenously, while the amount charged or discharged in each period remains a decision variable. This prevents the battery's policy from also determining, history by history, which actions are treated as charging and which are treated as discharging in the balance constraint. The restriction matches the usual daily arbitrage pattern when low- and high-net-demand hours are persistent: batteries charge in low net demand ``off-peak'' periods and discharge in high net demand ``peak'' periods.

\begin{assumption}\label{assump:aligned-charge-discharge}
Fix a partition $\mathcal T^+\cup \mathcal T^-=\{1,\dots,T\}$ with
$\mathcal T^+\cap \mathcal T^-=\varnothing$. For every battery $b=1,\ldots,n$, the periods in $\mathcal T^+$ are discharge periods and the periods in $\mathcal T^-$ are charge periods:
\[
z^{DA}_{b,t}\ge 0,\ z^{RT}_{b,t}(D_{1:t})\ge 0 \quad \forall t\in\mathcal T^+,
\qquad
z^{DA}_{b,t}\le 0,\ z^{RT}_{b,t}(D_{1:t})\le 0 \quad \forall t\in\mathcal T^-.
\]
\end{assumption}

Under Assumption~\ref{assump:aligned-charge-discharge}, efficiency losses enter the day-ahead and real-time balance constraints as linear weighted-balance equations:
\[
\sum_{t\in\mathcal T^+} z_{b,t}^{DA}
= \eta \sum_{t\in\mathcal T^-}(-z_{b,t}^{DA}),
\qquad
\sum_{t\in\mathcal T^+} z_{b,t}^{RT}(D_{1:t})
= \eta \sum_{t\in\mathcal T^-}(-z_{b,t}^{RT}(D_{1:t}))
\]
for every battery $b$ and every demand path $D$. Appendix~\ref{app:subsec:battery-inefficiency} shows that the optimal battery strategies can still be solved in closed form. The PoA bounds are the same as in Theorem~\ref{thm:competition}:
\[
1 + \frac{1}{n(n+1)(n^2+n+2)}
\le \poa \le
1 + \frac{1}{n(n+2)}.
\]

Operationally, inefficiency raises the hurdle for arbitrage, reduces the value of storage, and changes which intertemporal trades are attractive. However, because the same physical loss applies under both centralized and decentralized operations, it does not create a new source of market power.

\subsection{Virtual Bidders}\label{subsec:virtual-bidders}

We next consider virtual bidders, which change the market design rather than the physical storage technology. This case is useful because virtual bidders target predictable DA--RT price gaps, which are also central to the battery's timing distortion.

A virtual bidder has no physical storage capacity. Instead, in each period $t$, it takes an offsetting financial position across the two settlements: it sells $y_{v,t}$ in the day-ahead market and buys the same quantity back in real time, earning $(\lambda_t^{DA}-\lambda_t^{RT})y_{v,t}$, where $y_{v,t}$ may be positive or negative. The position $y_{v,t}$ therefore enters day-ahead and real-time residual demand with opposite signs. The position is financial and same-period: it arbitrages the DA--RT price gap for period $t$, but it does not move energy from one period to another.

Appendix~\ref{app:subsec:virtual-bidders} characterizes the equilibrium with $n$ strategic batteries and $m$ virtual bidders. The theorem gives the following tight PoA bounds:
\[
1+\frac{(m+1)^2}{n(n+m+1)(n^2+mn+n+2m+2)}
\le \poa \le
1+\frac{1}{n(n+2)} .
\]
Virtual bidders preserve the same upper bound as in the baseline competition result. More surprisingly, the lower bound is increasing in $m$ and converges to the same upper bound as $m\to\infty$. Adding many virtual bidders therefore does not make the decentralized storage outcome converge to centralized operations; instead, the tight PoA range collapses to the baseline upper bound.

The mechanism is a substitution in how batteries arbitrage. Virtual bidders compete in same-period financial DA--RT arbitrage, compressing predictable DA--RT price gaps and reducing the return to committing physical storage in the day-ahead market. Batteries therefore shift a larger share of their strategic response toward real time, where virtual bidders cannot condition on realized demand shocks. But this shift does not increase physical intertemporal arbitrage overall: the reduction in day-ahead physical arbitrage dominates the increase in predictable real-time discharge, so each battery's total expected physical discharge falls as the number of virtual bidders increases. Virtual bidders do not replace the physical storage service that batteries would otherwise provide. The decentralized system cost rises and PoA increases with the number of virtual bidders, so DA and RT prices can move closer together even while the system cost outcome worsens.

\section{Numerical Experiments and Nonlinear Supply Curves}\label{sec:calibration}

We now turn the theory into data-driven numerical experiments for California (CAISO) and Texas (ERCOT) markets. We first calibrate the linear-supply model and the capacity-constrained benchmark to observed price and net demand data. We then replace the linear inverse supply curve with convex cubic spline estimates. Across these experiments, the aim is to put the theoretical PoA predictions on a quantitative scale: PoA is moderate in the calibrated markets even without competition, and it falls rapidly as battery competition increases.

All experiments use hourly observations from 2024. For each market, we combine day-ahead and real-time energy prices with the corresponding system-wide net demand. We use energy-only wholesale prices, excluding location-specific congestion and loss components: prices from the same representative CAISO trading hub in both settlements and ERCOT's market-wide system prices. Subhourly real-time prices are averaged within each hour. Net demand is electricity demand minus wind and solar generation: the day-ahead measure uses the forecasts available before real time, while the real-time measure uses realized demand and renewable production. We exclude days without a complete set of 24 hourly observations, leaving 358 days (8,592 hours)\footnote{Figure~\ref{fig:demand_curve_caiso} uses all 8,781 matched California hours available for estimating the inverse supply curve; the daily experiments require complete 24-hour profiles and therefore use 8,592 hours.} for California and 364 days (8,736 hours) for Texas. This construction focuses on market-level energy conditions rather than location-specific transmission effects, consistent with the aggregate scope of our model, while preserving complete intraday demand profiles relevant for battery operations.

The calibration is the empirical analogue of the linear pricing system in \eqref{eqn:linear-pricing-system}, where the DA price is linear in DA net demand and the RT price equals the DA price plus an incremental real-time imbalance term. For each hourly observation $i$, let $d_i^{DA}$ and $d_i^{RT}$ denote day-ahead and real-time net demand levels, so that $d_i^{RT}-d_i^{DA}$ is the realized real-time imbalance, and let $\lambda_i^{DA}$ and $\lambda_i^{RT}$ denote day-ahead and real-time prices. For each market, we estimate $\alpha$, $\beta$, and the incremental real-time slope $\gamma=\beta/k_f$ by solving
\[
\min_{\alpha,\beta,\gamma}
\sum_i
\left(\lambda_i^{DA}-(\alpha+\beta d_i^{DA})\right)^2
+
\left(\lambda_i^{RT}-
\left(\alpha+\beta d_i^{DA}+\gamma(d_i^{RT}-d_i^{DA})\right)\right)^2 .
\]
The fitted day-ahead equation gives the linear inverse supply curve $p(q)=\alpha+\beta q$, while the fitted real-time imbalance coefficient identifies $k_f=\beta/\gamma$. This gives $k_f=0.89$ for California and $k_f=0.60$ for Texas. We estimate the joint demand distribution $\pi$ by fitting a multivariate normal distribution to complete daily 24-hour net demand vectors. Throughout this section, each market's estimated $k_f$ and $\pi$ are held fixed as calibrated parameters.

In each experiment, we compute three expected generation costs on the same demand model: the no-battery cost $\costnb$, the centralized-battery cost $\costcn$, and the decentralized $n$-battery cost $\costdcn$. We then report the same Price of Anarchy used in the theory,
\[
    \poa=\frac{\costnb-\costcn}{\costnb-\costdcn}.
\]
$\poa=1$ means that decentralized batteries recover the full centralized cost reduction, while larger values indicate that strategic operations recover a smaller fraction of the centralized storage value.

\subsection{Linear Supply Curves}\label{subsec:calibration-linear-capacity}

This subsection reports two linear-supply experiments: the unconstrained case in Table~\ref{tab:calibration-baseline-new} and the capacity-constrained case in Figure~\ref{fig:calibration-capacity-new}.

Table~\ref{tab:calibration-baseline-new} reports PoA as the number of competing batteries varies; these values can be computed in closed form. With one strategic battery, PoA is $1.177$ in California and $1.267$ in Texas. Moving from one battery to two lowers PoA to $1.047$ in California and $1.091$ in Texas, and by $n=5$ the values are $1.008$ and $1.019$, respectively. The larger Texas values reflect the lower estimated real-time flexibility share: when less generation can adjust cheaply in real time, the battery's real-time withholding distortion is more costly.

\begin{table}[t]
\centering
\footnotesize
\setlength{\tabcolsep}{8pt}
\begin{tabular}{lccccc}
\toprule
Market & $n=1$ & $n=2$ & $n=3$ & $n=4$ & $n=5$\\
\midrule
California & 1.177 & 1.047 & 1.021 & 1.012 & 1.008\\
Texas      & 1.267 & 1.091 & 1.046 & 1.028 & 1.019\\
\bottomrule
\end{tabular}
\caption{Baseline Price of Anarchy for California and Texas as a function of the number of competing batteries.}
\label{tab:calibration-baseline-new}
\end{table}

Next we add battery power and energy capacity constraints. These constraints break the closed-form solution, so we replace the fitted Gaussian distribution by a finite nonanticipative scenario tree. The tree is built by drawing daily demand paths from the fitted Gaussian distribution and recursively clustering histories; a day-ahead decision is common across all paths, while each real-time decision is attached to a history node and therefore depends only on information revealed up to that hour. For each capacity level and each $n$, we solve the centralized planner problem and compute the decentralized Cournot equilibrium, imposing the same within-day balance, power, and energy constraints in both cases. The centralized planner problem is a convex quadratic program. The Cournot equilibrium is computed by iterated best responses, with each best response also solved as a convex quadratic program. The maximum best-response residual across all reported capacity specifications is below $10^{-7}$.

Figure~\ref{fig:calibration-capacity-new} reports the resulting PoA over the number of competing batteries and aggregate battery-fleet power capacity, measured as a fraction of peak net demand in the corresponding market. For every value of $n$, total fleet capacity is held fixed and divided equally among the $n$ batteries. Energy capacity is set equal to duration times power capacity: 4 hours in California, reflecting the dominant CAISO storage duration, and 2 hours in Texas, reflecting ERCOT's shorter-duration battery fleet and recent movement toward 2-hour systems \citep{caiso-storage-2024,caiso-battery-special-report-2025,ercot-cdr-dec2024,modo-ercot-buildout-2025}. Capacity affects PoA through the size of the arbitrage opportunity. When batteries are very small, increasing capacity raises the amount of intraday arbitrage that storage can perform, which also raises the amount a strategic battery can withhold. Once capacity is large enough to cover the main within-day arbitrage opportunity, the single-battery PoA plateaus at about $1.168$ in California and $1.242$ in Texas. Competition again does most of the work. At the high-capacity plateau, moving from one to five batteries lowers PoA to $1.006$ in California and $1.015$ in Texas. In both calibrated markets and at every reported capacity, PoA decreases as $n$ increases, with the largest improvement coming from the first few competitors.

\begin{figure}[t]
\centering
\begin{subfigure}[t]{0.48\textwidth}
\centering
\footnotesize
\setlength{\tabcolsep}{3.5pt}
\begin{tabular}{lccccc}
\toprule
& \multicolumn{5}{c}{Fleet power / peak net demand}\\
\cmidrule(lr){2-6}
$n$ & $0.2$ & $0.4$ & $0.6$ & $0.8$ & $1.0$\\
\midrule
1 & 1.036 & 1.153 & 1.168 & 1.168 & 1.168\\
2 & 1.011 & 1.030 & 1.042 & 1.042 & 1.042\\
3 & 1.005 & 1.012 & 1.018 & 1.018 & 1.018\\
4 & 1.003 & 1.007 & 1.010 & 1.010 & 1.010\\
5 & 1.002 & 1.004 & 1.006 & 1.006 & 1.006\\
\bottomrule
\end{tabular}
\caption{California}
\label{fig:calibration-capacity-ca-new}
\end{subfigure}\hfill
\begin{subfigure}[t]{0.48\textwidth}
\centering
\footnotesize
\setlength{\tabcolsep}{3.5pt}
\begin{tabular}{lccccc}
\toprule
& \multicolumn{5}{c}{Fleet power / peak net demand}\\
\cmidrule(lr){2-6}
$n$ & $0.2$ & $0.4$ & $0.6$ & $0.8$ & $1.0$\\
\midrule
1 & 1.127 & 1.242 & 1.242 & 1.242 & 1.242\\
2 & 1.030 & 1.078 & 1.078 & 1.078 & 1.078\\
3 & 1.014 & 1.038 & 1.038 & 1.038 & 1.038\\
4 & 1.009 & 1.023 & 1.023 & 1.023 & 1.023\\
5 & 1.006 & 1.015 & 1.015 & 1.015 & 1.015\\
\bottomrule
\end{tabular}
\caption{Texas}
\label{fig:calibration-capacity-tx-new}
\end{subfigure}
\caption{Price of Anarchy with battery capacity constraints. Columns give aggregate battery-fleet power capacity as a fraction of peak net demand. For each $n$, total fleet power and energy capacity are held fixed and divided equally among the $n$ batteries. Energy capacity equals 4 hours of power capacity in California and 2 hours in Texas.}
\label{fig:calibration-capacity-new}
\end{figure}

\subsection{Nonlinear Supply Curves}\label{subsec:calibration-nonlinear-supply}

We now relax the linear inverse-supply approximation while keeping the same economic model. For each market, we fit the day-ahead inverse supply curve from hourly pairs $(q_i,p_i)$, where $q_i$ is day-ahead net demand and $p_i$ is the day-ahead price. Let
\[
    x_i=\frac{q_i-q_{\min}}{q_{\max}-q_{\min}}
\]
denote normalized net demand. For a given number of knots $K$, we place knots $\tau_j$ at empirical quantiles of $x_i$ and fit the convex cubic spline
\[
    p(x)=\alpha+\beta x+\sum_{j=1}^{K}\gamma_j (x-\tau_j)_+^3,
\]
We use this cubic basis because it gives a smooth monotone-convex curve: unlike a piecewise-linear fit, it avoids kinks at the knots while still allowing the inverse supply curve to steepen in high-demand states. The coefficients are estimated by constrained least squares:
\[
    \min_{\alpha,\beta,\gamma}
    \sum_i \left(p_i-\alpha-\beta x_i-\sum_{j=1}^K\gamma_j(x_i-\tau_j)_+^3\right)^2
    +\lambda\left(\beta^2+\sum_{j=1}^K\gamma_j^2\right)
    \quad\text{s.t.}\quad
    \beta\ge0,\ \gamma_j\ge0.
\]
The sign restrictions make the fitted curve monotone and convex, while the ridge penalty $\lambda$ reduces sensitivity to sparse high-demand tail observations. We choose a common penalty $\lambda=100$ using leave-one-month-out validation on the Texas data, where the nonlinear estimates are most sensitive. Specifically, for each penalty $\lambda\in\{0,0.3,1,3,10,30,100,300,1000,3000\}$, we fit the $K=8$ spline on eleven months, evaluate the held-out root-mean-square prediction error on observations above the training-sample 95th percentile of net demand, repeat this for each held-out month, and choose the penalty with the smallest median tail error. This rule selects $\lambda=100$ in Texas. California is much less sensitive to the penalty choice; we use the same $\lambda=100$ for comparability across markets.

After mapping the fitted spline back to the original net demand scale, we define generation cost by integrating the inverse supply curve, so that the curve gives marginal generation cost. For every $K\in\{0,1,2,4,8\}$, we compute the no-battery, centralized-battery, and decentralized $n$-battery costs on the same nonanticipative scenario tree used in Section~\ref{subsec:calibration-linear-capacity}, and evaluate PoA using the formula above. The centralized planner problem is convex. For the decentralized market, we compute a symmetric Cournot equilibrium by iterated best responses on the scenario tree. Given the common policy of the other $n-1$ batteries, a representative battery solves its profit-maximization problem subject to the day-ahead and pathwise real-time balance constraints. We initialize the best-response iteration at the linear-supply solution and use damping for numerical stability. The iteration converges for all nonlinear specifications reported in Figure~\ref{fig:calibration-spline-new}.

\begin{figure}[t]
\centering
\begin{subfigure}[t]{0.48\textwidth}
\centering
\footnotesize
\setlength{\tabcolsep}{3.5pt}
\begin{tabular}{lccccc}
\toprule
& \multicolumn{5}{c}{Number of knots $K$}\\
\cmidrule(lr){2-6}
$n$ & $0$ & $1$ & $2$ & $4$ & $8$\\
\midrule
1 & 1.164 & 1.164 & 1.164 & 1.164 & 1.165\\
2 & 1.040 & 1.040 & 1.040 & 1.040 & 1.040\\
3 & 1.017 & 1.017 & 1.017 & 1.017 & 1.017\\
4 & 1.009 & 1.009 & 1.009 & 1.009 & 1.009\\
5 & 1.006 & 1.006 & 1.006 & 1.006 & 1.006\\
\bottomrule
\end{tabular}
\caption{California}
\label{tab:calibration-spline-table-california-new}
\end{subfigure}\hfill
\begin{subfigure}[t]{0.48\textwidth}
\centering
\footnotesize
\setlength{\tabcolsep}{3.5pt}
\begin{tabular}{lccccc}
\toprule
& \multicolumn{5}{c}{Number of knots $K$}\\
\cmidrule(lr){2-6}
$n$ & $0$ & $1$ & $2$ & $4$ & $8$\\
\midrule
1 & 1.240 & 1.241 & 1.242 & 1.245 & 1.252\\
2 & 1.077 & 1.077 & 1.077 & 1.077 & 1.078\\
3 & 1.038 & 1.038 & 1.037 & 1.037 & 1.038\\
4 & 1.022 & 1.022 & 1.022 & 1.022 & 1.022\\
5 & 1.015 & 1.015 & 1.015 & 1.015 & 1.015\\
\bottomrule
\end{tabular}
\caption{Texas}
\label{tab:calibration-spline-table-texas-new}
\end{subfigure}
\caption{Price of Anarchy with convex cubic spline supply curves.}
\label{fig:calibration-spline-new}
\end{figure}

Figure~\ref{fig:calibration-spline-new} reports PoA for the convex supply curve specifications. The column $K=0$ reduces to the linear case, while larger values of $K$ allow progressively more curvature. In California, PoA is essentially unchanged as $K$ increases. Together with Figure~\ref{fig:demand_curve_caiso}, where the California inverse supply curve appears close to linear over most observed demand levels, this indicates that linear inverse supply provides a reasonable benchmark for the calibrated California setting. The fitted Texas inverse supply curve exhibits more curvature, but allowing this curvature still has only a small effect on PoA in the calibrated experiment: for a single battery, PoA rises from $1.240$ at $K=0$ to $1.252$ at $K=8$. Competition remains highly effective: under $K=8$, Texas PoA falls to $1.078$ with two batteries and $1.015$ with five batteries.

Overall, the numerical experiments reinforce the main message of the theory. Market power creates meaningful but moderate losses when a single battery acts strategically. Across the linear, capacity-constrained, and convex-spline specifications, battery competition rapidly alleviates these losses: moving from one battery to two removes most of the excess PoA, and a small fleet brings PoA close to one. The nonlinear experiments provide a calibrated robustness check: in California and Texas, allowing the fitted inverse supply curve to be convex changes the magnitude of PoA only modestly and leaves its rapid decline with competition intact. Thus, for the parameter values and demand distributions induced by these calibrations, the linear model provides a useful quantitative benchmark for the PoA comparisons reported here.

\section{Conclusion}\label{sec:conclusion}

This paper develops an analytically tractable model of battery operations in two-settlement electricity markets and uses it to compare centralized and decentralized operations. Relative to the centralized benchmark, a profit-maximizing battery distorts its operations in three ways: it withholds quantity, shifts participation from the day-ahead market to the real-time market, and responds too weakly to real-time demand fluctuations. These distortions are not arbitrary. Their composition depends systematically on the flexibility of the generation fleet: when a larger share of generators is fast-ramping, the distortion tilts toward delaying discharge into real time, whereas when generation is less flexible, it appears more as outright quantity withholding.

These distortions translate into bounded efficiency losses. Even in the worst case of a single strategic battery with unlimited capacity, strategic battery operations raise generation cost relative to centralized operations, but the loss is sharply bounded: the Price of Anarchy lies between $9/8$ and $4/3$. Competition rapidly curbs these distortions. With $n$ competing batteries, the Price of Anarchy converges to one at rate $1/n^2$. These conclusions are not an artifact of the baseline assumptions: the same upper-bound message continues to hold with capacity constraints, virtual bidders, battery inefficiency, and the additional extensions developed in the appendix. The market power mitigation results also show why policy design is subtle. Because the battery can exercise market power through several types of distortions, an intervention that targets one distortion can redirect behavior toward another and increase system cost.

The numerical experiments translate these bounds into empirical magnitudes. In the California and Texas calibrations, single-battery losses are non-trivial but not extreme and shrink sharply with competition. Capacity constraints and convex-spline supply curves do not change this message.

The analysis suggests two directions for future work. First, the model treats each market as a single node and abstracts from transmission constraints. This is a useful benchmark when congestion is limited or a congested region is sufficiently isolated, but an explicit network model with locational marginal pricing is needed to study how battery market power propagates across locations. Second, the model assumes a known demand distribution. In practice, batteries may preserve charge both to exercise market power and to manage forecast errors, price spikes, and operating contingencies. Distinguishing strategic withholding from prudent flexibility management remains an important research direction.
\clearpage
\bibliographystyle{plainnat}
\bibliography{references}

\clearpage
\setcounter{footnote}{0}

\begin{APPENDICES}
\AppendixTheoremNumbering
\AppendixTableOfContents
\AppendixContentsOn

\section{Generation Cost, Market Payments, and Welfare}\label{app:sec:welfare-accounting}

We provide the explicit accounting behind the generation-cost objective used in the main text. Fix a generator type $j\in\{s,f\}$. Because $G_j(\lambda)$ is the mass of type-$j$ generators with marginal cost at most $\lambda$, $G_j^{-1}(q)$ is the marginal cost of the unit at cumulative quantity $q$ in the merit order. Dispatching $q$ units of type $j$ therefore incurs physical production cost
\begin{align}\label{eqn:appendix-generation-cost-identity}
    \int_0^q G_j^{-1}(x)\,dx
    =
    \int_{\lambda\leq G_j^{-1}(q)} \lambda\,dG_j(\lambda).
\end{align}
The equality in \eqref{eqn:appendix-generation-cost-identity} follows from the monotone change of variables $x=G_j(\lambda)$, for which $dx=dG_j(\lambda)$.
The left-hand side is the area under the inverse supply curve up to quantity $q$. Equivalently, $dG_j(\lambda)$ is the infinitesimal mass of type-$j$ generators with marginal cost near $\lambda$, so $\lambda\,dG_j(\lambda)$ is the physical cost of that block and the right-hand side sums these costs over all dispatched generators. Applying \eqref{eqn:appendix-generation-cost-identity} to slow generation cleared day-ahead and fast generation after real-time adjustment, then summing across periods and taking expectations, gives the generation-cost expression in \eqref{eqn:gen-cost}.

This physical production cost is distinct from market payments. Let $U$ denote consumers' gross utility from the fixed quantity served, $P_L$ the total payments made by load, $P_G$ the total market payments received by generators, $C_G$ the physical production cost incurred by generators, and $P_B$ the battery's net market revenue. These quantities include both day-ahead and real-time settlements. Because settlement payments balance, $P_L=P_G+P_B$. Aggregate welfare is therefore
\begin{align*}
W
&=
\underbrace{(U-P_L)}_{\text{consumer surplus}}
+
\underbrace{(P_G-C_G)}_{\text{generator profit}}
+
\underbrace{P_B}_{\text{battery profit}}\\
&=
U-C_G+\underbrace{(P_G+P_B-P_L)}_{=\,0}\\
&=U-C_G.
\end{align*}
Thus market payments determine participant revenues and profits but cancel as transfers when welfare is aggregated. Because demand is perfectly inelastic and the served quantity is fixed, $U$ is constant, so maximizing welfare is equivalent to minimizing physical generation cost $C_G$.

\section{Additional Extensions}\label{app:sec:more-extensions}

For reasons of space, Section~\ref{sec:extensions} focuses on the three extensions most closely tied to the main robustness discussion. This appendix records three additional extensions: joint DA--RT energy balance, ramping costs, and strategic generators. They are considered one at a time. In each case, the discussion below explains why the extension matters, what modeling feature is added, and what implication follows for the equilibrium or PoA comparison; Appendix~\ref{app:sec:theorem-more-extensions} gives the full theorem statements and proofs.

\subsection{Joint Day-Ahead and Real-Time Energy Balance}\label{subsec:alternative-balance}

The baseline imposes separate day-ahead and real-time energy-balance constraints,
\[
\sum_{t=1}^{T} z_{b,t}^{DA}=0,
\qquad
\sum_{t=1}^{T} z_{b,t}^{RT}(D_{1:t})=0
\quad\text{for every battery }b\text{ and every demand path.}
\]
These constraints reflect the requirement that each battery submit a feasible DA operating plan, while RT actions represent incremental adjustments that must also balance over the day. A natural alternative is to impose only the physically realized pathwise balance requirement.

In this extension, we replace the two separate balance constraints by
\begin{equation}\label{eq:joint-balance}
\sum_{t=1}^{T}\bigl(z_{b,t}^{DA}+z_{b,t}^{RT}(D_{1:t})\bigr)=0
\quad\text{for every battery }b\text{ and every demand path.}
\end{equation}
This joint condition allows each battery to shift a constant quantity between its
DA schedule and its RT adjustment while keeping the same physical dispatch.
Theorem~\ref{thm:joint-balance} shows that this extra degree of freedom is not used at optimum. The centralized aggregate optimum is unchanged and can be implemented with separate DA and RT balance; every decentralized $n$-battery equilibrium also endogenously satisfies separate DA and RT balance. Thus the centralized and decentralized outcomes coincide with their baseline counterparts, and the PoA bounds
\[
1 + \frac{1}{n(n+1)(n^2+n+2)}
\le \poa \le
1 + \frac{1}{n(n+2)}
\]
are unchanged. The separate DA and RT balance constraints are therefore without loss in the linear model.

\subsection{Ramping Costs}\label{subsec:ramping-costs}

In the baseline model, generation cost is separable across periods: the cost in each period depends on that period's generation level, but not on how quickly generation changes from one period to the next. In this extension, we introduce ramping costs, which break this separability by making generation cost depend on changes in physical generation across periods. We use the following quadratic specification as a tractable convex proxy for the cost of changing generation output, consistent with work that incorporates ramping costs into dispatch and price formation \citep{kuang2017pricing,scaglione2016continuous}:
\[
\frac{c}{2}\,\bE\!\left[\sum_{t=1}^{T}(g_{t+1}-g_t)^2\right],
\]
where $g_t$ denotes physical generation in period $t$ and $g_{T+1}=g_1$ closes the representative-day horizon cyclically.

This term is part of the generation cost used throughout the PoA comparison. It affects the centralized problem directly: the planner minimizes energy production costs plus ramping costs. In the decentralized game, it enters through price formation. With ramping costs, prices require a more careful interpretation than in the separable baseline: DA and RT prices are the marginal values of serving demand under the augmented generation-cost function, so the price in period $t$ also reflects how changing generation in that period affects ramping costs between adjacent periods. The resulting decentralized dispatch is then evaluated using the same augmented generation cost. Appendix~\ref{app:subsec:ramping-costs} gives the formal statement and proof.

Theorem~\ref{thm:ramping-costs} shows that adding ramping costs leaves the main competition result unchanged. The centralized day-ahead policy still flattens predictable demand, the decentralized game still has a unique symmetric equilibrium, and the same PoA bounds continue to hold:
\[
1+\frac{1}{n(n+1)(n^2+n+2)}\le \poa \le 1+\frac{1}{n(n+2)}.
\]

\subsection{Strategic Generators}\label{subsec:strategic-generators}

The baseline treats conventional generators as non-strategic so that the analysis can isolate battery market power. As discussed in Section~\ref{subsec:discussion-assumptions}, this is a reasonable benchmark because conventional generator marginal costs and operating characteristics are relatively observable and already subject to market power mitigation, whereas battery bids depend on intertemporal opportunity costs. This extension asks whether relaxing that benchmark matters for the battery distortion: if generators also bid strategically, does generator market power change the battery's optimal smoothing rule or amplify battery withholding?

As a focused robustness check, we study a deterministic one-settlement model with one strategic battery and heterogeneous generators. This simplified setting isolates the interaction between generator bid shading and battery withholding, rather than introducing the full settlement- and state-dependent generator strategies that would arise in the stochastic two-settlement model. Formally, there are periods $t=1,\ldots,T$ with deterministic demand $\mu_t$, one battery, and $n_g\ge3$ generators indexed by $i=1,\ldots,n_g$. The requirement $n_g\ge3$ is the condition under which this unrestricted bid-slope game has a finite pure-strategy equilibrium. The battery chooses a balanced schedule $z=(z_t)_{t=1}^T$ with $\sum_t z_t=0$, while each generator simultaneously chooses a bid slope $\hat c_i>0$. Generator $i$ has true cost $C_i(x)=c_i x^2$, with $c_i>0$, and submits the quadratic bid cost $\widehat C_i(x)=\hat c_i x^2$. Given these choices, generator net demand is $D_t=\mu_t-z_t$, and the market clears in each period against submitted bid costs:
\[
\min_{x_{1,t},\ldots,x_{n_g,t}}
\sum_{i=1}^{n_g}\hat c_i x_{i,t}^2
\qquad
\text{s.t.}\quad
\sum_{i=1}^{n_g}x_{i,t}=D_t .
\]
The clearing price $p_t$ is the multiplier on this constraint. The battery's payoff is $\sum_t p_tz_t$, generator $i$'s payoff is $\sum_t(p_tx_{i,t}-c_i x_{i,t}^2)$, and system cost is evaluated using true generation costs, $\sum_{t,i}c_i x_{i,t}^2$.

Theorem~\ref{thm:strategic-generators} establishes a unique pure-strategy equilibrium. Allowing generator bid shading raises true generation cost, but does not change the battery schedule. Conditional on the aggregate bid slope chosen by generators, the battery's objective is only rescaled, so it still smooths demand by half exactly as in the truthful generator benchmark. Generators inflate their bids in equilibrium. Thus generator market power adds its own cost distortion, but it does not amplify the battery's withholding distortion.

\section{Proofs for Section~\ref{sec:three-modes-battery}}

\subsection{Proofs for Section~\ref{subsec:main-no-battery}}\label{app:subsec:main-no-battery}

\begin{proof}[Proof of Theorem~\ref{thm:no-battery}.]

With no battery,
\[
\tilde d_t^{DA}=\mu_t,
\qquad
\tilde d_t^{RT}=\mu_t+\frac{D_t-\mu_t}{k_f}.
\]
For each $t$, we compute
\begin{align*}
    \bE[(\tilde{d}_t^{RT})] &= \bE\left[ \mu_t + \frac{D_t - \mu_t}{k_f}\right] = \mu_t \\
    \bE[(\tilde{d}_t^{RT})^2] &= \bE\left[ \left( \mu_t + \frac{D_t - \mu_t}{k_f} \right)^2 \right] = \mu_t^2 + \frac{2\mu_t}{k_f} \bE[(D_t- \mu_t)] + \frac{1}{k_f^2} \bE[(D_t - \mu_t)^2] = \mu_t^2 + \frac{\Var(D_t)}{k_f^2} 
\end{align*}

Substituting these moments into \eqref{eqn:gen-cost-linear} gives
\begin{align*}
    \costnb
    =
    \sum_{t=1}^{T}
    \left[
    k_s \left( \alpha \mu_t + \frac{\beta}{2} \mu_t^2 \right)
    + k_f  \left( \alpha \mu_t + \frac{\beta}{2} \left( \mu_t^2 + \frac{\Var(D_t)}{k_f^2} \right)  \right)
    \right],
\end{align*}
which simplifies to the given expression.

\end{proof}

\subsection{Proofs for Section~\ref{subsec:main-fb-battery}}\label{app:sec:fb-battery}

\begin{proof}[Proof of Theorem~\ref{thm:fb-battery}.]
The proof separates the predictable intertemporal smoothing problem from the centered real-time shock-smoothing problem.
Let $\epsilon_t:=D_t-\mu_t$. Write the day-ahead and real-time battery actions as
$x_t:=z_t^{DA}$ and $u_t:=z_t^{RT}(D_{1:t})$, and decompose the real-time action as
\[
m_t:=\bE[u_t],
\qquad
v_t:=u_t-m_t.
\]
Let $y_t:=x_t+m_t$ be the expected physical dispatch. The balance constraints imply
$\sum_t x_t=0$, $\sum_t m_t=0$, $\sum_t y_t=0$, and $\sum_t v_t=0$ pathwise. Also,
$x_t=y_t-m_t$ and $u_t=m_t+v_t$.

Substituting this decomposition into the generation cost in \eqref{eqn:gen-cost-linear}, the
day-ahead demand and modified real-time demand are
\[
d_t^{DA}=\mu_t-y_t+m_t,
\qquad
d_t^{DA}+\frac1{k_f}d_t^{RT}
=\mu_t-y_t+\left(1-\frac1{k_f}\right)m_t+\frac1{k_f}(\epsilon_t-v_t).
\]
The linear terms reduce to the constant $\alpha\sum_t\mu_t$ by balance. For the quadratic
terms, using $\bE[\epsilon_t]=\bE[v_t]=0$, we have
\begin{align*}
&(1-k_f)(d_t^{DA})^2
+k_f\bE\left[\left(d_t^{DA}+\frac1{k_f}d_t^{RT}\right)^2\right] \\
&\qquad
=
(\mu_t-y_t)^2
+\left(\frac1{k_f}-1\right)m_t^2
+\frac1{k_f}\bE[(\epsilon_t-v_t)^2].
\end{align*}
Dropping the constant $\alpha\sum_t\mu_t$ and the positive factor $\beta/2$, the part of the
objective that depends on the battery policy is
\begin{align}
\Phi(y,m,v)
=
\sum_{t=1}^T(\mu_t-y_t)^2
+\left(\frac{1}{k_f}-1\right)\sum_{t=1}^T m_t^2
+\frac{1}{k_f}\bE\left[\sum_{t=1}^T(\epsilon_t-v_t)^2\right].
\label{eq:centralized-proof-objective}
\end{align}

The deterministic terms therefore separate from the centered real-time response. Since
$\sum_t y_t=0$, the term $\sum_t(\mu_t-y_t)^2$ is minimized by
$y_t=\mu_t-\bar\mu$. The term involving $m$ is minimized by $m_t=0$ for all $t$. Hence
\[
z_t^{DA,CN}=x_t=y_t-m_t=\mu_t-\bar\mu,
\qquad
\bE[z_t^{RT,CN}]=0.
\]
It remains to solve the centered real-time smoothing problem
\begin{align}
\min_{(u_t)_{t=1}^T}
\bE\left[\sum_{t=1}^T(\epsilon_t-u_t(D_{1:t}))^2\right]
\quad
\text{s.t.}\quad
\sum_{t=1}^T u_t(D_{1:t})=0
\quad\text{pathwise}.
\label{eq:centralized-rt-subproblem}
\end{align}
The solution derived below has mean zero in every period, and hence is consistent with the
choice $m_t=0$ above.

We solve \eqref{eq:centralized-rt-subproblem} by dynamic programming. At the beginning of
period $t$, after the history $D_{1:t-1}$ and before choosing $u_t$, define the remaining
real-time balance requirement
\[
B_t(D_{1:t-1}):=-\sum_{s=1}^{t-1}u_s(D_{1:s}).
\]
After observing $D_{1:t}$, let
\[
S_t(D_{1:t})
:=
\epsilon_t+\sum_{i=t+1}^T(\mu_{i\mid D_{1:t}}-\mu_i),
\qquad
N_t:=T-t+1 .
\]
Here $S_t(D_{1:t})$ is the conditional expected total remaining demand shock from periods
$t,\ldots,T$.

For the continuation problem from period $t$ onward with remaining balance requirement $B$,
define
\[
V_t(B,D_{1:t})
:=
\min_{(u_s)_{s=t}^T}
\bE\left[
\sum_{s=t}^T(\epsilon_s-u_s(D_{1:s}))^2
\mid D_{1:t}
\right],
\]
subject to the pathwise remaining-balance constraint
\[
\sum_{s=t}^T u_s(D_{1:s})=B .
\]
We claim that this value function has the form
\[
V_t(B,D_{1:t})
=
C_t(D_{1:t})+\frac{1}{N_t}\bigl(S_t(D_{1:t})-B\bigr)^2,
\]
where $C_t(D_{1:t})$ does not depend on $B$. At $t=T$, the balance constraint forces
$u_T=B$, so
\[
V_T(B,D_{1:T})=(\epsilon_T-B)^2
=\bigl(S_T(D_{1:T})-B\bigr)^2,
\]
which proves the claim with $C_T=0$. If the claim holds at $t+1$, then, conditional on
$D_{1:t}$, choosing $z:=u_t(D_{1:t})$ leaves remaining balance requirement $B-z$ for the
future periods. Dropping terms independent of $z$, the continuation objective is
\[
(\epsilon_t-z)^2
+\frac{1}{N_t-1}
\bE\left[
\left(S_{t+1}(D_{1:t+1})-(B-z)\right)^2
\mid D_{1:t}
\right].
\]
Let
\[
M_t:=
\bE[S_{t+1}(D_{1:t+1})\mid D_{1:t}]
=
\sum_{i=t+1}^T(\mu_{i\mid D_{1:t}}-\mu_i).
\]
The first-order condition for $z$ is
\[
0=(z-\epsilon_t)+\frac{1}{N_t-1}(z-B+M_t),
\]
and therefore
\[
z
=
\frac{N_t-1}{N_t}\epsilon_t
+\frac{1}{N_t}B-\frac{1}{N_t}M_t .
\]
Substituting this optimizer back gives the same value-function form at $t$, since
$S_t(D_{1:t})=\epsilon_t+M_t$ and all terms independent of $B$ can be absorbed into
$C_t(D_{1:t})$.
Substituting $N_t=T-t+1$ and $B=B_t(D_{1:t-1})$ yields
\eqref{eqn:z-RT-CN} for $t\le T-1$. Equivalently, this recursion can be written as
\begin{align}
(T-t+1)z_t^{RT,CN}(D_{1:t})+\sum_{s=1}^{t-1}z_s^{RT,CN}(D_{1:s})
=
(T-t)(D_t-\mu_t)-\sum_{i=t+1}^T(\mu_{i\mid D_{1:t}}-\mu_i).
\label{eqn:diff-ziRT-fb-new}
\end{align}
Taking expectations in \eqref{eqn:z-RT-CN} and using induction on $t$ gives
$\bE[z_t^{RT,CN}]=0$ for every $t\le T-1$; the terminal balance condition then gives
$\bE[z_T^{RT,CN}]=0$ as well.
In the last period, the balance constraint pins down
\[
z_T^{RT,CN}(D_{1:T})
=
-\sum_{s=1}^{T-1}z_s^{RT,CN}(D_{1:s}).
\]

\end{proof}

\subsection{Proofs for Section~\ref{subsec:main-sb-battery}}\label{app:sec:sb-battery}

\begin{proof}[Proof of Theorem~\ref{thm:sb-battery}.]
The proof follows the same decomposition as the centralized proof, but now for a profit-maximizing battery. Let $\epsilon_t:=D_t-\mu_t$. Write
\[
x_t:=z_t^{DA},\qquad
u_t:=z_t^{RT}(D_{1:t}),\qquad
r_t:=\bE[u_t],\qquad
\tilde u_t:=u_t-r_t,
\]
and let $p_t:=x_t+r_t$ denote expected physical dispatch. The balance constraints imply that
$p$ and $r$ both sum to zero over $t$, and that $\tilde u$ has mean zero in every period and
sums to zero pathwise. Since $x_t=p_t-r_t$ and $u_t=r_t+\tilde u_t$, the DA and RT prices are
\[
\lambda_t^{DA}
=\alpha+\beta(\mu_t-p_t+r_t),
\qquad
\lambda_t^{RT}
=\alpha+\beta(\mu_t-p_t+r_t)
+\frac{\beta}{k_f}(\epsilon_t-r_t-\tilde u_t).
\]
The intercept term drops out of profit by balance. After dividing by $\beta$, the battery's
profit can be written as
\begin{align*}
\Pi/\beta
&=
\sum_{t=1}^{T}p_t(\mu_t-p_t+r_t)
-\frac1{k_f}\sum_{t=1}^{T}r_t^2
+\frac1{k_f}\bE\left[\sum_{t=1}^{T}\tilde u_t(\epsilon_t-\tilde u_t)\right].
\end{align*}
Thus the predictable variables $(p,r)$ separate from the centered real-time response
$\tilde u$.

We first solve the predictable part. The first-order condition for $p_t$, projected onto the
zero-sum subspace, is
\[
\mu_t-\bar\mu-2p_t+r_t=0,
\]
and the first-order condition for $r_t$ is
\[
k_f p_t-2r_t=0.
\]
Solving these two equations gives
\[
p_t=\frac{2}{4-k_f}(\mu_t-\bar\mu),
\qquad
r_t=\frac{k_f}{4-k_f}(\mu_t-\bar\mu).
\]
Therefore
\[
z_t^{DA,DCN}=x_t=p_t-r_t
=\frac{2-k_f}{4-k_f}(\mu_t-\bar\mu).
\]

It remains to solve the centered real-time part. Let
\[
\mathcal H
:=\left\{
h:\ h_t\text{ is adapted to }D_{1:t},\ \bE[h_t]=0,\ 
\sum_{t=1}^{T}h_t=0\ \text{pathwise}
\right\}.
\]
Using the inner product $\langle a,b\rangle:=\sum_{t=1}^{T}\bE[a_tb_t]$, the centered part of
profit is $k_f^{-1}\langle\tilde u,\epsilon-\tilde u\rangle$. Its first-order condition is
\[
\left\langle \epsilon-2\tilde u,\ h\right\rangle=0
\qquad\text{for all }h\in\mathcal H.
\]
The centralized real-time policy $z^{RT,CN}$ is the orthogonal projection of $\epsilon$ onto
$\mathcal H$, by the centralized analysis in the proof of Theorem~\ref{thm:fb-battery}. Hence
\[
\tilde u_t=\frac12 z_t^{RT,CN}(D_{1:t}).
\]
Combining this with the expression for $r_t$ gives
\[
z_t^{RT,DCN}(D_{1:t})
=r_t+\tilde u_t
=\frac{k_f}{4-k_f}(\mu_t-\bar\mu)
+\frac12 z_t^{RT,CN}(D_{1:t}).
\]
The objective is strictly concave in $(p,r,\tilde u)$ on the feasible subspace, so this first-order
solution is the unique maximizer.

\end{proof}

\section{Proofs for Section~\ref{sec:compare-gencost}}\label{app:sec:compare-gencost}

\begin{proof}[Proof of Theorem~\ref{thm:competition}.]
Let $\epsilon_t:=D_t-\mu_t$. For battery $b$, write
\[
x_{b,t}:=z_{b,t}^{DA},\qquad
u_{b,t}:=z_{b,t}^{RT}(D_{1:t}),\qquad
r_{b,t}:=\bE[u_{b,t}],\qquad
\tilde u_{b,t}:=u_{b,t}-r_{b,t},
\]
and let $p_{b,t}:=x_{b,t}+r_{b,t}$ denote battery $b$'s expected physical dispatch. Define
\[
P_t:=\sum_{b=1}^n p_{b,t},\qquad
R_t:=\sum_{b=1}^n r_{b,t},\qquad
\widetilde U_t:=\sum_{b=1}^n \tilde u_{b,t}.
\]
The separate DA and RT balance constraints imply that each $p_b$ and each $r_b$ sums to zero
over $t$, and that each $\tilde u_b$ has mean zero in every period and sums to zero pathwise.
The modified DA and RT demands are
\[
\tilde d_t^{DA}=\mu_t-P_t+R_t,
\qquad
\tilde d_t^{RT}
=\mu_t-P_t+R_t+\frac{\epsilon_t-R_t-\widetilde U_t}{k_f}.
\]

Fix a battery $b$. After dropping the intercept term, which is zero by balance, battery $b$'s
expected profit divided by $\beta$ can be written as
\[
\sum_{t=1}^T p_{b,t}(\mu_t-P_t+R_t)
-\frac1{k_f}\sum_{t=1}^T r_{b,t}R_t
+\frac1{k_f}\bE\left[\sum_{t=1}^T\tilde u_{b,t}(\epsilon_t-\widetilde U_t)\right].
\]
The first-order condition for $p_{b,t}$, projected onto the zero-sum subspace, is
\begin{equation}\label{eq:competition-p-foc}
\mu_t-\bar\mu-P_t+R_t-p_{b,t}=0.
\end{equation}
The first-order condition for $r_{b,t}$ is
\begin{equation}\label{eq:competition-r-foc}
k_f p_{b,t}-R_t-r_{b,t}=0.
\end{equation}
Equations \eqref{eq:competition-p-foc}--\eqref{eq:competition-r-foc} imply that $p_{b,t}$ and
$r_{b,t}$ are the same for every battery $b$ in each period $t$. Let these common values be
$p_t$ and $r_t$. Then $P_t=np_t$ and $R_t=nr_t$, so
\[
\mu_t-\bar\mu-(n+1)p_t+nr_t=0,
\qquad
k_f p_t-(n+1)r_t=0.
\]
Solving these two equations with
\[
D_n:=(n+1)^2-nk_f
\]
gives
\[
p_t=\frac{n+1}{D_n}(\mu_t-\bar\mu),
\qquad
r_t=\frac{k_f}{D_n}(\mu_t-\bar\mu).
\]
Therefore
\[
z_{b,t}^{DA,DCN}=x_{b,t}=p_t-r_t
=\frac{n+1-k_f}{D_n}(\mu_t-\bar\mu).
\]

It remains to characterize the centered real-time response. Let
\[
\mathcal H
:=\left\{
h:\ h_t\text{ is adapted to }D_{1:t},\ \bE[h_t]=0,\ 
\sum_{t=1}^T h_t=0\ \text{pathwise}
\right\}.
\]
Using the inner product $\langle a,b\rangle:=\sum_{t=1}^T\bE[a_tb_t]$, the centered part of
battery $b$'s payoff is $k_f^{-1}\langle \tilde u_b,\epsilon-\widetilde U\rangle$. Its
first-order condition is
\[
\left\langle \epsilon-\widetilde U-\tilde u_b,\ h\right\rangle=0
\qquad\text{for all }h\in\mathcal H.
\]
Because $\widetilde U+\tilde u_b\in\mathcal H$, this condition says that
$\widetilde U+\tilde u_b$ is the orthogonal projection of $\epsilon$ onto $\mathcal H$.
By the centralized analysis in the proof of Theorem~\ref{thm:fb-battery}, this projection is
$z^{RT,CN}$. Thus, for every battery $b$,
\[
\widetilde U+\tilde u_b=z^{RT,CN}.
\]
Comparing this equation across any two batteries shows that the centered real-time responses
are the same for all batteries. Writing the common response as $\tilde u$, we have
$\widetilde U=n\tilde u$ and therefore
\[
\tilde u_{b,t}=\frac{1}{n+1}z_t^{RT,CN}(D_{1:t}),
\]
which gives
\[
z_{b,t}^{RT,DCN}(D_{1:t})
=r_t+\tilde u_{b,t}
=\frac{k_f}{D_n}(\mu_t-\bar\mu)
+\frac{1}{n+1}z_t^{RT,CN}(D_{1:t}).
\]
The displayed policies satisfy the balance constraints and the first-order conditions. The
predictable payoff is concave in each battery's own $(p_b,r_b)$, and the centered payoff is
concave in $\tilde u_b$ on $\mathcal H$, so these first-order conditions are sufficient for a
best response. Hence an equilibrium exists. Conversely, any equilibrium must satisfy the same
first-order conditions; the arguments above force the displayed policies, so the equilibrium is
unique and symmetric.

Define
\[
A:=\sum_{t=1}^{T}(\mu_t-\bar\mu)^2,
\qquad
S:=\sum_{t=1}^{T}\bE\left[(D_t-\mu_t)z_t^{RT,CN}(D_{1:t})\right]
=\sum_{t=1}^{T}\bE\left[\left(z_t^{RT,CN}(D_{1:t})\right)^2\right],
\]
where the second equality is \eqref{eqn:centralized-rt-orthogonality}. The equilibrium formulas
imply
\begin{align*}
\tilde d_t^{DA}
&=
\bar\mu+\frac{n+1}{D_n}(\mu_t-\bar\mu),\\
\tilde d_t^{RT}
&=
\bar\mu+\frac{1}{D_n}(\mu_t-\bar\mu)
+\frac{1}{k_f}\left(D_t-\mu_t-\frac{n}{n+1}z_t^{RT,CN}\right).
\end{align*}
Substituting into \eqref{eqn:gen-cost-linear} and using
\eqref{eqn:centralized-rt-orthogonality} gives
\begin{align*}
\costnb-\costcn
&=
\beta\left[\frac12A+\frac{1}{2k_f}S\right],\\
\costnb-\costdcn
&=
\beta\left[
\frac{H_n(k_f)}{2D_n^2}A
+\frac{1}{2k_f}\frac{n(n+2)}{(n+1)^2}S
\right],
\end{align*}
where
\[
H_n(k_f)
:=
n(n+2)(n+1)^2-n^2(2n+3)k_f+n^2k_f^2.
\]
When $\costnb>\costdcn$, $\poa$ is a weighted average of
\[
r_\mu(k_f):=\frac{D_n^2}{H_n(k_f)}
\qquad\text{and}\qquad
r_S:=1+\frac{1}{n(n+2)}.
\]
The first ratio is decreasing in $k_f$ because
\[
r_\mu'(k_f)
=
-\frac{n^2\bigl(n^2+(2-k_f)n+1\bigr)\bigl((n+1)^2+(n+2)k_f\bigr)}{H_n(k_f)^2}
\le0.
\]
Taking limits at the endpoints,
\[
\lim_{k_f\uparrow 1}r_\mu(k_f)
=1+\frac{1}{n(n+1)(n^2+n+2)},
\qquad
\lim_{k_f\downarrow 0}r_\mu(k_f)
=r_S=1+\frac{1}{n(n+2)}.
\]
Thus
\[
1+\frac{1}{n(n+1)(n^2+n+2)}
\le \poa \le
1+\frac{1}{n(n+2)}.
\]
\end{proof}

\section{Proofs for Section~\ref{sec:market-power-mitigation}}\label{app:sec:market-power-mitigation}

We begin with the proof of Theorem~\ref{thm:da-rt-backfire}, and then prove Theorem~\ref{thm:subsidy-discharge}.

\subsection{Regulating Day-Ahead versus Real-Time Discrepancy}\label{app:subsec:da-vs-rt}

\begin{proof}[Proof of Theorem~\ref{thm:da-rt-backfire}.]

Let $\epsilon_t:=D_t-\mu_t$. For battery $b$, write
\[
x_{b,t}:=z_{b,t}^{DA},\qquad
u_{b,t}:=z_{b,t}^{RT}(D_{1:t}),\qquad
r_{b,t}:=\bE[u_{b,t}],\qquad
\tilde u_{b,t}:=u_{b,t}-r_{b,t},
\]
and let $p_{b,t}:=x_{b,t}+r_{b,t}$ denote battery $b$'s expected physical dispatch. Define
\[
P_t:=\sum_{b=1}^n p_{b,t},\qquad
R_t:=\sum_{b=1}^n r_{b,t},\qquad
\widetilde U_t:=\sum_{b=1}^n \tilde u_{b,t}.
\]
The intervention imposes $r_{b,t}=0$ for every $b$ and $t$. Hence $R=0$ and
$p_b=x_b$. After dropping the intercept term and dividing by $\beta$, battery $b$'s payoff
under the intervention is
\[
\sum_{t=1}^T p_{b,t}(\mu_t-P_t)
+\frac1{k_f}\bE\left[\sum_{t=1}^T\tilde u_{b,t}(\epsilon_t-\widetilde U_t)\right].
\]

The first-order condition for $p_{b,t}$, projected onto the zero-sum subspace, is
\[
\mu_t-\bar\mu-P_t-p_{b,t}=0.
\]
This equation implies that $p_{b,t}$ is the same for every battery $b$ in each period $t$. Writing
the common value as $p_t$, we have $P_t=np_t$, so
\[
p_t=\frac{1}{n+1}(\mu_t-\bar\mu).
\]

It remains to solve the centered real-time part. Let
\[
\mathcal H
:=\left\{
h:\ h_t\text{ is adapted to }D_{1:t},\ \bE[h_t]=0,\ 
\sum_{t=1}^{T}h_t=0\ \text{pathwise}
\right\}.
\]
Using the inner product $\langle a,b\rangle:=\sum_t\bE[a_tb_t]$, the first-order condition for
battery $b$'s centered real-time response is
\[
\left\langle \epsilon-\widetilde U-\tilde u_b,\ h\right\rangle=0
\qquad\text{for all }h\in\mathcal H.
\]
As in the proof of Theorem~\ref{thm:competition}, this implies
$\widetilde U+\tilde u_b=z^{RT,CN}$ for every battery $b$. Therefore the centered real-time
responses are symmetric and
\[
\tilde u_{b,t}=\frac{1}{n+1}z_t^{RT,CN}(D_{1:t}).
\]
The displayed policy satisfies the regulated balance constraints and the first-order conditions;
concavity of each best-response problem under the linear restriction $r_b=0$ makes these
conditions sufficient. Conversely, any regulated equilibrium must satisfy the same conditions,
which force the displayed symmetric policy.

Thus the intervention eliminates each battery's predictable real-time component, but it also lowers
aggregate expected physical discharge. Under the intervention, the quantity-withholding measure is
\[
1-\frac{\sum_{b=1}^n\left(z_{b,t}^{DA,Reg}+\bE[z_{b,t}^{RT,Reg}]\right)}
{z_t^{DA,CN}+\bE[z_t^{RT,CN}]}
=
1-\frac{\frac{n}{n+1}(\mu_t-\bar\mu)}{\mu_t-\bar\mu}
=\frac{1}{n+1},
\]
whenever the ratio is defined. In the unregulated $n$-battery equilibrium, the corresponding
quantity-withholding measure is
\[
\frac{n+1-nk_f}{(n+1)^2-nk_f}.
\]
Hence the intervention increases quantity withholding by
\[
\frac{1}{n+1}
-\frac{n+1-nk_f}{(n+1)^2-nk_f}
=
\frac{n^2k_f}{(n+1)((n+1)^2-nk_f)}
\ge0.
\]

It remains to compare system costs. Let
\[
A:=\sum_{t=1}^{T}(\mu_t-\bar\mu)^2,
\qquad
S:=\sum_{t=1}^{T}\E\!\left[\epsilon_t z_t^{RT,CN}(D_{1:t})\right].
\]
As shown in the proof of Theorem~\ref{thm:competition}, with
$D_n:=(n+1)^2-nk_f$,
\[
\costnb-\costdcn
=
\beta\left[
\frac{H_n(k_f)}{2D_n^2}A
+\frac{1}{2k_f}\frac{n(n+2)}{(n+1)^2}S
\right]
\]
where
\[
H_n(k_f)
:=
n(n+2)(n+1)^2-n^2(2n+3)k_f+n^2k_f^2.
\]
Substituting the regulated policy into the same cost expression gives
\[
\costnb-\textnormal{Cost(DCN-Reg)}
=
\beta\left[
\frac12\frac{n(n+2)}{(n+1)^2}A
+\frac{1}{2k_f}\frac{n(n+2)}{(n+1)^2}S
\right].
\]
Therefore
\[
\textnormal{Cost(DCN-Reg)}-\costdcn
=
\beta\left[
\frac{H_n(k_f)}{2D_n^2}
-\frac12\frac{n(n+2)}{(n+1)^2}
\right]A
=
\beta\,
\frac{n^2k_f((n+1)^2+k_f)}{2(n+1)^2D_n^2}\,A
\ge0.
\]
Finally, the intervention lowers each battery's profit. At the unregulated equilibrium, each
battery's profit is
\[
\beta\left[
\frac{1}{D_n}A+\frac{1}{k_f(n+1)^2}S
\right],
\]
while at the regulated equilibrium it is
\[
\beta\left[
\frac{1}{(n+1)^2}A+\frac{1}{k_f(n+1)^2}S
\right].
\]
The profit loss is therefore
\[
\beta\left[
\frac{1}{D_n}-\frac{1}{(n+1)^2}
\right]A
=
\beta\,\frac{nk_f}{D_n(n+1)^2}\,A
\ge0.
\]

\end{proof}

\subsection{Battery Discharge Subsidy}\label{app:subsec:discharge-subsidy}

\begin{proof}[Proof of Theorem~\ref{thm:subsidy-discharge}.]
Fix an arbitrary equilibrium of the subsidized game.
If $T=1$, the separate balance constraints force all
battery quantities to be zero, and the claim is immediate. If $\beta=0$, both market prices
equal $\alpha$, generation cost is invariant to the battery policies by the balance constraints,
and subsidy payments are nonnegative. Thus the claim is again immediate. In the remainder,
suppose $T\ge2$ and $\beta>0$.

Use the decomposition from the proof of Theorem~\ref{thm:competition}. For battery $b$, let
\[
p_{b,t}:=z_{b,t}^{DA}+\bE[z_{b,t}^{RT}],
\qquad
r_{b,t}:=\bE[z_{b,t}^{RT}],
\qquad
\tilde u_{b,t}:=z_{b,t}^{RT}-r_{b,t},
\]
and define the aggregate vectors
\[
P:=\sum_{b=1}^n p_b,
\qquad
R:=\sum_{b=1}^n r_b,
\qquad
\widetilde U:=\sum_{b=1}^n\tilde u_b.
\]
Each $p_b$ and $r_b$ sums to zero across periods. Let
\[
\Delta:=\mu-\bar\mu\mathbf 1,
\qquad
\epsilon:=D-\mu,
\qquad
D_n:=(n+1)^2-nk_f.
\]
After dropping the intercept term and dividing by $\beta$, battery $b$'s payoff is
\[
\langle p_b,\mu-P+R\rangle
-\frac1{k_f}\langle r_b,R\rangle
+\frac1{k_f}\bE\!\left[\langle\tilde u_b,\epsilon-\widetilde U\rangle\right]
+\frac1\beta\sum_{t=1}^T s_t[p_{b,t}]_+,
\]
where $\langle a,b\rangle:=\sum_t a_tb_t$.

We first address the kink in the subsidy payment. At an equilibrium,
\begin{equation}\label{eq:subsidy-positive-dispatch-nonzero}
s_t>0\quad\Longrightarrow\quad p_{b,t}\ne0
\qquad\text{for every battery $b$ and period $t$}.
\end{equation}
To see this, suppose instead that $p_{b,t}=0<s_t$. Choose $j\ne t$ and let
$d=e_t-e_j$. Holding $r_b$ and $\tilde u_b$ fixed, the two perturbations
$p_b+\varepsilon d$ and $p_b-\varepsilon d$ are feasible. The smooth part of the
normalized payoff has Hessian $-2I$ in $p_b$, so the sum of its changes under these two
perturbations is $-2\varepsilon^2\|d\|^2=-4\varepsilon^2$. By convexity of
$[\,\cdot\,]_+$, the combined change in the normalized subsidy payment is at least
$s_t\varepsilon/\beta$. For all sufficiently small $\varepsilon>0$, the sum of the two
payoff changes is therefore positive. At least one perturbation is profitable, contradicting
optimality and proving \eqref{eq:subsidy-positive-dispatch-nonzero}.

Consequently, the subsidy payment is differentiable at every equilibrium strategy. Define
\[
\sigma_{b,t}:=s_t\mathbf 1\{p_{b,t}>0\},
\qquad
\overline\sigma_b:=\frac1T\sum_{t=1}^T\sigma_{b,t},
\qquad
\tau_b:=\frac{\sigma_b-\overline\sigma_b\mathbf 1}{\beta}.
\]
If $p_{b,t}=0$, then $s_t=0$ by \eqref{eq:subsidy-positive-dispatch-nonzero}, so the
definition is unambiguous. Moreover,
\begin{equation}\label{eq:subsidy-payment-identity}
\sigma_{b,t}p_{b,t}=s_t[p_{b,t}]_+.
\end{equation}

The first-order condition for $p_b$, projected onto the zero-sum subspace, is
\begin{equation}\label{eq:subsidy-general-p-foc}
\Delta+\tau_b-P+R-p_b=0.
\end{equation}
Indeed, the unconstrained gradient is
$\mu-P+R-p_b+\sigma_b/\beta$; it must be constant across periods, and subtracting
its time average gives \eqref{eq:subsidy-general-p-foc}. Similarly, the gradient with respect
to $r_b$ is $p_b-(R+r_b)/k_f$. It must be constant across periods, but its time average is
zero, so
\begin{equation}\label{eq:subsidy-general-r-foc}
k_fp_b-R-r_b=0.
\end{equation}
These are necessary equilibrium conditions; no symmetry has been imposed.

The subsidy does not affect the centered real-time problem. Its first-order condition is
\[
\bE\!\left[\langle\epsilon-\widetilde U-\tilde u_b,h\rangle\right]=0
\qquad\text{for every feasible centered perturbation $h$}.
\]
As in the proof of Theorem~\ref{thm:competition}, this implies
$\widetilde U+\tilde u_b=z^{RT,CN}$ for every $b$. Hence
\begin{equation}\label{eq:subsidy-general-centered-policy}
\tilde u_b=\frac1{n+1}z^{RT,CN}
\qquad\text{for every battery $b$}.
\end{equation}
Thus the centered real-time component is the same with and without the subsidy.

Now define battery averages
\[
\widehat p:=\frac1n\sum_{b=1}^n p_b,
\qquad
\widehat r:=\frac1n\sum_{b=1}^n r_b,
\qquad
\widehat\tau:=\frac1n\sum_{b=1}^n\tau_b,
\qquad
\theta:=\Delta+\widehat\tau.
\]
Averaging \eqref{eq:subsidy-general-p-foc}--\eqref{eq:subsidy-general-r-foc}
across batteries and solving gives
\begin{equation}\label{eq:subsidy-general-averages}
\widehat p=\frac{n+1}{D_n}\theta,
\qquad
\widehat r=\frac{k_f}{D_n}\theta.
\end{equation}
Subtracting the battery averages from the individual first-order conditions gives
\begin{equation}\label{eq:subsidy-general-deviations}
p_b-\widehat p=\tau_b-\widehat\tau,
\qquad
r_b-\widehat r=k_f(\tau_b-\widehat\tau).
\end{equation}
The predictable components need not be symmetric; equations
\eqref{eq:subsidy-general-averages}--\eqref{eq:subsidy-general-deviations} characterize
exactly how any asymmetry enters.

Let
\[
\delta_b:=\tau_b-\widehat\tau,
\qquad
V:=\sum_{b=1}^n\|\delta_b\|^2,
\qquad
M:=\frac{n(n+1)}{D_n},
\qquad
B:=\frac{nk_f}{D_n}.
\]
Then \eqref{eq:subsidy-general-averages} implies $P=M\theta$ and $R=B\theta$.
Let
\[
\textnormal{Payment}_s:=\sum_{b=1}^n\sum_{t=1}^T s_t[p_{b,t}]_+.
\]
Using \eqref{eq:subsidy-payment-identity}, the zero-sum property of $p_b$, and
\eqref{eq:subsidy-general-deviations}, the aggregate subsidy payment is
\begin{align}
\frac{\textnormal{Payment}_s}{\beta}
&=\frac1\beta\sum_{b=1}^n\langle\sigma_b,p_b\rangle \nonumber\\
&=\sum_{b=1}^n\langle\tau_b,p_b\rangle \nonumber\\
&=M\langle\widehat\tau,\theta\rangle+V.
\label{eq:subsidy-general-payment}
\end{align}
In particular, the right-hand side is nonnegative.

It remains to compare generation costs. Up to terms that do not depend on battery policies,
the generation cost in \eqref{eqn:gen-cost-linear} can be written as
\begin{equation}\label{eq:subsidy-general-generation-cost}
\frac\beta2\left[
\|\Delta-P\|^2
+\left(\frac1{k_f}-1\right)\|R\|^2
+\frac1{k_f}\bE\|\epsilon-\widetilde U\|^2
\right].
\end{equation}
The final term is the same with and without the subsidy by
\eqref{eq:subsidy-general-centered-policy}. For aggregate predictable policies of the form
$P=Mq$ and $R=Bq$, the remaining part of \eqref{eq:subsidy-general-generation-cost} is
\[
\beta\left[
\frac12\|\Delta\|^2
-M\langle\Delta,q\rangle
+C_n\|q\|^2
\right],
\qquad
C_n:=\frac{n^2\left((n+1)^2+k_f(1-k_f)\right)}{2D_n^2}.
\]
The unsubsidized equilibrium corresponds to $q=\Delta$, while the subsidized equilibrium
corresponds to $q=\theta=\Delta+\widehat\tau$. Expanding their difference gives
\begin{equation}\label{eq:subsidy-general-generation-difference}
\frac{\textnormal{GenerationCost}_s-\costdcn}{\beta}
=(2C_n-M)\langle\theta,\widehat\tau\rangle
+(M-C_n)\|\widehat\tau\|^2.
\end{equation}
Adding \eqref{eq:subsidy-general-payment} to
\eqref{eq:subsidy-general-generation-difference} and then using
\eqref{eq:subsidy-general-payment} once more yields the exact decomposition
\begin{align}
\frac{\textnormal{Cost}(\textnormal{DCN-}s)-\costdcn}{\beta}
&=\frac{2C_n}{M}\frac{\textnormal{Payment}_s}{\beta}
+\left(1-\frac{2C_n}{M}\right)V
+(M-C_n)\|\widehat\tau\|^2.
\label{eq:subsidy-general-financial-difference}
\end{align}
Every term on the right-hand side is nonnegative. The payment is nonnegative because
$s_t\ge0$, and $V\ge0$. Moreover,
\[
M-2C_n
=
\frac{n\left[n+1+n(1-k_f)(n+1-k_f)\right]}{D_n^2}
>0.
\]
Thus $M>2C_n>0$, which also implies $M-C_n>0$. Equation
\eqref{eq:subsidy-general-financial-difference} therefore proves
$\textnormal{Cost}(\textnormal{DCN-}s)\ge\costdcn$ at every equilibrium.
\end{proof}

\section{Theorem Statements and Proofs for Section~\ref{sec:extensions}}\label{app:sec:theorem-extensions}

This appendix gives the formal theorem statements and proofs for the three extensions in Section~\ref{sec:extensions}, in the same order as the main text. Appendix~\ref{app:subsec:battery-capacity} proves equilibrium existence and uniqueness and the heterogeneous-capacity bound for Section~\ref{subsec:battery-capacity}; Appendix~\ref{app:subsec:battery-inefficiency} proves the closed-form characterization and PoA bounds for Section~\ref{subsec:battery-inefficiency}; and Appendix~\ref{app:subsec:virtual-bidders} characterizes the equilibrium and PoA bounds for the virtual bidder extension in Section~\ref{subsec:virtual-bidders}.

\subsection{Battery Capacity}\label{app:subsec:battery-capacity}

\begin{theorem}[Heterogeneous Capacity Bound]\label{thm:hetero-capacity-bound}
Consider $n$ competing batteries with capacities $C_1,\ldots,C_n>0$ in the capacity-constrained model of Section~\ref{subsec:battery-capacity}. The game has a unique equilibrium DA/RT dispatch profile, and
$\costnb\ge \costdcn$ at that equilibrium. If $\costnb>\costdcn$, then
\[
\poa \le \frac{2n+3}{2n+4}
+\frac{1}{2}\sum_{b=1}^n \left(\frac{C_b}{\sum_{j=1}^n C_j}\right)^2.
\]
\end{theorem}

\begin{proof}[Proof of Theorem~\ref{thm:hetero-capacity-bound}]
Define $\varepsilon_t:=D_t-\mu_t$. For any DA/RT policy, its realized physical dispatch is
\[
q_t:=z_t^{DA}+z_t^{RT}(D_{1:t}).
\]
This is the quantity constrained by the battery's state of charge. More generally, call a
physical dispatch policy $q=(q_t)_{t=1}^T$ history-adapted if $q_t$ depends only on $D_{1:t}$.
For any such policy, define
\[
\bar q_t:=\bE[q_t],
\qquad
\tilde q_t:=q_t-\bar q_t.
\]
For $C\ge 0$, let $\mathcal Q(C)$ denote the set of feasible realized physical dispatch policies
for a battery with capacity $C$:
\[
\mathcal Q(C)
:=\left\{
q:\begin{array}{l}
q\ \text{is history-adapted},\\[1mm]
\sum_{t=1}^T q_t=0\ \text{pathwise},\\[1mm]
\exists s_0\in[0,C]\ \text{such that}\ 
0\le s_0-\sum_{\tau=1}^t q_\tau\le C
\ \text{for all }t\text{ and every demand path}
\end{array}
\right\}.
\]
The set $\mathcal Q(C)$ captures the physical dispatch and
state-of-charge restrictions; it is convex, contains $0$, and satisfies
$\mathcal Q(C)=C\mathcal Q(1)$.

To recover a DA/RT policy from a physical dispatch policy, it remains to specify how the
expected dispatch is split between the two settlements. Let $m_t$ denote expected real-time
dispatch in period $t$. Given any $q\in\mathcal Q(C)$ and any deterministic $m\in\R^T$
satisfying $\sum_t m_t=0$, the DA/RT policy
\[
z_t^{RT}=m_t+\tilde q_t,
\qquad
z_t^{DA}=\bar q_t-m_t
\]
satisfies the separate DA and RT balance constraints and induces physical dispatch $q$.
Conversely, every feasible DA/RT policy induces such a pair by taking
$q_t=z_t^{DA}+z_t^{RT}(D_{1:t})$ and $m_t=\bE[z_t^{RT}(D_{1:t})]$. The two
representations are therefore equivalent: feasible DA/RT policies are in one-to-one
correspondence with pairs $(q,m)$ such that $q\in\mathcal Q(C)$ and
$\sum_t m_t=0$; the physical dispatch $q$ is what enters the capacity constraint, while
$m$ records the expected RT component.

Throughout this proof, use the inner product
\[
\ip{x}{y}:=\sum_{t=1}^T \bE[x_t y_t],
\]
with the expectation omitted when both arguments are deterministic. If an aggregate policy has
physical dispatch $q$ and expected real-time dispatch $m$, write, for readability,
$y:=\bar q$ and $v:=\tilde q$. The induced aggregate DA and RT dispatches are $y-m$
and $m+v$, respectively. Thus the modified DA and RT demands in
\eqref{eqn:gen-cost-linear} are
\[
\tilde d_t^{DA}=\mu_t-y_t+m_t,
\qquad
\tilde d_t^{RT}
=\mu_t-y_t+m_t+\frac{\varepsilon_t-m_t-v_t}{k_f}.
\]
Therefore the quadratic part of the cost reduction relative to no battery is
\[
\begin{aligned}
\frac{\costnb-\cost(q,m)}{\beta}
&=
\frac{1-k_f}{2}\left(\norm{\mu}^2-\norm{\mu-y+m}^2\right)\\
&\quad
+\frac{k_f}{2}\left(
\norm{\mu+\varepsilon/k_f}^2
-\norm{\mu-y+m+(\varepsilon-m-v)/k_f}^2
\right)\\
&=
\ip{\mu}{y}-\frac12\norm{y}^2
-\frac12\left(\frac1{k_f}-1\right)\norm{m}^2
+\frac1{k_f}\left(\ip{\varepsilon}{v}-\frac12\norm{v}^2\right).
\end{aligned}
\]
The simplification uses $\bE[\varepsilon_t]=\bE[v_t]=0$ and deterministic $m$. Hence
\[
\costnb-\cost(q,m)=\beta W(q,m),
\]
where
\begin{align}
W(q,m)
&:=
\ip{\mu}{y}-\frac12\norm{y}^2
-\frac12\left(\frac1{k_f}-1\right)\norm{m}^2
+\frac1{k_f}\left(\ip{\varepsilon}{v}-\frac12\norm{v}^2\right).
\label{eq:capacity_full_value_function}
\end{align}
The intercept $\alpha$ drops out because all feasible DA, RT, and physical dispatch policies are
balanced over the horizon.

We first identify the centralized benchmark. For the centralized planner, $n$ batteries
with capacities $C_1,\ldots,C_n$ are equivalent
to one battery with total capacity
\[
C_{\mathrm{tot}}:=\sum_{b=1}^n C_b.
\]
Indeed, aggregate feasible physical dispatches can be added across batteries, and any
$q\in\mathcal Q(C_{\mathrm{tot}})$ can be implemented by assigning the capacity share
$w_b q$ to battery $b$, where
\[
w_b:=\frac{C_b}{C_{\mathrm{tot}}}.
\]
The centralized benchmark therefore chooses
\[
q^*\in\argmax_{q\in\mathcal Q(C_{\mathrm{tot}})} W(q,0).
\]
For any fixed $q$, the only term in $W(q,m)$ that depends on $m$ is
\[
-\frac12\left(\frac1{k_f}-1\right)\norm{m}^2,
\]
so $m=0$ is optimal.

We next analyze the decentralized game. For any feasible profile
$(z_b^{DA},z_b^{RT})_{b=1}^n$, set
\[
q_{b,t}:=z_{b,t}^{DA}+z_{b,t}^{RT}(D_{1:t}),
\qquad
m_{b,t}:=\bE[z_{b,t}^{RT}(D_{1:t})].
\]
Then $q_b\in\mathcal Q(C_b)$ and $\sum_t m_{b,t}=0$.
Write
\[
y_b:=\bar q_b,\qquad v_b:=\tilde q_b,\qquad
y:=\sum_b y_b,\qquad v:=\sum_b v_b,\qquad m:=\sum_b m_b,
\]
and also
\[
\widetilde y:=\sum_b w_b y_b,\qquad \widetilde v:=\sum_b w_b v_b.
\]
These capacity-weighted averages appear because battery $b$ will be compared to the feasible
deviation that uses its capacity share of the centralized physical dispatch.
In the $(q_b,m_b)$ representation, the best-response conditions separate into an
unconstrained first-order condition for the expected RT split $m_b$ and a variational
inequality for the capacity-constrained physical dispatch $q_b$.
After dividing profit by $\beta$, battery $b$'s payoff can be written as
\[
\begin{aligned}
\Pi_b/\beta
&=
\ip{\mu+m-y}{y_b-m_b}
+\ip{\mu+m-y+(\varepsilon-m-v)/k_f}{m_b+v_b}\\
&=
\ip{\mu+m-y}{y_b}
+\frac1{k_f}\ip{\varepsilon-v}{v_b}
-\frac1{k_f}\ip{m}{m_b}.
\end{aligned}
\]
The first line is the sum of DA revenue and RT revenue after substituting
$z_b^{DA}=y_b-m_b$ and $z_b^{RT}=m_b+v_b$; the second line uses
$\bE[v_b]=0$ and deterministic $m_b$. The intercept term drops out by balance.

We first establish existence and uniqueness of the equilibrium dispatch. Identify policies
that agree almost surely and place the square-integrable adapted physical dispatches in their
usual $L^2$ Hilbert space. Because the horizon is finite, $\mathcal Q(C_b)$ is a nonempty,
closed, and convex subset of that space: the zero dispatch is feasible, and adaptation,
the almost-sure balance condition, and the state-of-charge inequalities are preserved under limits. Let
\[
\mathcal M:=\{m\in\mathbb R^T:\mathbf 1^\top m=0\},
\qquad
\mathcal K:=\prod_{b=1}^n\bigl(\mathcal Q(C_b)\times\mathcal M\bigr).
\]
Thus $\mathcal K$ is a nonempty, closed, convex subset of a Hilbert space.

Let $F$ be the negative pseudo-gradient of the batteries' payoffs. In the orthogonal
coordinates $q_b=y_b+v_b$, where $y_b=\bE[q_b]$ and $\bE[v_b]=0$, its components are
\[
F_b^y=y+y_b-m-\mu,
\qquad
F_b^v=\frac1{k_f}(v+v_b-\varepsilon),
\qquad
F_b^m=\frac1{k_f}(m+m_b)-y_b.
\]
To verify strong monotonicity, consider two profiles and let $\delta y_b$, $\delta v_b$,
and $\delta m_b$ denote the differences between their components. Write
$Y:=\sum_b\delta y_b$, $V:=\sum_b\delta v_b$, and $M:=\sum_b\delta m_b$.
The inner product of the profile difference with the corresponding pseudo-gradient difference is
\[
\begin{aligned}
&\|Y\|^2+\sum_b\|\delta y_b\|^2-\ip{Y}{M}\\
&\quad+\frac1{k_f}\left(
\|V\|^2+\sum_b\|\delta v_b\|^2
+\|M\|^2+\sum_b\|\delta m_b\|^2
\right)
-\sum_b\ip{\delta y_b}{\delta m_b}.
\end{aligned}
\]
Using $2\ip{a}{b}\le\|a\|^2+\|b\|^2$ and $k_f\le1$, this expression is at least
\[
\frac12\sum_b\left(
\|\delta y_b\|^2+\|\delta v_b\|^2+\|\delta m_b\|^2
\right).
\]
Hence $F$ is strongly monotone; it is also affine and Lipschitz continuous. The standard
projection argument for strongly monotone variational inequalities therefore gives a unique
solution of the variational inequality on $\mathcal K$. Each battery's payoff is a strictly
concave quadratic in its own $(q_b,m_b)$, so this variational-inequality solution is exactly
the unique equilibrium dispatch profile. The relations
$z_b^{RT}=m_b+v_b$ and $z_b^{DA}=y_b-m_b$ recover its unique DA/RT representation.

We now use the equilibrium conditions.
The variable $m_b$ is unconstrained except for $\sum_t m_{b,t}=0$, and the payoff is strictly
concave in $m_b$. The capacity constraint is imposed on $q_b$, not on this settlement
split, so the $m_b$ part gives an equality first-order condition. Hence the first-order condition in $m_b$ is
\begin{equation}\label{eq:capacity_m_foc}
m+m_b=k_f y_b,\qquad b=1,\ldots,n.
\end{equation}
The zero-sum constraint introduces no additional constant term because both sides of
\eqref{eq:capacity_m_foc} have zero total over time.
Summing over $b$ gives
\begin{equation}\label{eq:capacity_m_aggregate}
m=\frac{k_f}{n+1}y.
\end{equation}

It remains to derive the first-order condition for the capacity-constrained physical
policy. Holding $m_b$ fixed and varying only $q_b$, write
$y=y_{-b}+y_b$ and $v=v_{-b}+v_b$. The part of battery $b$'s payoff that depends on
$q_b$ is
\[
\ip{\mu+m-y_{-b}-y_b}{y_b}
+\frac1{k_f}\ip{\varepsilon-v_{-b}-v_b}{v_b}.
\]
The directional derivative from $q_b$ toward an alternative
$x_b\in\mathcal Q(C_b)$ is therefore
\[
\ip{\mu+m-y_{-b}-2y_b}{\bar x_b-y_b}
+\frac1{k_f}\ip{\varepsilon-v_{-b}-2v_b}{\tilde x_b-v_b}.
\]
Since $y=y_{-b}+y_b$ and $v=v_{-b}+v_b$, this derivative is
\[
\ip{\mu+m-y-y_b}{\bar x_b-y_b}
+\frac1{k_f}\ip{\varepsilon-v-v_b}{\tilde x_b-v_b}.
\]
At an equilibrium, this derivative must be nonpositive for every feasible direction in
the convex set $\mathcal Q(C_b)$.
The first-order variational inequality for the physical policy $q_b$ says that, for every
$x_b\in\mathcal Q(C_b)$,
\begin{align}
\ip{\mu+m-y-y_b}{\bar x_b-y_b}
+\frac1{k_f}\ip{\varepsilon-v-v_b}{\tilde x_b-v_b}
\le 0 .
\label{eq:capacity_physical_vi}
\end{align}
Let $q^*$ be a centralized optimal physical policy, and write $y^*:=\bar q^*$ and
$v^*:=\tilde q^*$.
Fix $\theta\in[0,1]$ to be chosen below and test \eqref{eq:capacity_physical_vi} with
\[
x_b=\theta w_b q^*,\qquad b=1,\ldots,n.
\]
This is feasible because $q^*\in\mathcal Q(C_{\mathrm{tot}})$ implies
$w_b q^*\in\mathcal Q(C_b)$ and $\mathcal Q(C_b)$ is convex and contains $0$.
For this deviation, \eqref{eq:capacity_physical_vi} becomes
\[
\begin{aligned}
\theta w_b\left[
\ip{\mu+m-y-y_b}{y^*}
+\frac1{k_f}\ip{\varepsilon-v-v_b}{v^*}
\right]
&\le
\ip{\mu+m-y-y_b}{y_b}
+\frac1{k_f}\ip{\varepsilon-v-v_b}{v_b}.
\end{aligned}
\]
The terms weighted by $w_b$ are what produce the capacity-weighted averages
$\widetilde y$ and $\widetilde v$ after summing over batteries. Summing
\eqref{eq:capacity_physical_vi} over $b$ gives
\begin{align}
&\theta\left[
\ip{\mu+m-y-\widetilde y}{y^*}
+\frac1{k_f}\ip{\varepsilon-v-\widetilde v}{v^*}
\right] \nonumber\\
&\qquad\le
\ip{\mu+m-y}{y}-\sum_b\norm{y_b}^2
+\frac1{k_f}\left(\ip{\varepsilon-v}{v}-\sum_b\norm{v_b}^2\right).
\label{eq:capacity_summed_vi}
\end{align}

Let $W^*:=W(q^*,0)$ and $W^{NE}:=W(\sum_b q_b,m)$. By
\eqref{eq:capacity_full_value_function}, the centralized value is
\[
W^*
=
\ip{\mu}{y^*}-\frac12\norm{y^*}^2
+\frac1{k_f}\left(\ip{\varepsilon}{v^*}-\frac12\norm{v^*}^2\right).
\]
Equation~\eqref{eq:capacity_summed_vi} bounds the linear terms in
$\theta W^*$. To see this, rearrange the left-hand side of
\eqref{eq:capacity_summed_vi} as
\[
\begin{aligned}
\theta\ip{\mu}{y^*}+\frac{\theta}{k_f}\ip{\varepsilon}{v^*}
&
-\theta\ip{y+\widetilde y-m}{y^*}
-\frac{\theta}{k_f}\ip{v+\widetilde v}{v^*}.
\end{aligned}
\]
Hence \eqref{eq:capacity_summed_vi} implies
\[
\begin{aligned}
\theta\ip{\mu}{y^*}+\frac{\theta}{k_f}\ip{\varepsilon}{v^*}
&\le
R+\theta\ip{y+\widetilde y-m}{y^*}
+\frac{\theta}{k_f}\ip{v+\widetilde v}{v^*},
\end{aligned}
\]
where
\[
R:=
\ip{\mu+m-y}{y}-\sum_b\norm{y_b}^2
+\frac1{k_f}\left(\ip{\varepsilon-v}{v}-\sum_b\norm{v_b}^2\right).
\]
Now set
\[
r_y:=y+\widetilde y-m,\qquad r_v:=v+\widetilde v.
\]
Substituting this bound on the linear terms into $\theta W^*$ gives
\begin{align*}
\theta W^*
\le
R+\theta\ip{r_y}{y^*}-\frac{\theta}{2}\norm{y^*}^2
+\frac1{k_f}\left(\theta\ip{r_v}{v^*}-\frac{\theta}{2}\norm{v^*}^2\right).
\end{align*}
Completing squares in $y^*$ and $v^*$ gives
\[
\theta W^*
\le
R+\frac{\theta}{2}\norm{r_y}^2
+\frac{\theta}{2k_f}\norm{r_v}^2.
\]
At the equilibrium aggregate policy,
\[
W^{NE}
=
\ip{\mu}{y}-\frac12\norm{y}^2
-\frac12\left(\frac1{k_f}-1\right)\norm{m}^2
+\frac1{k_f}\left(\ip{\varepsilon}{v}-\frac12\norm{v}^2\right).
\]
Thus, subtracting $W^{NE}$ from the right-hand side of the previous display isolates
exactly the two quadratic remainders
\[
\begin{aligned}
R+\frac{\theta}{2}\norm{r_y}^2+\frac{\theta}{2k_f}\norm{r_v}^2-W^{NE}
&=
-\left[
\frac12\norm{y}^2+\sum_b\norm{y_b}^2-\ip{m}{y}
-\frac12\left(\frac1{k_f}-1\right)\norm{m}^2
-\frac{\theta}{2}\norm{y+\widetilde y-m}^2
\right]\\
&\quad
-\frac1{k_f}\left[
\frac12\norm{v}^2+\sum_b\norm{v_b}^2
-\frac{\theta}{2}\norm{v+\widetilde v}^2
\right].
\end{aligned}
\]
Combining this identity with the completed-square bound gives
\begin{align}
\theta W^*
\le
W^{NE}
&-\left[
\frac12\norm{y}^2+\sum_b\norm{y_b}^2-\ip{m}{y}
-\frac12\left(\frac1{k_f}-1\right)\norm{m}^2
-\frac{\theta}{2}\norm{y+\widetilde y-m}^2
\right]\nonumber\\
&-\frac1{k_f}\left[
\frac12\norm{v}^2+\sum_b\norm{v_b}^2
-\frac{\theta}{2}\norm{v+\widetilde v}^2
\right].
\label{eq:capacity_after_squares}
\end{align}
It remains to choose $\theta$ so that the two bracketed terms are nonnegative. This follows
from two pointwise matrix inequalities. We use the fact that, for any positive definite matrix
$M$ and vector $c$,
\begin{equation}\label{eq:capacity_matrix_cs}
(c^\top a)^2\le (c^\top M^{-1}c)\,a^\top M a
\qquad\text{for all vectors }a.
\end{equation}
This is Cauchy--Schwarz applied to
$c^\top a=(M^{-1/2}c)^\top(M^{1/2}a)$.
Let $\one=(1,\ldots,1)^\top$ and let $w=(w_1,\ldots,w_n)^\top$ be the vector of
capacity shares, $w_b=C_b/C_{\mathrm{tot}}$. Define
\[
\lambda_C:=
\frac{2n+3}{2n+4}+\frac12\sum_{b=1}^n w_b^2.
\]
We claim that, for every $a\in\R^n$,
\begin{align}
\lambda_C^{-1}\bigl((\one+w)^\top a\bigr)^2
&\le
(\one^\top a)^2+2\norm{a}^2,
\label{eq:capacity_matrix_stoch}\\
\lambda_C^{-1}
\left(\left(1-\frac{k_f}{n+1}\right)\one^\top a+w^\top a\right)^2
&\le
\left(1-\frac{2k_f}{n+1}-\frac{k_f(1-k_f)}{(n+1)^2}\right)(\one^\top a)^2
+2\norm{a}^2.
\label{eq:capacity_matrix_det}
\end{align}
We first prove \eqref{eq:capacity_matrix_stoch}. Let $J:=\one\one^\top$. The right-hand
side of \eqref{eq:capacity_matrix_stoch} is $a^\top(J+2I)a$. Applying
\eqref{eq:capacity_matrix_cs} with $M=J+2I$ and $c=\one+w$, it suffices to compute
$(\one+w)^\top(J+2I)^{-1}(\one+w)$. Since $J=\one\one^\top$, the
Sherman--Morrison formula gives
\[
(J+2I)^{-1}=\frac12 I-\frac{1}{2(n+2)}J.
\]
Also, $w$ is the vector of capacity shares, so $\one^\top w=1$. Therefore
\[
\norm{\one+w}^2=n+2+\sum_{b=1}^n w_b^2,
\qquad
\one^\top(\one+w)=n+1,
\]
and hence
\[
\begin{aligned}
(\one+w)^\top(J+2I)^{-1}(\one+w)
&=\frac12\norm{\one+w}^2-\frac{1}{2(n+2)}
\bigl(\one^\top(\one+w)\bigr)^2  \\
&=\frac12\left(n+2+\sum_{b=1}^n w_b^2\right)
-\frac{(n+1)^2}{2(n+2)} \\
&=\frac{2n+3}{2n+4}+\frac12\sum_{b=1}^n w_b^2
=\lambda_C.
\end{aligned}
\]
This proves \eqref{eq:capacity_matrix_stoch}.

We next prove \eqref{eq:capacity_matrix_det}. Define
\[
d(k_f):=
1-\frac{2k_f}{n+1}-\frac{k_f(1-k_f)}{(n+1)^2}.
\]
Because $d'(k_f)<0$, we have $d(k_f)>(n-1)/(n+1)\ge0$ for
$k_f\in(0,1)$, so the matrix $d(k_f)J+2I$ is positive definite. Applying
\eqref{eq:capacity_matrix_cs} with
\[
M=d(k_f)J+2I,\qquad
c=\left(1-\frac{k_f}{n+1}\right)\one+w,
\]
it suffices to show $c^\top M^{-1}c\le\lambda_C$. Again using the
Sherman--Morrison formula,
\[
M^{-1}=\frac12 I-\frac{d(k_f)}{2(2+n d(k_f))}J.
\]
Write
\[
A:=1-\frac{k_f}{n+1},
\qquad c=A\one+w.
\]
Since $\one^\top w=1$,
\[
\norm{c}^2=nA^2+2A+\sum_{b=1}^n w_b^2,
\qquad
\one^\top c=nA+1.
\]
Thus the constant $c^\top M^{-1}c$ is
\[
\lambda_{\det}(k_f)
=
\frac12\left(nA^2+2A+\sum_{b=1}^n w_b^2\right)
-\frac{d(k_f)}{2(2+n d(k_f))}
\left(nA+1\right)^2,
\]
and, substituting the definition of $A$, a direct simplification gives
\[
\lambda_C-\lambda_{\det}(k_f)
=
\frac{k_f\bigl(k_f+(n+1)^2\bigr)}
{(n+2)(n+1)^2\bigl(2+n d(k_f)\bigr)}
\ge 0.
\]
Hence $\lambda_{\det}(k_f)\le\lambda_C$ for $k_f\in(0,1)$, proving
\eqref{eq:capacity_matrix_det}.

Set $\theta:=\lambda_C^{-1}$. Since $\sum_b w_b^2\ge 1/n$, we have $\lambda_C\ge1$ and
therefore $\theta\in[0,1]$. Applying \eqref{eq:capacity_matrix_stoch} pointwise to
$a=(v_{1,t},\ldots,v_{n,t})$ and then summing over $t$ and taking expectations over
demand realizations gives
\[
\frac12\norm{v}^2+\sum_b\norm{v_b}^2-\frac{\theta}{2}\norm{v+\widetilde v}^2\ge 0.
\]
Indeed, for this choice of $a$,
\[
\one^\top a=v_t,\qquad
w^\top a=\widetilde v_t,\qquad
\norm{a}^2=\sum_b v_{b,t}^2,
\]
so \eqref{eq:capacity_matrix_stoch}, divided by $2$, is exactly the pointwise
version of the displayed inequality.
Similarly, applying \eqref{eq:capacity_matrix_det} pointwise to
$a=(y_{1,t},\ldots,y_{n,t})$ and using \eqref{eq:capacity_m_aggregate} gives
\[
\frac12\norm{y}^2+\sum_b\norm{y_b}^2-\ip{m}{y}
-\frac12\left(\frac1{k_f}-1\right)\norm{m}^2
-\frac{\theta}{2}\norm{y+\widetilde y-m}^2
\ge 0.
\]
Here \eqref{eq:capacity_m_aggregate} implies
$m_t=k_f y_t/(n+1)$, so
\[
y_t+\widetilde y_t-m_t
=
\left(1-\frac{k_f}{n+1}\right)y_t+\widetilde y_t
\]
and
\[
\begin{aligned}
\frac12 y_t^2-y_tm_t
-\frac12\left(\frac1{k_f}-1\right)m_t^2+\sum_b y_{b,t}^2
&=
\frac12\left(1-\frac{2k_f}{n+1}
-\frac{k_f(1-k_f)}{(n+1)^2}\right)y_t^2
+\sum_b y_{b,t}^2.
\end{aligned}
\]
Thus \eqref{eq:capacity_matrix_det}, again divided by $2$, gives the deterministic
bracket after summing over $t$.
Therefore \eqref{eq:capacity_after_squares} yields
\[
\theta W^*\le W^{NE}.
\]

We now translate this value comparison into the PoA bound. Because the zero policy is
feasible for the centralized problem, $W^*\ge 0$. Hence
$W^{NE}\ge 0$, and therefore $\costnb-\costdcn=\beta W^{NE}\ge0$.

For any fixed aggregate physical policy, setting $m=0$ weakly increases $W$ because the only
$m$-dependent term in \eqref{eq:capacity_full_value_function} is
$-\frac12(1/k_f-1)\norm{m}^2\le0$. Since the decentralized aggregate physical policy is feasible
for the centralized problem,
\[
W^{NE}=W\left(\sum_b q_b,m\right)
\le W\left(\sum_b q_b,0\right)
\le W^*.
\]
Thus, whenever $W^{NE}>0$,
\[
\poa=\frac{W^*}{W^{NE}}\le \theta^{-1}
=\frac{2n+3}{2n+4}+\frac12\sum_{b=1}^n w_b^2.
\]
\end{proof}

\subsection{Battery Inefficiency}\label{app:subsec:battery-inefficiency}

\begin{theorem}[Battery Inefficiency]
\label{thm:inefficiency}
Consider $n\ge1$ competing batteries in the battery-inefficiency model of
Section~\ref{subsec:battery-inefficiency}, and suppose
Assumption~\ref{assump:aligned-charge-discharge} holds. Define
\[
w_t \equiv
\begin{cases}
1, & t\in \mathcal T^+,\\
\eta, & t\in \mathcal T^-,
\end{cases}
\qquad
W_t \equiv \sum_{i=t}^T w_i^2.
\]
Let
\[
\epsilon_t \equiv D_t-\mu_t,
\qquad
\theta_t \equiv \mu_t+\frac{\alpha}{\beta},
\qquad
\bar\theta_w \equiv \frac{\sum_{t=1}^T w_t\theta_t}{\sum_{t=1}^T w_t^2},
\qquad
\Delta_t \equiv \theta_t-w_t\bar\theta_w .
\]
Assume the primitives are such that the policies displayed below satisfy the sign restrictions in
Assumption~\ref{assump:aligned-charge-discharge}.

A centralized optimal policy is
\[
z_t^{DA,CN}=\Delta_t,\qquad t=1,\dots,T,
\]
and, for $t=1,\dots,T-1$,
\[
z_t^{RT,CN}(D_{1:t})
=
\frac{W_{t+1}}{W_t}\,\epsilon_t
-\frac{w_t}{W_t}\sum_{i=t+1}^T w_i\bigl(\E[D_i\mid D_{1:t}]-\mu_i\bigr)
-\frac{w_t}{W_t}\sum_{s=1}^{t-1} w_s z_s^{RT,CN}(D_{1:s}),
\]
with terminal value
\[
z_T^{RT,CN}(D_{1:T})
=
-\frac{1}{w_T}\sum_{t=1}^{T-1} w_t z_t^{RT,CN}(D_{1:t}).
\]

The game has a unique equilibrium, which is symmetric. In this equilibrium, each battery uses
\[
\begin{aligned}
z_{b,t}^{DA,DCN}
&= \frac{n+1-k_f}{(n+1)^2-nk_f}\,\Delta_t,\\
z_{b,t}^{RT,DCN}(D_{1:t})
&=
\frac{k_f}{(n+1)^2-nk_f}\,\Delta_t
+\frac{1}{n+1}z_t^{RT,CN}(D_{1:t}),
\qquad t=1,\dots,T.
\end{aligned}
\]

The corresponding price of anarchy satisfies
\[
1+\frac{1}{n(n+1)(n^2+n+2)}
\;\le\;
\poa
\;\le\;
1+\frac{1}{n(n+2)}.
\]
\end{theorem}

\begin{proof}[Proof of Theorem~\ref{thm:inefficiency}]
Within the sign region specified by Assumption~\ref{assump:aligned-charge-discharge}, the
round-trip-efficiency constraints reduce to the weighted balance constraints
\[
\sum_{t=1}^T w_t z_{b,t}^{DA}=0,
\qquad
\sum_{t=1}^T w_t z_{b,t}^{RT}(D_{1:t})=0
\quad\text{a.s.}
\]
In particular,
\[
\sum_{t=1}^T w_t\Delta_t
=
\sum_{t=1}^T w_t\theta_t-\bar\theta_w\sum_{t=1}^T w_t^2
=0.
\]

Consider first the centralized problem. Let $x_t$ denote the aggregate day-ahead battery action and
$u_t(D_{1:t})$ the aggregate real-time battery action. The centralized objective is
\[
\sum_{t=1}^T
\left[
(1-k_f)\left(\alpha(\mu_t-x_t)+\frac{\beta}{2}(\mu_t-x_t)^2\right)
+k_f\,\E\left[
\alpha\left(\mu_t-x_t+\frac{\epsilon_t-u_t}{k_f}\right)
+\frac{\beta}{2}\left(\mu_t-x_t+\frac{\epsilon_t-u_t}{k_f}\right)^2
\right]
\right],
\]
subject to
\[
\sum_{t=1}^T w_t x_t=0,
\qquad
\sum_{t=1}^T w_t u_t(D_{1:t})=0
\quad\text{a.s.}
\]

Write $m_t\equiv \E[u_t]$, $v_t\equiv u_t-m_t$, and $y_t\equiv x_t+m_t$. Then
\[
\E[v_t]=0,
\qquad
\sum_{t=1}^T w_t x_t=0,
\qquad
\sum_{t=1}^T w_t m_t=0,
\qquad
\sum_{t=1}^T w_t v_t=0
\quad\text{a.s.}
\]
Also, $x_t=y_t-m_t$ and $u_t=m_t+v_t$. Substituting these identities into the objective and using
$\E[\epsilon_t]=\E[v_t]=0$ gives the period-$t$ cost as
\[
\alpha(\mu_t-y_t)
+\frac{\beta}{2}(\mu_t-y_t)^2
+\frac{\beta}{2}\Bigl(\frac1{k_f}-1\Bigr)m_t^2
+\frac{\beta}{2k_f}\E[(\epsilon_t-v_t)^2].
\]
Therefore the centralized problem separates into
\[
\min_{y\in H_w}
\sum_{t=1}^T\left[\alpha(\mu_t-y_t)+\frac{\beta}{2}(\mu_t-y_t)^2\right],
\qquad
\min_{m\in H_w}
\frac{\beta}{2}\Bigl(\frac1{k_f}-1\Bigr)\sum_{t=1}^T m_t^2,
\]
and
\[
\min_{v\in\mathcal H_w}
\sum_{t=1}^T \E[(\epsilon_t-v_t)^2],
\]
where
\[
H_w\equiv \left\{z\in\mathbb R^T:\ \sum_{t=1}^T w_t z_t=0\right\},
\qquad
\mathcal H_w
\equiv
\left\{
v=(v_t)_{t=1}^T:
v_t \text{ is adapted to } D_{1:t},\ \E[v_t]=0,\ \sum_{t=1}^T w_t v_t=0 \ \text{a.s.}
\right\}.
\]

For the $y$-problem, completing the square gives
\[
\alpha(\mu_t-y_t)+\frac{\beta}{2}(\mu_t-y_t)^2
=
\frac{\beta}{2}(y_t-\theta_t)^2+\text{constant}.
\]
Thus $y$ is the Euclidean projection of $\theta=(\theta_t)_{t=1}^T$ onto $H_w$, so
\[
y_t=\theta_t-w_t\bar\theta_w=\Delta_t.
\]
For the $m$-problem, the minimizer is $m_t=0$ for all $t$.
Hence the displayed centralized policy has $z_t^{DA,CN}=x_t=y_t=\Delta_t$.

It remains to solve the $v$-problem. The mean-zero restriction is without loss: if an adapted
policy satisfies the pathwise weighted balance constraint, subtracting its componentwise means
preserves that constraint and weakly lowers $\sum_t\E[(\epsilon_t-v_t)^2]$. Thus we can solve the
pathwise projection problem and obtain an element of $\mathcal H_w$. For
$t=1,\dots,T$, let $\E_t[\cdot]\equiv \E[\cdot\mid D_{1:t}]$. For a residual weighted-balance
target $r$ that is measurable with respect to the current history, define
\[
V_t(r)\equiv
\inf\left\{
\E_t\!\left[\sum_{i=t}^T (\epsilon_i-u_i)^2\right]:
\begin{array}{l}
u_i \text{ is adapted to } D_{1:i} \text{ for } i=t,\dots,T,\\[1mm]
\sum_{i=t}^T w_i u_i=r \ \text{a.s.}
\end{array}
\right\}.
\]
We claim that
\[
V_t(r)=C_t+\frac{(r-\Gamma_t)^2}{W_t},
\qquad
\Gamma_t\equiv \E_t\!\left[\sum_{i=t}^T w_i\epsilon_i\right],
\]
where $C_t$ may depend on the current history but is independent of $r$.

For $t=T$, feasibility forces $u_T=r/w_T$, so
\[
V_T(r)=\left(\epsilon_T-\frac{r}{w_T}\right)^2
=\frac{(r-w_T\epsilon_T)^2}{w_T^2}
=\frac{(r-\Gamma_T)^2}{W_T}.
\]
Now suppose the claim holds at time $t+1$. For brevity, define
\[
B_t(D_{1:t})\equiv \sum_{i=t+1}^T w_i\bigl(\E[D_i\mid D_{1:t}]-\mu_i\bigr).
\]
The induction hypothesis gives, up to a term independent of $r$ and $u_t$,
\[
\E_t[V_{t+1}(r-w_tu_t)]
=
\frac{1}{W_{t+1}}\E_t[(r-w_tu_t-\Gamma_{t+1})^2].
\]
Since $r$ and $u_t$ are measurable with respect to $D_{1:t}$ and
$\E_t[\Gamma_{t+1}]=B_t(D_{1:t})$, the conditional variance decomposition gives
\[
\E_t[(r-w_tu_t-\Gamma_{t+1})^2]
=
(r-w_tu_t-B_t(D_{1:t}))^2+\text{constant},
\]
where the constant is independent of $r$ and $u_t$. Therefore, over adapted $u_t$, the minimizer
solves the strictly convex
quadratic problem
\[
\inf_{u_t}
\left\{
(\epsilon_t-u_t)^2
+\frac{(r-w_tu_t-B_t(D_{1:t}))^2}{W_{t+1}}
\right\}.
\]
The first-order condition is
\[
-2(\epsilon_t-u_t)-\frac{2w_t}{W_{t+1}}(r-w_tu_t-B_t(D_{1:t}))=0.
\]
Using $W_t=W_{t+1}+w_t^2$, we obtain
\[
u_t^*(r)=\frac{W_{t+1}}{W_t}\epsilon_t+\frac{w_t}{W_t}\bigl(r-B_t(D_{1:t})\bigr).
\]
Substituting back shows that $V_t(r)$ again has the claimed form, with a new constant $C_t$
independent of $r$. This proves the claim by backward induction.

Now let
\[
s_{t-1}\equiv \sum_{i=1}^{t-1} w_i z_i^{RT,CN}(D_{1:i}),
\qquad s_0=0.
\]
At time $t$, the residual weighted balance from periods $t,\dots,T$ must equal $-s_{t-1}$, so
\[
z_t^{RT,CN}(D_{1:t})
=
u_t^*(-s_{t-1})
=
\frac{W_{t+1}}{W_t}\epsilon_t
-\frac{w_t}{W_t}B_t(D_{1:t})
-\frac{w_t}{W_t}\sum_{s=1}^{t-1} w_s z_s^{RT,CN}(D_{1:s}),
\]
for $t=1,\dots,T-1$, and the terminal action is forced by feasibility:
\[
z_T^{RT,CN}(D_{1:T})
=
-\frac{1}{w_T}\sum_{t=1}^{T-1} w_t z_t^{RT,CN}(D_{1:t}).
\]
This gives the centralized real-time policy stated in the theorem.

We now turn to the decentralized game. Consider an arbitrary decentralized equilibrium.
For each battery $b$, write its day-ahead action as $x_b\in H_w$ and decompose its real-time
action as
\[
u_b=m_b+v_b,\qquad m_b:=\E[u_b]\in H_w,\qquad v_b\in\mathcal H_w.
\]
Let
\[
X:=\sum_{b=1}^n x_b,\qquad
M:=\sum_{b=1}^n m_b,\qquad
V:=\sum_{b=1}^n v_b.
\]
After dividing by $\beta$, battery $b$'s expected profit can be written as
\[
\sum_{t=1}^T
\left[
\theta_t(x_{b,t}+m_{b,t})
-X_t(x_{b,t}+m_{b,t})
-\frac1{k_f}M_t m_{b,t}
\right]
+\frac1{k_f}\sum_{t=1}^T\E[(\epsilon_t-V_t)v_{b,t}].
\]
The deterministic part is strictly concave in battery $b$'s own pair $(x_b,m_b)$, and the
stochastic part is strictly concave in $v_b$. Therefore the first-order conditions below are
necessary and sufficient for a best response.

For the deterministic part, the first-order conditions on the weighted zero-sum subspace give
multipliers $\lambda_b$ and $\nu_b$ such that, for all $t$,
\[
\theta_t-X_t-x_{b,t}-m_{b,t}=\lambda_b w_t,
\qquad
\theta_t-X_t-\frac{M_t+m_{b,t}}{k_f}=\nu_b w_t.
\]
Comparing two batteries $b$ and $c$ gives
\[
(x_{b,t}-x_{c,t})+(m_{b,t}-m_{c,t})=-(\lambda_b-\lambda_c)w_t,
\qquad
m_{b,t}-m_{c,t}=-k_f(\nu_b-\nu_c)w_t.
\]
Because $x_b-x_c$ and $m_b-m_c$ both belong to $H_w$, multiplying by $w_t$ and summing over
$t$ implies $\nu_b=\nu_c$ and then $\lambda_b=\lambda_c$. Hence
\[
x_b=x_c,\qquad m_b=m_c
\]
for all batteries $b$ and $c$.

For the stochastic part, the first-order condition is
\[
\sum_{t=1}^T\E[(\epsilon_t-V_t-v_{b,t})h_t]=0
\qquad\forall\, h\in\mathcal H_w.
\]
Since $V+v_b\in\mathcal H_w$, this says that $V+v_b$ is the orthogonal projection of
$\epsilon$ onto $\mathcal H_w$. By the centralized analysis above, that projection is
$z^{RT,CN}$. Thus
\[
V+v_b=z^{RT,CN}
\]
for every battery $b$, and comparing across batteries gives $v_b=v_c$ for all $b,c$.

We have shown that any equilibrium must be symmetric. Let the common components be denoted by
$(x,m,v)$. For convenience, define
\[
D_n\equiv (n+1)^2-nk_f,
\qquad
a_n\equiv \frac{n+1-k_f}{D_n},
\qquad
b_n\equiv \frac{k_f}{D_n}.
\]
Introduce multipliers $\lambda,\nu$ for the constraints
\[
\sum_{t=1}^T w_t x_t=0,
\qquad
\sum_{t=1}^T w_t m_t=0.
\]
The deterministic first-order conditions reduce to
\[
\alpha+\beta\bigl(\mu_t-(n+1)x_t-m_t\bigr)=\lambda w_t,
\qquad
\alpha+\beta\left(\mu_t-nx_t-\frac{n+1}{k_f}m_t\right)=\nu w_t.
\]
Equivalently, writing $\tilde\lambda=\lambda/\beta$ and $\tilde\nu=\nu/\beta$,
\[
(n+1)x_t+m_t=\theta_t-\tilde\lambda w_t,
\qquad
nx_t+\frac{n+1}{k_f}m_t=\theta_t-\tilde\nu w_t.
\]
Solving this system yields
\[
x_t
=
\frac{(n+1-k_f)\theta_t-\bigl((n+1)\tilde\lambda-k_f\tilde\nu\bigr)w_t}{D_n},
\qquad
m_t
=
\frac{k_f\theta_t-k_f\bigl((n+1)\tilde\nu-n\tilde\lambda\bigr)w_t}{D_n}.
\]
Imposing the two weighted-balance constraints gives
\[
(n+1)\tilde\lambda-k_f\tilde\nu=(n+1-k_f)\bar\theta_w,
\qquad
(n+1)\tilde\nu-n\tilde\lambda=\bar\theta_w.
\]
The unique solution is
\[
\tilde\lambda=\tilde\nu=\bar\theta_w,
\]
and therefore
\[
x_t=a_n\Delta_t,
\qquad
m_t=b_n\Delta_t.
\]

For the stochastic part, symmetry gives $(n+1)v=z^{RT,CN}$, hence
\[
v_t=\frac{1}{n+1}z_t^{RT,CN}(D_{1:t}).
\]
Combining the deterministic and stochastic parts gives
\[
z_{b,t}^{DA,DCN}=a_n\Delta_t,
\qquad
z_{b,t}^{RT,DCN}(D_{1:t})
=
b_n\Delta_t+\frac{1}{n+1}z_t^{RT,CN}(D_{1:t}),
\]
which is the decentralized policy stated in the theorem.
This policy is feasible because $\Delta\in H_w$ and $z^{RT,CN}\in\mathcal H_w$.
Since it satisfies the first-order conditions and each best-response problem is strictly concave, an
equilibrium exists. The arguments above show that every equilibrium is symmetric and must solve the
same common first-order conditions, so the equilibrium is unique.

It remains to establish the PoA bounds. Define the aggregate cost functional
\[
\mathcal C(y,m,v)
\equiv
\sum_{t=1}^T
\left[
\alpha(\mu_t-y_t)+\frac{\beta}{2}(\mu_t-y_t)^2
+\frac{\beta}{2}\Bigl(\frac1{k_f}-1\Bigr)m_t^2
+\frac{\beta}{2k_f}\E[(\epsilon_t-v_t)^2]
\right].
\]
By the decomposition above, this is exactly the expected system cost generated by the separated variables
$(y,m,v)$. The no-battery benchmark corresponds to $(y,m,v)=(0,0,0)$, so
\[
\costnb=\mathcal C(0,0,0).
\]

For the centralized policy, we have $y_t=\Delta_t$, $m_t=0$, and $v_t=z_t^{RT,CN}$. Define
\[
R_\alpha\equiv \frac12\sum_{t=1}^T \Delta_t^2,
\qquad
S_\eta\equiv \sum_{t=1}^T \E[\epsilon_t z_t^{RT,CN}].
\]
Since $\theta_t=\Delta_t+w_t\bar\theta_w$ and $\sum_{t=1}^T w_t\Delta_t=0$, we have
\[
\sum_{t=1}^T \theta_t\Delta_t=\sum_{t=1}^T \Delta_t^2.
\]
Because $\theta_t=\mu_t+\alpha/\beta$, it follows that
\[
\sum_{t=1}^T \left[\alpha\Delta_t+\beta\mu_t\Delta_t-\frac{\beta}{2}\Delta_t^2\right]
=
\beta R_\alpha.
\]
Also, $z^{RT,CN}$ is the orthogonal projection of $\epsilon$ onto $\mathcal H_w$ and belongs to
$\mathcal H_w$, so
\[
\sum_{t=1}^T \E\bigl[(\epsilon_t-z_t^{RT,CN})z_t^{RT,CN}\bigr]=0,
\]
which implies
\[
S_\eta=\sum_{t=1}^T \E\bigl[(z_t^{RT,CN})^2\bigr].
\]
Therefore
\[
\costnb-\costcn
=
\beta\left(R_\alpha+\frac{S_\eta}{2k_f}\right).
\]

For the symmetric decentralized equilibrium, the aggregate separated variables are
\[
y_t^{DCN}
=
\sum_{b=1}^n \bigl(z_{b,t}^{DA,DCN}+\E[z_{b,t}^{RT,DCN}]\bigr)
=
n(a_n+b_n)\Delta_t
=
\frac{n(n+1)}{D_n}\Delta_t,
\]
\[
m_t^{DCN}
=
\sum_{b=1}^n \E[z_{b,t}^{RT,DCN}]
=
n b_n\Delta_t
=
\frac{nk_f}{D_n}\Delta_t,
\]
and
\[
v_t^{DCN}
=
\sum_{b=1}^n \bigl(z_{b,t}^{RT,DCN}-\E[z_{b,t}^{RT,DCN}]\bigr)
=
\frac{n}{n+1}z_t^{RT,CN}.
\]
Define
\[
\gamma_n\equiv \frac{n(n+1)}{D_n},
\qquad
\rho_n\equiv \frac{n}{n+1}.
\]
Then
\[
y^{DCN}=\gamma_n\Delta,
\qquad
m^{DCN}=\frac{nk_f}{D_n}\Delta,
\qquad
v^{DCN}=\rho_n z^{RT,CN}.
\]
Substituting into $\mathcal C(y,m,v)$ and subtracting from $\costnb$ yields
\[
\costnb-\costdcn
=
\beta\left(
c_n(k_f)R_\alpha
+
\frac{n(n+2)}{(n+1)^2}\frac{S_\eta}{2k_f}
\right),
\]
where
\[
c_n(k_f)
\equiv
\frac{n(n+2)(n+1)^2-n^2(2n+3)k_f+n^2k_f^2}
     {\bigl((n+1)^2-nk_f\bigr)^2}.
\]
Indeed, the deterministic coefficient is
\[
2\gamma_n-\gamma_n^2-\Bigl(\frac1{k_f}-1\Bigr)\left(\frac{nk_f}{D_n}\right)^2
=
c_n(k_f),
\]
and the stochastic coefficient is
\[
2\rho_n-\rho_n^2
=
\frac{n(n+2)}{(n+1)^2}.
\]

Let
\[
e_n\equiv \frac{n(n+2)}{(n+1)^2}.
\]
If $S_\eta=0$, then
\[
\poa=\frac{\costnb-\costcn}{\costnb-\costdcn}=\frac{1}{c_n(k_f)}.
\]
If $S_\eta>0$, define
\[
x\equiv \frac{2k_f R_\alpha}{S_\eta}\in[0,\infty).
\]
Then
\[
\poa=\frac{\costnb-\costcn}{\costnb-\costdcn}
=\frac{x+1}{c_n(k_f)x+e_n}.
\]
Differentiating with respect to $x$ gives
\[
\frac{d}{dx}\poa
=
\frac{e_n-c_n(k_f)}{(c_n(k_f)x+e_n)^2}.
\]
A direct calculation shows that
\[
c_n(k_f)-e_n
=
\frac{k_f n^2\bigl(k_f+n^2+2n+1\bigr)}
{(n+1)^2\bigl((n+1)^2-nk_f\bigr)^2}
\ge 0,
\]
so $\poa$ is decreasing in $x$. Hence
\[
\frac{1}{c_n(k_f)}
\le
\poa
\le
\frac{1}{e_n}
=
1+\frac{1}{n(n+2)}.
\]
Finally,
\[
c_n'(k_f)
=
\frac{n^2\bigl(k_f n+2k_f+n^2+2n+1\bigr)}
{\bigl((n+1)^2-nk_f\bigr)^3}
>0,
\]
so $c_n(k_f)$ is increasing in $k_f$, and therefore $1/c_n(k_f)$ is decreasing in $k_f$. Taking
the limit as the fast-generator share approaches one,
\[
\frac{1}{c_n(k_f)}
\ge
\lim_{\kappa\uparrow 1}\frac{1}{c_n(\kappa)}
=
1+\frac{1}{n(n+1)(n^2+n+2)}.
\]
Combining the last two displays yields
\[
1+\frac{1}{n(n+1)(n^2+n+2)}
\le
\poa
\le
1+\frac{1}{n(n+2)}.
\]
\end{proof}

\subsection{Virtual Bidders}\label{app:subsec:virtual-bidders}

\begin{theorem}[Virtual Bidders]\label{thm:sb-battery-competition}
Consider the virtual bidder model with $n$ batteries and $m$ virtual bidders. The game has a unique equilibrium, which is symmetric. In this equilibrium, each battery $b$'s DA and RT discharges in period $t$ are
\begin{align*}
    z_{b,t}^{DA,DCN} &= \frac{(n+m+1)-(m+1)k_f}{ (n+m+1)(n+1) - n k_f } (\mu_{t} - \bar{\mu})  \\
    z_{b,t}^{RT,DCN}(D_{1:t}) &= \frac{(m+1)k_f}{ (n+m+1)(n+1) - n k_f } (\mu_{t} - \bar{\mu}) + \frac{1}{(n+1)} z_{t}^{RT,CN}(D_{1:t}).
\end{align*}

Each virtual bidder $v$'s financial position in period $t$ is
\begin{align*}
    y_{v,t} &= \frac{n k_f}{ (n+m+1)(n+1) - n k_f} (\mu_{t} - \bar{\mu}).
\end{align*} 

The corresponding Price of Anarchy satisfies
\begin{align*}
    1 + \frac{(m+1)^2}{ n(n+m+1)(n^2+mn+n+2m+2) } \leq \poa \leq 1 + \frac{1}{n(n+2)} .
\end{align*}

\end{theorem}

\begin{proof}[Proof of Theorem~\ref{thm:sb-battery-competition}]
The proof parallels the proof of Theorem~\ref{thm:competition}, with the aggregate virtual bidder position added to the predictable DA--RT arbitrage equations.
Let $\epsilon_t:=D_t-\mu_t$. For battery $b$, write
\[
x_{b,t}:=z_{b,t}^{DA},\qquad
u_{b,t}:=z_{b,t}^{RT}(D_{1:t}),\qquad
r_{b,t}:=\bE[u_{b,t}],\qquad
\tilde u_{b,t}:=u_{b,t}-r_{b,t},
\]
and let $p_{b,t}:=x_{b,t}+r_{b,t}$ denote battery $b$'s expected physical dispatch. Define the
aggregates
\[
P_t:=\sum_{b=1}^n p_{b,t},\qquad
R_t:=\sum_{b=1}^n r_{b,t},\qquad
\widetilde U_t:=\sum_{b=1}^n \tilde u_{b,t},\qquad
Y_t:=\sum_{v=1}^m y_{v,t}.
\]
The separate DA and RT balance constraints imply that each $p_b$ and each $r_b$ sums to zero
over $t$, and that each $\tilde u_b$ has mean zero in every period and sums to zero pathwise.
The modified DA and RT demands are
\[
\tilde d_t^{DA}=\mu_t-P_t+R_t-Y_t,
\qquad
\tilde d_t^{RT}
=\mu_t-P_t+R_t-Y_t+\frac{\epsilon_t-R_t-\widetilde U_t+Y_t}{k_f}.
\]

First consider a virtual bidder $v$. Since
\[
\lambda_t^{DA}-\lambda_t^{RT}
=-\frac{\beta}{k_f}\bigl(\epsilon_t-R_t-\widetilde U_t+Y_t\bigr),
\]
the first-order condition for $y_{v,t}$ is
\[
Y_t+y_{v,t}=R_t,\qquad t=1,\ldots,T.
\]
Thus all virtual bidders choose the same position in each period and
\begin{equation}\label{eq:virtual-bidder-position-foc}
y_{v,t}=\frac{R_t}{m+1}.
\end{equation}

Now consider battery $b$. After dropping the intercept term, which is zero by balance, battery
$b$'s expected profit divided by $\beta$ can be written as
\[
\sum_{t=1}^T p_{b,t}(\mu_t-P_t+R_t-Y_t)
+\frac1{k_f}\sum_{t=1}^T r_{b,t}(Y_t-R_t)
+\frac1{k_f}\bE\left[\sum_{t=1}^T
\tilde u_{b,t}(\epsilon_t-\widetilde U_t)\right].
\]
The first-order condition for $p_{b,t}$, projected onto the zero-sum subspace, is
\begin{equation}\label{eq:virtual-p-foc}
\mu_t-\bar\mu-P_t+R_t-Y_t-p_{b,t}=0.
\end{equation}
The first-order condition for $r_{b,t}$ is
\begin{equation}\label{eq:virtual-r-foc}
k_f p_{b,t}+Y_t-R_t-r_{b,t}=0.
\end{equation}
Equations \eqref{eq:virtual-p-foc} and \eqref{eq:virtual-r-foc} imply that $p_{b,t}$ and
$r_{b,t}$ are the same for every battery $b$ in each period $t$. Let these common values be
$p_t$ and $r_t$. Then $P_t=np_t$, $R_t=nr_t$, and, by
\eqref{eq:virtual-bidder-position-foc}, each virtual bidder chooses
$y_{v,t}=nr_t/(m+1)$. Substituting these identities into
\eqref{eq:virtual-p-foc}--\eqref{eq:virtual-r-foc} gives
\[
\mu_t-\bar\mu-(n+1)p_t+\frac{n}{m+1}r_t=0,
\qquad
k_f p_t-\frac{n+m+1}{m+1}r_t=0.
\]
Solving these two equations, with
\[
D_m:=(n+m+1)(n+1)-nk_f,
\]
yields
\[
p_t=\frac{n+m+1}{D_m}(\mu_t-\bar\mu),
\qquad
r_t=\frac{(m+1)k_f}{D_m}(\mu_t-\bar\mu).
\]
Therefore
\[
z_{b,t}^{DA,DCN}=x_{b,t}=p_t-r_t
=\frac{(n+m+1)-(m+1)k_f}{D_m}(\mu_t-\bar\mu),
\]
and
\[
y_{v,t}=\frac{nr_t}{m+1}
=\frac{nk_f}{D_m}(\mu_t-\bar\mu).
\]

It remains to characterize the centered real-time response. Let
\[
\mathcal H
:=\left\{
h:\ h_t\text{ is adapted to }D_{1:t},\ \bE[h_t]=0,\ 
\sum_{t=1}^T h_t=0\ \text{pathwise}
\right\}.
\]
Using the inner product $\langle a,b\rangle:=\sum_{t=1}^T\bE[a_tb_t]$, the centered part of
battery $b$'s payoff is $k_f^{-1}\langle \tilde u_b,\epsilon-\widetilde U\rangle$. Its
first-order condition is
\[
\left\langle \epsilon-\widetilde U-\tilde u_b,\ h\right\rangle=0
\qquad\text{for all }h\in\mathcal H.
\]
Because $\widetilde U+\tilde u_b\in\mathcal H$, this condition says that
$\widetilde U+\tilde u_b$ is the orthogonal projection of $\epsilon$ onto $\mathcal H$.
By the centralized analysis in the proof of Theorem~\ref{thm:fb-battery}, this projection is
$z^{RT,CN}$. Thus, for every battery $b$,
\[
\widetilde U+\tilde u_b=z^{RT,CN}.
\]
Comparing this equation across any two batteries shows that the centered real-time responses
are the same for all batteries. Writing the common response as $\tilde u$, we have
$\widetilde U=n\tilde u$ and therefore
\[
\tilde u_{b,t}=\frac{1}{n+1}z_t^{RT,CN}(D_{1:t}),
\]
which gives
\[
z_{b,t}^{RT,DCN}(D_{1:t})
=r_t+\tilde u_{b,t}
=\frac{(m+1)k_f}{D_m}(\mu_t-\bar\mu)
+\frac{1}{n+1}z_t^{RT,CN}(D_{1:t}).
\]
This proves the equilibrium formulas.
The displayed battery and virtual-bidder policies satisfy the balance constraints and the
first-order conditions. Each battery's payoff is concave in its own $(p_b,r_b,\tilde u_b)$,
and each virtual bidder's payoff is concave in its own position, so these first-order
conditions are sufficient for best responses and an equilibrium exists. Conversely, any
equilibrium must satisfy the same first-order conditions; the arguments above force the
displayed policies, so the equilibrium is unique and symmetric.

We next compute the cost and PoA. Recall from the proof of Theorem~\ref{thm:cost-comparison} that
\begin{align*}
    V &= \sum_{t=1}^{T} \Var(D_t) \\
    S &= \sum_{t=1}^{T} \bE\left[ (D_t-\mu_t) z_{t}^{RT,CN} \right] = \sum_{t=1}^{T} \bE\left[ \left( z_{t}^{RT,CN} \right)^2 \right].
\end{align*}
The equilibrium formulas imply the aggregate modified demands
\begin{align*}
    \tilde{d}_t^{DA}
    &= \bar{\mu} + \frac{n+m+1}{D_m}(\mu_t-\bar{\mu}), \\
    \tilde{d}_t^{RT}
    &= \bar{\mu} + \frac{m+1}{D_m}(\mu_t-\bar{\mu})
    + \frac{1}{k_f}\left(D_t-\mu_t-\frac{n}{n+1}z_t^{RT,CN}\right).
\end{align*}
Define $\sigma_\mu^2:=T^{-1}\sum_{t=1}^T(\mu_t-\bar\mu)^2$.
Substituting these expressions into \eqref{eqn:gen-cost-linear} gives
\begin{align*}
\costdcn
=
\alpha\sum_{t=1}^T\mu_t
+\beta\left[
\frac12T\bar\mu^2
+\frac{(n+m+1)^2-n(n+2m+2)k_f}{2D_m^2}T\sigma_\mu^2
+\frac{1}{2k_f}\left(V-\frac{n(n+2)}{(n+1)^2}S\right)
\right].
\end{align*}
The only term in $\costdcn$ that depends on $m$ is the coefficient of $T\sigma_\mu^2$, and
\begin{align*}
\frac{\partial}{\partial m}
\frac{(n+m+1)^2-n(n+2m+2)k_f}{2D_m^2}
=
\frac{k_f(m+k_f)n^2}{D_m^3}
\ge 0.
\end{align*}
Thus $\costdcn$ is increasing in the number of virtual bidders. Since $\costnb$ and $\costcn$
do not depend on $m$, this also implies that $\poa=(\costnb-\costcn)/(\costnb-\costdcn)$ is
increasing in $m$ whenever the denominator is positive.

Using the no-battery and centralized costs from Theorem~\ref{thm:cost-comparison},
\[
\costnb-\costcn
=
\beta\left[\frac12T\sigma_\mu^2+\frac{1}{2k_f}S\right],
\]
while
\begin{align*}
\costnb-\costdcn
=
\beta\left[
\frac{H_m(k_f)}{2D_m^2}T\sigma_\mu^2
+\frac{1}{2k_f}\frac{n(n+2)}{(n+1)^2}S
\right],
\end{align*}
where
\[
H_m(k_f)
:=
n(n+2)(n+m+1)^2
-n^2(2n+2m+3)k_f
+n^2k_f^2.
\]
Therefore $\poa$ is a weighted average of the two componentwise ratios
\[
R_m(k_f):=\frac{D_m^2}{H_m(k_f)}
\qquad\text{and}\qquad
R_S:=\frac{(n+1)^2}{n(n+2)}=1+\frac{1}{n(n+2)}.
\]
The second ratio is constant. The first ratio is decreasing in $k_f$ because
\begin{align*}
\frac{\partial R_m(k_f)}{\partial k_f}
=
-\frac{n^2\left(n^2+m+1+(2+m-k_f)n \right)
\left((n+m+1)(n+2m+1)+(n+2m+2)k_f\right)}
{H_m(k_f)^2}
\le 0.
\end{align*}
Moreover,
\[
R_m(0)=1+\frac{1}{n(n+2)},
\qquad
R_m(1)=1+\frac{(m+1)^2}{n(n+m+1)(n^2+mn+n+2m+2)}.
\]
Since $R_S=R_m(0)$, the weighted-average representation gives
\begin{align*}
    1 + \frac{(m+1)^2}{ n(n+m+1)(n^2+mn+n+2m+2) } \leq \poa \leq 1 + \frac{1}{n(n+2)} .
\end{align*}
The lower-bound term is increasing in $m$ because
\[
\frac{\partial}{\partial m}
\frac{(m+1)^2}{n(n+m+1)(n^2+mn+n+2m+2)}
=
\frac{2(m+1)(mn+m+n^2+n+1)}
{(m+n+1)^2(mn+2m+n^2+n+2)^2}
\ge 0,
\]
and it converges to $1/[n(n+2)]$ as $m\to\infty$.

Finally, the equilibrium formulas imply
\[
z_{b,t}^{DA,DCN}+\bE[z_{b,t}^{RT,DCN}]
=
\frac{n+m+1}{D_m}(\mu_t-\bar\mu),
\]
and the derivative of the coefficient with respect to $m$ is $-nk_f/D_m^2\le0$. This proves
the claimed monotonicity of each battery's expected physical dispatch and completes the proof.
    
\end{proof}

\section{Theorem Statements and Proofs for Appendix~\ref{app:sec:more-extensions}}\label{app:sec:theorem-more-extensions}

\subsection{Joint Day-Ahead and Real-Time Energy Balance}\label{app:subsec:joint-balance}

\begin{theorem}[Joint balance is without loss in the linear model]\label{thm:joint-balance}
Impose \eqref{eq:joint-balance} in the model with $n$ competing batteries. The centralized aggregate optimum is the same as in the baseline model and can be implemented by policies satisfying the separate balance constraints in \eqref{eqn:da-rt-balance} battery by battery. Every decentralized equilibrium satisfies these separate balance constraints battery by battery. Consequently, the decentralized equilibrium, generation costs, and PoA bounds are the same as in Theorem~\ref{thm:competition}.
\end{theorem}

\begin{proof}[Proof of Theorem~\ref{thm:joint-balance}]
Fix any feasible policy profile under \eqref{eq:joint-balance}. For each battery $b$, let
\[
a_b:=\frac{1}{T}\sum_{t=1}^{T}z_{b,t}^{DA},
\qquad
A:=\sum_{b=1}^n a_b
\]
be the battery's average day-ahead position and the aggregate average day-ahead position. Shift each battery's average position from day ahead to real time by setting
\[
\tilde z_{b,t}^{DA}:=z_{b,t}^{DA}-a_b,
\qquad
\tilde z_{b,t}^{RT}(D_{1:t}):=z_{b,t}^{RT}(D_{1:t})+a_b.
\]
The shifted profile has the same realized physical dispatch for every battery because
$\tilde z_{b,t}^{DA}+\tilde z_{b,t}^{RT}\equiv z_{b,t}^{DA}+z_{b,t}^{RT}$ for every $b,t$, and every history.
It also satisfies separate DA and RT balance battery by battery: $\sum_t\tilde z_{b,t}^{DA}=0$, and \eqref{eq:joint-balance} implies $\sum_t\tilde z_{b,t}^{RT}(D_{1:t})=0$ for every $b$ and every demand path.

For the centralized planner, a direct expansion of the linear generation cost in \eqref{eqn:gen-cost-linear} gives
\[
\cost(z^{DA},z^{RT})-\cost(\tilde z^{DA},\tilde z^{RT})
=
\frac{\beta T}{2}\cdot \frac{k_s}{k_f}\,A^2.
\]
Thus any centralized optimum must have $A=0$ at the aggregate level; and when $A=0$, the shifted separate-balance implementation has exactly the same aggregate DA and RT quantities and therefore the same cost. Hence the centralized aggregate optimum is unchanged and can be implemented with separate DA and RT balance.

It remains to show that decentralized equilibria also eliminate the settlement-shift degree of freedom. Fix the physical dispatches and the settlement shifts of the other batteries, and let battery $b$ choose an average day-ahead position $r$ while keeping its physical dispatch fixed. If $A_{-b}:=\sum_{j\ne b}a_j$, then a direct expansion of battery $b$'s expected profit gives, up to terms independent of $r$,
\[
\Pi_b(r)
=
\textnormal{constant}
-\frac{\beta T}{k_f}\bigl(r^2+A_{-b}r\bigr).
\]
This scalar objective is strictly concave, so at equilibrium
\[
2a_b+A_{-b}=0.
\]
Equivalently, $a_b+A=0$ for every $b$, where $A=\sum_j a_j$. Summing over $b$ gives $(n+1)A=0$, so $A=0$ and then $a_b=0$ for every battery $b$. The joint balance condition then implies $\sum_t z_{b,t}^{RT}(D_{1:t})=0$ for every $b$ and every demand path. Therefore every decentralized equilibrium under joint balance lies in the baseline separate-balance feasible set. The baseline competition result, Theorem~\ref{thm:competition}, applies directly.
\end{proof}

\subsection{Ramping Costs}\label{app:subsec:ramping-costs}

\begin{theorem}[Ramping Costs]\label{thm:ramping-costs}
Consider $n\ge1$ competing batteries in the ramping-cost extension of
Appendix~\ref{subsec:ramping-costs}. The centralized day-ahead action remains
\[
z^{DA,CN}=\mu-\bar\mu\,\mathbf 1.
\]
The game has a unique equilibrium, which is symmetric. The corresponding Price of Anarchy satisfies
\[
1+\frac{1}{n(n+1)(n^2+n+2)}
\le
\poa
\le
1+\frac{1}{n(n+2)}.
\]
\end{theorem}

\begin{proof}[Proof of Theorem~\ref{thm:ramping-costs}]
Let $\varepsilon:=D-\mu$. Let $\Delta$ be the cyclic difference operator,
$(\Delta q)_t=q_{t+1}-q_t$ with $q_{T+1}=q_1$, and let $L:=\Delta^\top\Delta$ be the cycle
Laplacian. Then the ramping term is $\frac{c}{2}\,\bE[q^\top Lq]$.

First write the centralized problem in terms of deterministic day-ahead net demand $d^{DA}:=\mu-z^{DA}$ and real-time residual $r:=\varepsilon-z^{RT}$, so physical generation is $q=d^{DA}+r$. Up to constants independent of the decision, the objective is
\[
J(d^{DA},r)
=\frac{\beta}{2}\Big(\|d^{DA}\|^2+\frac{1}{k_f}\,\bE\|r\|^2+2(d^{DA})^\top \bE[r]\Big)
+\frac{c}{2}\,\bE\big[(d^{DA}+r)^\top L(d^{DA}+r)\big].
\]
For any feasible pair, let $m:=\bE[r]$. Because $\bE[\varepsilon]=0$ and the real-time
battery action is pathwise balanced, $\mathbf 1^\top m=0$. Define
$d^{DA\prime}:=d^{DA}+m$ and $r':=r-m$. Then $d^{DA\prime}+r'=q$ pathwise, so the ramping
term is unchanged. The day-ahead action induced by $d^{DA\prime}$ remains balanced, and the
real-time action induced by $r'$ remains adapted and pathwise balanced. Moreover,
\[
J(d^{DA},r)-J(d^{DA\prime},r')
=\frac{\beta}{2}\Big(\frac{1}{k_f}-1\Big)\|m\|^2\ge 0.
\]
Hence an optimum can be chosen with $\bE[r]=0$. The day-ahead problem therefore reduces to
\[
\min_{d\in\mathbb R^T}\ \frac12 d^\top(\beta I+cL)d
\qquad\text{s.t.}\qquad \mathbf 1^\top d=\mathbf 1^\top\mu.
\]
Because $(\beta I+cL)\mathbf 1=\beta\mathbf 1$, the unique minimizer is $d=\bar\mu\,\mathbf 1$, which is equivalent to $z^{DA,CN}=\mu-\bar\mu\,\mathbf 1$.

For the real-time problem, decompose every feasible real-time action into its deterministic mean and mean-zero stochastic parts. Because the stochastic parts have zero expectation, the expected quadratic objective separates additively into a deterministic mean game and a mean-zero stochastic game. We analyze these two pieces in turn.

For the stochastic game, let $\mathcal A$ denote the linear space of adapted processes $z$ with $\bE[z]=0$ and $\mathbf 1^\top z=0$ pathwise. Define the weighted inner product
\[
\langle x,y\rangle_Q:=\bE[x^\top Qy],
\qquad
Q:=\frac{\beta}{k_f}I+cL\succ 0,
\]
and write $\|x\|_Q^2:=\langle x,x\rangle_Q$. After the centralized day-ahead action has been characterized, the centralized stochastic welfare gain from an aggregate real-time battery action $z\in\mathcal A$ is
\[
\Delta^{CN}_{\mathrm{stoch}}(z)
=\langle \varepsilon,z\rangle_Q-\frac12\|z\|_Q^2.
\]
Therefore $Z_{\mathrm{stoch}}^{RT,CN}$ is the $Q$-orthogonal projection of $\varepsilon$ onto $\mathcal A$, characterized by
\[
\langle \varepsilon-Z_{\mathrm{stoch}}^{RT,CN},h\rangle_Q=0
\qquad \forall h\in\mathcal A.
\]
In particular,
\[
\Delta^{CN}_{\mathrm{stoch}}
=\frac12\big\|Z_{\mathrm{stoch}}^{RT,CN}\big\|_Q^2.
\]

Now consider an arbitrary equilibrium of the $n$-battery stochastic game. Let
$z_b\in\mathcal A$ be battery $b$'s mean-zero stochastic real-time action, let
$Z:=\sum_{b=1}^n z_b$, and let $z_{-b}:=\sum_{j\ne b}z_j$. Up to constants independent of
$z_b$, battery $b$'s stochastic profit is
\[
\pi_b(z_b;z_{-b})
=\langle \varepsilon-z_{-b},z_b\rangle_Q-\|z_b\|_Q^2.
\]
This objective is strictly concave in $z_b$, so the first-order condition is necessary and
sufficient for a best response. At any equilibrium,
\[
\langle \varepsilon-Z-z_b,h\rangle_Q=0
\qquad \forall h\in\mathcal A.
\]
Since $Z+z_b\in\mathcal A$, this says that $Z+z_b$ is the $Q$-orthogonal projection of
$\varepsilon$ onto $\mathcal A$. The projection is unique and equals
$Z_{\mathrm{stoch}}^{RT,CN}$, so
\[
Z+z_b=Z_{\mathrm{stoch}}^{RT,CN}
\qquad \text{for every battery }b.
\]
Comparing two batteries gives $z_b=z_c$ for all $b,c$. Writing the common action as $x$, we
have $Z=nx$, and therefore
\[
x=\frac{1}{n+1}Z_{\mathrm{stoch}}^{RT,CN},
\qquad
Z_{\mathrm{stoch}}^{RT,DCN}=nx=\frac{n}{n+1}Z_{\mathrm{stoch}}^{RT,CN},
\]
which proves existence, uniqueness, and symmetry for the stochastic equilibrium and gives the
stochastic-response scaling. Substituting this scaling into the quadratic gain formula yields
\[
\Delta^{DCN}_{\mathrm{stoch}}
=\Big(\frac{n}{n+1}-\frac12\Big(\frac{n}{n+1}\Big)^2\Big)
\big\|Z_{\mathrm{stoch}}^{RT,CN}\big\|_Q^2
=\frac{n(n+2)}{2(n+1)^2}\big\|Z_{\mathrm{stoch}}^{RT,CN}\big\|_Q^2,
\]
whereas
\[
\Delta^{CN}_{\mathrm{stoch}}
=\frac12\big\|Z_{\mathrm{stoch}}^{RT,CN}\big\|_Q^2.
\]
Therefore
\[
\Delta^{CN}_{\mathrm{stoch}}
=\frac{(n+1)^2}{n(n+2)}\,\Delta^{DCN}_{\mathrm{stoch}}
=\Bigl(1+\frac{1}{n(n+2)}\Bigr)\Delta^{DCN}_{\mathrm{stoch}},
\]
which remains valid even when both stochastic gains are zero.

For the deterministic mean part, restrict attention to the zero-sum subspace $\mathcal S:=\{x\in\mathbb R^T:\mathbf 1^\top x=0\}$. Let $u^{(0)},\dots,u^{(T-1)}$ be an orthonormal eigenbasis of $L$ with eigenvalues $0=\lambda_0<\lambda_1\le\cdots\le\lambda_{T-1}$. The operators
\[
H_0:=\beta I+cL,
\qquad
H_1:=\frac{\beta}{k_f}I+cL
\]
are both polynomials in $L$, so they are simultaneously diagonalized by this basis. On a
nonconstant mode $u^{(j)}$, write the predictable-demand coefficient as $\mu_j$, and let
$p_{b,j}$ and $r_{b,j}$ be battery $b$'s coefficients of expected physical dispatch and
expected real-time dispatch, respectively. Define the corresponding aggregate coefficients
$P_j:=\sum_b p_{b,j}$ and $R_j:=\sum_b r_{b,j}$. The deterministic cost-reduction term on
this mode is, up to the positive factor $\beta+c\lambda_j$,
\[
\mu_jP_j-\frac12 P_j^2
-\frac12\left(\frac{1}{\kappa_j}-1\right)R_j^2,
\qquad
\kappa_j:=\frac{\beta+c\lambda_j}{\beta/k_f+c\lambda_j}\in(0,1].
\]
Thus the eigenmode decomposition changes only the relative curvature of the predictable physical
dispatch and expected real-time terms. Up to the same positive factor, battery $b$'s payoff on
this mode is
\[
p_{b,j}(\mu_j-P_j+R_j)-\frac{1}{\kappa_j}r_{b,j}R_j.
\]
Its first-order conditions are
\[
\mu_j-P_j+R_j-p_{b,j}=0,
\qquad
\kappa_jp_{b,j}-R_j-r_{b,j}=0.
\]
Comparing these conditions across batteries shows that the equilibrium coefficients are equal.
Writing their common values as $p_j$ and $r_j$, we obtain
\[
\mu_j-(n+1)p_j+nr_j=0,
\qquad
\kappa_jp_j-(n+1)r_j=0.
\]
Therefore, with $D_n(\kappa):=(n+1)^2-n\kappa$,
\[
p_j=\frac{n+1}{D_n(\kappa_j)}\mu_j,
\qquad
r_j=\frac{\kappa_j}{D_n(\kappa_j)}\mu_j.
\]
Each battery's mode payoff is strictly concave in its own pair $(p_{b,j},r_{b,j})$, so these
conditions give the unique equilibrium on that mode. Define
\[
H_n(\kappa)
:=n(n+2)(n+1)^2-n^2(2n+3)\kappa+n^2\kappa^2.
\]
Substitution into the cost-reduction expression gives
\[
\Delta^{CN}_{j,\mathrm{mean}}
=\frac{\beta+c\lambda_j}{2}\mu_j^2,
\qquad
\Delta^{DCN}_{j,\mathrm{mean}}
=\frac{\beta+c\lambda_j}{2}
\frac{H_n(\kappa_j)}{D_n(\kappa_j)^2}\mu_j^2.
\]
When $\mu_j\ne0$, the ratio of the two gains is
$D_n(\kappa_j)^2/H_n(\kappa_j)$; when $\mu_j=0$, both gains are zero. The bounds for this
ratio established in the proof of Theorem~\ref{thm:competition} therefore give, for every mode,
\[
\Big(1+\frac{1}{n(n+1)(n^2+n+2)}\Big)\Delta^{DCN}_{j,\mathrm{mean}}
\le
\Delta^{CN}_{j,\mathrm{mean}}
\le
\Big(1+\frac{1}{n(n+2)}\Big)\Delta^{DCN}_{j,\mathrm{mean}}.
\]
Summing over modes yields
\[
\Big(1+\frac{1}{n(n+1)(n^2+n+2)}\Big)\Delta^{DCN}_{\mathrm{mean}}
\le
\Delta^{CN}_{\mathrm{mean}}
\le
\Big(1+\frac{1}{n(n+2)}\Big)\Delta^{DCN}_{\mathrm{mean}}.
\]
Because the full game decomposes additively into the independent mean and stochastic pieces analyzed above, combining the unique equilibria of those pieces gives the unique equilibrium of the full ramping game, which is symmetric. Since PoA is the ratio of centralized cost reduction to decentralized cost reduction, adding the mean and stochastic inequalities gives
\[
1+\frac{1}{n(n+1)(n^2+n+2)}\le \poa \le 1+\frac{1}{n(n+2)},
\]
which proves the PoA bounds.
\end{proof}

\subsection{Strategic Generators}\label{app:subsec:strategic-generators}

\begin{theorem}[Strategic Generators]\label{thm:strategic-generators}
Consider the strategic-generator model described in Appendix~\ref{subsec:strategic-generators}.
The game has a unique pure-strategy Nash equilibrium. Write
$\bar\mu:=T^{-1}\sum_t\mu_t$. At the equilibrium, the battery schedule is
\[
z_t^*=\frac{\mu_t-\bar\mu}{2}.
\]
Thus the battery half-smooths demand exactly as in the truthful-generator benchmark.
Moreover, every generator inflates its bid, $\hat c_i^*>c_i$, and true generation cost is
weakly higher than under truthful generator bidding:
\[
\costsg\ge \costdcn.
\]
\end{theorem}

\begin{proof}[Proof of Theorem~\ref{thm:strategic-generators}]
Fix bids, and write $h_i:=1/\hat c_i$ and $H:=\sum_i h_i$. If net demand in period $t$ is
$D_t$, market clearing solves
\[
\min_{x_{1,t},\dots,x_{n_g,t}}\ \sum_{i=1}^{n_g}\hat c_i x_{i,t}^2
\quad\text{s.t.}\quad \sum_{i=1}^{n_g}x_{i,t}=D_t.
\]
The first-order conditions give $2\hat c_i x_{i,t}=p_t$ for every $i$. Hence
\[
x_{i,t}=\frac{h_i}{H}D_t,
\qquad
p_t=\frac{2D_t}{H}.
\]

Given bids, the battery chooses $(z_t)$ with $\sum_t z_t=0$ to maximize
\[
\sum_{t=1}^T p_tz_t
=\frac{2}{H}\sum_{t=1}^T(\mu_t-z_t)z_t.
\]
The factor $2/H$ is positive and independent of the battery schedule, so the battery solves
\[
\max_{\sum_t z_t=0}\ \sum_{t=1}^T(\mu_tz_t-z_t^2).
\]
This is a strictly concave quadratic problem. The first-order conditions are
$\mu_t-2z_t-\eta=0$, where $\eta$ is the multiplier on $\sum_t z_t=0$. Imposing the balance
constraint gives $\eta=\bar\mu$, and therefore
\[
z_t^*=\frac{\mu_t-\bar\mu}{2},
\qquad
D_t^*=\frac{\mu_t+\bar\mu}{2}.
\]
Because $\mu_t>0$ for every $t$, we also have $D_t^*>0$ for every $t$.

It remains to characterize generator bids. Fix the battery schedule and the other generators'
bids. Let $H_{-i}:=\sum_{j\ne i}h_j$. Generator $i$'s true profit, as a function of its own
bid slope $h_i$, is
\[
\Pi_i(h_i)
=\sum_{t=1}^T
\frac{D_t^2}{(h_i+H_{-i})^2}\bigl(2h_i-c_i h_i^2\bigr)
=K\,\frac{2h_i-c_i h_i^2}{(h_i+H_{-i})^2},
\]
where $K:=\sum_t D_t^2$ is independent of $h_i$. Differentiating the last expression gives
\[
\frac{d}{dh_i}
\left[
\frac{2h_i-c_i h_i^2}{(h_i+H_{-i})^2}
\right]
=
\frac{2\bigl(H_{-i}-(1+c_iH_{-i})h_i\bigr)}
{(h_i+H_{-i})^3}.
\]
Thus generator $i$'s unique best response is
\[
h_i=\frac{H_{-i}}{1+c_iH_{-i}}.
\]
Since $H_{-i}>0$ in an interior equilibrium, $c_i h_i<1$, and therefore
$\hat c_i=1/h_i>c_i$.

In equilibrium, $H= h_i+H_{-i}$, so the best-response condition can be written as
\[
h_i=\frac{H-h_i}{1+c_i(H-h_i)}.
\]
Equivalently,
\[
c_i h_i^2-(c_iH+2)h_i+H=0.
\]
The root in $(0,1/c_i)$ is
\[
h_i(H):=
\frac{(c_i H+2)-\sqrt{(c_i H+2)^2-4c_i H}}{2c_i}.
\]
For $H>0$, rationalizing the numerator gives
\[
\frac{h_i(H)}{H}
=
\frac{2}{c_iH+2+\sqrt{c_i^2H^2+4}}.
\]
This ratio is continuous and strictly decreasing in $H$, with
\[
\lim_{H\downarrow0}\frac{h_i(H)}{H}=\frac12,
\qquad
\lim_{H\to\infty}\frac{h_i(H)}{H}=0.
\]
Since $n_g\ge3$, the function
\[
\Phi(H):=\sum_{i=1}^{n_g}\frac{h_i(H)}{H}
\]
is continuous and strictly decreasing from $n_g/2>1$ to $0$. There is therefore a unique
$H^*>0$ such that $\Phi(H^*)=1$. Setting $h_i^*:=h_i(H^*)$ gives
$H^*=\sum_i h_i^*$, and the defining quadratic equation implies
\[
h_i^*=\frac{H^*-h_i^*}{1+c_i(H^*-h_i^*)},
\]
so every $h_i^*$ is the unique best response to the other generators. Conversely, every
equilibrium must satisfy the same scalar equation $\Phi(H)=1$. The generator-bidding
equilibrium is therefore unique. Together with the battery's unique best response derived
above, this proves existence and uniqueness of the pure-strategy Nash equilibrium. Moreover,
$D_t^*>0$ and $h_i^*>0$, so every generator produces a positive amount in every period.

Under strategic generator bidding, equilibrium dispatch weights are $w_i^*=h_i^*/H^*$, so
$x_{i,t}^*=w_i^*D_t^*$. Hence
\[
\costsg
=\sum_{t=1}^T\sum_{i=1}^{n_g} c_i(x_{i,t}^*)^2
=\Bigl(\sum_{t=1}^T(D_t^*)^2\Bigr)\sum_{i=1}^{n_g}c_i(w_i^*)^2.
\]
Under truthful generator bidding, dispatch is true-cost minimizing in each period:
\[
x_{i,t}^{DCN}=\frac{(1/c_i)}{\sum_j(1/c_j)}D_t^*,
\qquad
\sum_{i=1}^{n_g}c_i(x_{i,t}^{DCN})^2
=\frac{(D_t^*)^2}{\sum_i(1/c_i)}.
\]
Thus
\[
\costdcn=\sum_{t=1}^T \frac{(D_t^*)^2}{\sum_i(1/c_i)}.
\]
Finally, for any weights $w_i\ge0$ with $\sum_iw_i=1$, Cauchy--Schwarz gives
\[
\left(\sum_{i=1}^{n_g}c_iw_i^2\right)
\left(\sum_{i=1}^{n_g}\frac1{c_i}\right)
\ge
\left(\sum_{i=1}^{n_g}w_i\right)^2
=1.
\]
Applying this inequality to $w_i=w_i^*$ yields
$\costsg\ge \costdcn$, with equality if and only if
$w_i^*\propto 1/c_i$.
\end{proof}

\end{APPENDICES}

\end{document}